\documentclass[prx,twocolumn,amssymb,floatfix,superscriptaddress,notitlepage]{revtex4-2}
\usepackage{float}

\usepackage{amsmath}
\usepackage{amstext}
\usepackage{amssymb}
\usepackage{amsfonts}
\usepackage{graphicx}
\usepackage{fancybox}

\usepackage{xfrac}
\usepackage{physics}
\usepackage{adjustbox}
\usepackage{color}
\usepackage{subfigure}
\usepackage{ulem}
\usepackage{caption}  
\usepackage{subcaption}

\usepackage[colorlinks,citecolor=blue]{hyperref}

\usepackage{url}
\usepackage{tikz}
\usetikzlibrary{calc}
\usetikzlibrary{decorations.markings}

\usepackage{subfigure}
\usepackage{subcaption}
\usepackage{dsfont}
\usepackage{booktabs}
\usepackage{amsbsy}
\usepackage{dcolumn}
\usepackage{amsthm}
\usepackage{bm}
\usepackage{esint}
\usepackage{multirow}
\usepackage{hyperref}
\hypersetup{
    colorlinks=true,
    linkcolor=magenta,
    filecolor=magenta,
    urlcolor=magenta,
}
\usepackage{cleveref}
\usepackage{ulem}

\usepackage{mathrsfs}
\usepackage{amsfonts}
\usepackage{amsbsy}
\usepackage{dcolumn}
\usepackage{bm}
\usepackage{multirow}
\usepackage{color}

 \renewcommand{\vec}[1]{{\mathbf #1}}
\newcommand{\comments}[1]{}
\def\Z{\mathbb{Z}}

  \usepackage{extarrows}
\usepackage{datetime}

\usepackage{comment}
\usepackage[super]{nth}

 \usepackage{varwidth}

\usepackage{adjustbox}

\newcommand{\bpm}{\begin{pmatrix}}
\newcommand{\epm}{\end{pmatrix}}

\def\Z{\mathbb{Z}}

\begin{document}
\title{Classification of Interacting Topological Crystalline Superconductors in Three Dimensions and Beyond}

\author{Shang-Qiang Ning}
\thanks{These authors contribute equally.}
\affiliation{Department of Physics, The Hong Kong University of Science and Technology, Clear Water Bay, Kowloon, Hong Kong, China}
\affiliation{Department of Physics, The Chinese University of Hong Kong, Shatin, New Territories, Hong Kong, China}

\author{Xing-Yu Ren}
\thanks{These authors contribute equally.}
\affiliation{Department of Physics, The Chinese University of Hong Kong, Shatin, New Territories, Hong Kong, China}

\author{Qing-Rui Wang}
\email{wangqr@mail.tsinghua.edu.cn}
\affiliation{Yau Mathematical Sciences Center, Tsinghua University, Haidian, Beijing 100084, China}

\author{Yang Qi}
\email{qiyang@fudan.edu.cn}
\affiliation{Center for Field Theory and Particle Physics, Department of Physics, Fudan University, Shanghai 200433, China}
\affiliation{State Key Laboratory of Surface Physics, Fudan University, Shanghai 200433, China}

\author{Zheng-Cheng Gu}
\email{zcgu@phy.cuhk.edu.hk}
\affiliation{Department of Physics, The Chinese University of Hong Kong, Shatin, New Territories, Hong Kong, China}

\begin{abstract}

Although classification for free-fermion topological superconductors (TSC) is established, systematically understanding the classification of 3D interacting TSCs remains difficult, especially those protected by crystalline symmetries like the 230 space groups. We build up a general framework for systematically classifying 3D interacting TSCs protected by crystalline symmetries together with discrete internal symmetries. We first establish a complete classification for fermionic symmetry protected topological phases (FSPT) with purely discrete internal symmetries, which determines the crystalline case via the crystalline equivalence principle. Using domain wall decoration, we obtain classification data and formulas for generic FSPTs, what are suitable for systematic computation. The four layers of decoration data $(n_1, n_2, n_3, \nu_4)$ characterize a 3D FSPT with symmetry $G_b\times_{\omega_2}\Z_2^f$, corresponding to $p+ip$, Kitaev chain, complex fermion, and bosonic SPT layers. Inspired by previous works, a crucial aspect is the $p+ip$ layer, where classification involves two possibilities: anti-unitary and infinite-order symmetries (e.g., translation). We show the former maps to some mirror FSPT classification with the mirror plane decorated by a $p+ip$ superconductor, while the latter is determined by the free part of $H^1(G_b, \Z_T)$, corresponding to weak TSCs. Another key point is the Kitaev chain decoration for anti-unitary symmetries, which differs essentially from unitary ones. We explicitly obtain formulas for all three layers of decoration $(n_2, n_3, \nu_4)$, which are amenable to automatic computation. As an application, we classify the 230 space-group topological crystalline superconductors in interacting electronic systems.  
  \end{abstract}

\date{\today}
\maketitle
\tableofcontents

\section{Introduction}
\subsection{The goal of this paper}

Symmetry-protected topological (SPT) phases \cite{gu09,chenScience2012,chen13} have emerged as a cornerstone in the landscape of quantum phases of matter, representing a paradigm of interplay between topology and symmetry in quantum matter. While these phases are short-range entangled in the bulk, they manifest non-trivial, symmetry-anomaly-driven features on the boundary. This framework has successfully categorized celebrated examples such as topological insulators \cite{hasan10,qi11} and the Haldane phase \cite{haldane83}. Rigorously, SPT phases are defined as equivalence classes of symmetric local unitary transformations. The classification of SPT phases protected by internal symmetries has been established through powerful frameworks ranging from cohomology  and cobordism theory \cite{chenScience2012, chen13, Kitaev_15_vedio,kapustin14,Kapustin2017} to topological field theory \cite{freed14,freed16}, which yield largely consistent results.

A natural and profound generalization of this paradigm is to incorporate crystalline symmetries, giving rise to crystalline SPT phases, or topological crystalline phases (TCP) 
\cite{Teo_parity,Fu_11_TCI,Judaun_13_space_TCI,Slager_2013NatPh_space,Shinsei_Reflection_13,Hemele2017,Else2018}.
Given the ubiquity of space group symmetries in solid-state materials, TCPs offer a rich landscape for experimental realization \cite{TCI_experiment1,TCI_experiment2,TCI_experiment3,TCI_experiment4,TCI_experiment5,TCI_experiment6,TCI_experiment7} and have garnered widespread attention \cite{Teo_C4_MZM,Yuji13_TCS_SrRuO, Zhang_Fan_TMSc_13, TCP_13_clifford, Free_TI_Point_Bernevig_14, TCI_nonsymmorphic_CXLiu_14, Fang_TCI_nonsymmorphic_15, RXZhang_TMCI_15, Huang_Real_space_17PRB, Langbehn_PRL_reflection_17, Luka_BBC_TCP_PRX_19, Higher_SPT_boson_YMLu_2020, Zhida_Topo_crystal_ScienceAdv_2019, Else_TCP_defect_net_19, Song_Real_recip_NC_2020, Roy_higher_TCP_anticomm_matrix_19, Shiozaki_2023_generalized_AHSS, Meng2018, Zhang25_PRX_point, TCI_review1, Fu_Kane_Mele_3D_PRL_07,K_theory1_Freed_2013,K_theory2_Thiang_2015,Ken_14_TCP,Ken_TCM_equi_K_theory_17,Ken_nonsymmorphic_15,Ken_TCI_TSc_nonsymmorphic_16,Shiozaki_2022_AHSS_momemtum_space_band_topology,YuRui_equivlant_expre_Topo_invairant_PRB_11,Fang_12_invariant_Point,Fang_invariant_nonsymmor_15,Zhida_Cn_PRL_17,Fang_2019_rotation_ScienAdv,Higher_order_TI_science_bernevig_19,Wang_Hourglass_2016,Wieder_2018,Bradlyn_2017_Topo_quan_Chem, Po_2017_symmetry_indicators_TCI, Po_symm_indi_PRX_17,Slager_PRX_17_crytalline,Zhida_NC_mapping_18,Po_SciAdv_18_MSG,ZZY_2022,ZZY_2407,ZZY_2501} . Unlike internal symmetries, crystalline SPT intertwine spatial geometry with quantum topology, leading to a classification problem that is significantly more intricate.

While the classification of free-fermion TCPs is largely settled—thanks to tools like K-theory \cite{K_theory1_Freed_2013,K_theory2_Thiang_2015,Ken_14_TCP,Ken_TCM_equi_K_theory_17,Ken_nonsymmorphic_15,Ken_TCI_TSc_nonsymmorphic_16,Shiozaki_2022_AHSS_momemtum_space_band_topology} and symmetry indicators \cite{YuRui_equivlant_expre_Topo_invairant_PRB_11,Fang_12_invariant_Point,Fang_invariant_nonsymmor_15,Zhida_Cn_PRL_17,Fang_2019_rotation_ScienAdv,Higher_order_TI_science_bernevig_19,Wang_Hourglass_2016,Wieder_2018,Bradlyn_2017_Topo_quan_Chem, Po_2017_symmetry_indicators_TCI, Po_symm_indi_PRX_17,Slager_PRX_17_crytalline,Zhida_NC_mapping_18,Po_SciAdv_18_MSG}—the landscape of interacting crystalline electronic systems, particularly in three dimensions, remains far from complete. The challenge is twofold: First, the  combinatorial complexity of the 230 space groups makes case-by-case analysis a daunting task. Second, and more fundamentally, interactions invalidate the single-particle band topology description, demanding a many-body treatment that can handle the interplay between strong correlations and spatial symmetries.

To navigate this challenge, two complementary strategies have emerged from various attempts. The real-space construction \cite{Song_Real_recip_NC_2020, Zhang2020} offers geometric intuition by decorating lower-dimensional SPTs onto symmetry-invariant subspaces; however, it becomes technically cumbersome when dealing with complex space groups. A more versatile alternative is the Crystalline Equivalence Principle (CEP) \cite{Else2018, Meng2018,Zhang2020,Zhang2023,Zhang25_PRX_point,ren2023stacking,Yu_CEP_25}. The CEP maps crystalline SPT phases to SPTs protected by an effective internal symmetry group, enabling the use of various computational tools for classification. For simple symmetries, spin cobordism and spectral sequences \cite{Debray2021} can be elegantly applied, though they often require a case-by-case analysis. However, for complex and extensive symmetry sets—such as the 230 space groups, the 1651 magnetic space groups, and spin groups—these methods are often impractical. In contrast, the Domain-Wall-Decoration (DWD) framework established in Refs.\cite{Wang2018,Wang2020}, combined with state-of-the-art algebraic algorithms \cite{Ouyang2021}, appears to be an ideal strategy for handling these complex symmetries.

However, the problem is not fully resolved. The effective internal symmetries induced by space groups are often intricate, generically involving anti-unitary operators and non-trivial central extensions by fermion parity $\mathbb{Z}_2^f$. Here lies a critical gap in the DWD framework: while the classification for 3D fermionic SPTs (FSPTs) with finite unitary symmetries is complete \cite{Wang2018,Wang2020}, the formulation for general case involving anti-unitary symmetries remains not incomplete. This  incompleteness prevents the full application of the DWD framework—even when aided by automated algorithms—to thoroughly classify the 3D topological crystalline superconductors.

In this work, we bridge this gap by developing a comprehensive classification based on DWD framework for to generic anti-unitary groups, which requires overcoming two specific, long-standing obstructions:
\begin{enumerate}
\item The $p+ip$ Layer Classification: We resolve the classification of the $p+ip$ decoration layer in the presence of generic anti-unitary elements and infinite discrete subgroups (e.g., translations)—a scenario previously lacking a general understanding and computable treatment (Sec.\ref{sec:decoupling_pplusip}).
\item The Super-coherence Obstruction ($O_5$): We derive a closed, computable expression for the $O_5$ obstruction arising from the Kitaev-chain layer. This missing piece is essential for verifying the super-coherence condition $\mathrm{d}\nu_4 = O_5$, thereby completing the algebraic consistency of the theory (Sec.\ref{sec:antiuni_SPT}).
\end{enumerate}

By solving these fundamental problems, we provide a complete classification of 3D FSPT phases protected by generic discrete symmetries. Leveraging this general formalism via the CEP and the powerful algorithm \cite{Ouyang2021}, we systematically compute the interacting classification for all 230 crystallographic space in electronic crystals. This represents a complete tabulation of interacting topological crystalline superconductors in electronic crystals. Our results not only provide a rigorous foundation for identifying topological phases in strongly correlated materials such as $\mathrm{UPt}_3$ \cite{RMP_UPt3} and $\mathrm{UTe}_2$ \cite{Science_UTe2_19} but also establish a universal toolkit for exploring topological crystalline materials in three dimensions and beyond.

\subsection{Some generalities on fermionic symmetry}
\label{sec:fermion_symm}
\textit{Description of fermionic symmetry}
In fermionic systems, the fermion parity  $P_f$ is always a symmetry operation, namely it always commutes with  Hamiltonian which consists of operators with an even number of fermionic operators.  The fermionic parity detects the parity of the occupation number of fermion particle(s) in quantum states. Besides, there may be more symmetry operations, for all of which, we collect them as the symmetry group $G_f$, where the subscript $f$ indicates the fermion parity is contained.  For all $G_f$, we only consider the symmetry operations that preserve the fermion parity of quantum states, namely those that commute with $P_f$, which means the group $\Z_2^f$ generated by $P_f$ is always in the center of $G_f$.  So the total symmetry group $G_{f}$ has a nice characterization as the central extension of the quotient group $G_f/\Z_2^f=G_b$, which we call bosonic symmetry group, by the fermion parity subgroup $\mathbb Z_{2}^{f}$. We  consider the group $G_{b}$ is a discrete group, which can be infinite, like space groups.   This central extension is defined by a 2-cocycle $\omega_2 \in H^2(G_b, \mathbb Z_2)$ and fits into the following short exact sequence. 
\begin{equation}
1 \rightarrow \mathbb Z_{2}^{f} \rightarrow G_{f} \rightarrow G_{b} \rightarrow 1. \label{eq:ses_Gf}
\end{equation}
If the system considered contains anti-unitary symmetry operators, we use the group homomorphism $s_{1} \in H^1(G_b, \mathbb Z_2)$ mapping from $G_{b}$ to $\mathbb Z_{2}$ to distinguish between unitary ($s_1 = 0$) and anti-unitary ($s_1 = 1$) group elements. 
Therefore, the symmetry group of an internal FSPT can be uniquely specified by a triple $(G_b, \omega_2, s_1)$. We note that two 2-cocycle $\omega_2$ and  $\tilde \omega_2$ differ by a coboundary $\tilde \omega_2=\omega_2+du_1$ mod 2 will give two isomorphic fermionic symmetry groups, namely $G_f\cong \tilde G_f$ where $G_f$ is specified by $(G_b,\omega_2,s_1)$ and $\tilde G_f$ by $(G_b,\tilde \omega_2,s_1)$. The isomorphism map is nothing more than relabeling  elements of the quotient group $G_b$ as inclusion in $G_f$ by adding or removing fermion parity $P_f$.  In fact, choosing the representative of $G_f$ to form $G_b$ is an arbitrary choice and should not cause any physical consequence. So, the classification of FSPT with such two isomorphic fermionic symmetry groups should be identical. 
The characterization of $G_f$ by $(G_b, \omega_2,s_1)$ in Eq.(\ref{eq:ses_Gf}) applies not only to internal symmetry but also to spatial symmetry.   

\textit{Fermionic crystalline equivalence principle}
According to the CEP~\cite{Else2018} for bosonic SPT, crystalline topological liquids with symmetry group $G$ are in one-to-one correspondence with topological phases protected by the same symmetry $G$, but act
internally, where if an element of $G$ is orientation-reversing, it is realized as an anti-unitary symmetry in the internal symmetry group.

However, the fermionic CEP is somewhat more intricate~\cite{Debray2021,Manjunath2023}. 
 The simplest example is that the time reversal symmetry with $T^2= 1$ or $P_f$ will be mapped to the mirror symmetry $M^2= P_f$ or 1 \cite{witten15,Else2018}, respectively. Moreover, a lot of examples have confirmed this correspondence 
 \cite{Zhang2020, Zhang2022,Manjunath2023,Zhang25_PRX_point}, including the 2D wallpaper group \cite{Zhang2022} and 3D point groups \cite{Zhang25_PRX_point}. With this assistance, once we obtain the  classification formulas for internal symmetries that amenable for automatic algorithm, we can apply the CEP for computing the classification of crystalline topological phases at scale.  More general mappings are discussed in Sec.\ref{sec:framework_for_spatial}.

\subsection{Main results}
Below we summarize our main results in this paper.

\begin{enumerate}
\item \textit{Classification framework.} We establish the complete framework for classifying the interacting topological superconductors protected by discrete symmetries, involving spatial symmetries. This framework is perfectly suitable for computation for large-scale discrete symmetries, such as space groups, magnetic space groups, and mixed spatial-internal symmetries. For explicit computation of the classification for generic discrete groups, a state-of-the-art algorithm for the computation is provided. 
\item
\textit{Compatibility and classification of $p{+}ip$ layers.}
For internal symmetries, the classification of the $p{+}ip$ layer is based on the $n_1\in H^1(G_b, \Z_T)$ which usually contains two parts: a free part $\Z^r$ and a torsion part $\Z_2$. We find that (1) the free part will eventually lead to a $\Z^r$ classification, and (2) whether the torsion part contributes to the classification or not depends on whether  $[\omega_2 + s_1 \cup s_1]=0$ in $H^2(G_b, \Z_T)$ or not.  
 
More explicitly, for space groups, we can further show that weak $p{+}ip$-stack classifications are governed purely by the action of the point group on translations: a translation direction contributes a $\mathbb{Z}$ factor iff it is flipped to its inverse by exactly the orientation-reversing elements (such as mirror, inversion), leading to simple $\mathbb{Z}$ weak indices in some of monoclinic, tetragonal, trigonal, and hexagonal systems, and  $\mathbb{Z}^3$ for triclinic system.
\item
\textit{New anti-unitary $O_5$ structure.}
We derive an explicit decomposition
\begin{equation}
\,\,\,O_5 = O_5^{\mathrm{sym}}[n_2,n_3]\; O_5^{c}[n_3]\; O_5^{c\gamma}[\mathrm{d}n_3]\; O_5^{\gamma}[n_2], \nonumber
\end{equation}
where $n_2$ and $n_3$ characterize the Kitaev chain and complex fermion decoration. The genuinely new component $O_5^{\gamma}$ encodes the quantum-averaged Majorana-dimer phase and generically produces $e^{i\pi/4}$ factors. The $O_5$  can be applied to generic discrete internal symmetries and complete the framework for classifying the interacting topological superconductors (TSC) for both internal and spatial symmetry (via CEP). 
\item
\textit{Systematic Classification for interacting crystalline TSC.}
Using the fermionic CEP and our DWD framework, we obtain the systematic  classification of electronic interacting topological crystalline superconductors (iTCS)  protected by the 230 space groups. The main results are listed in Table.\ref{tab:TCSc} and \ref{tab:TCSc2}.
\end{enumerate}

\subsection{Organization of the paper}

\begin{itemize}
\item Section~II: Framework for classifying the 3D iTCS (FSPT) protected by internal discrete symmetries and also those involving the spatial symmetries.
\item Section~III: DWD framework in $3$D with anti-unitary symmetries; classification of the $p{+}ip$ layer; weak indices from translations; obstruction functions $O_3$, $O_4$, $O_5$;  AH spectral sequence interpretation.

\item Section~IV: Full interacting classifications for all 230 space groups and representative examples.

\item Section~V: Summary and Discussion. 
\item Appendix: Present some mathematical and technical details and some reviews on the relevant physical aspects.
\end{itemize}

\section{General framework for classification of 3+1D interacting topological superconductor}

Here we describe the general framework for classification of 3+1D iTSC  protected discrete  symmetries that can apply to three cases: (1) purely internal symmetries, (2) purely spatial symmetry and (3) mixed spatial-internal symmetries.  With the concept of fermionic CEP, the central key is the classification of iTSC (FSPT)  protected by purely discrete internal symmetries. We leave the relevant derivation in Sec.\ref{sec:decoupling_pplusip} and \ref{sec:antiuni_SPT}.

\subsection{3+1D  FSPT protected by internal discrete symmetries}
\label{sec:framework_for_internal}

We primarily employ the domain-wall decoration framework to classify FSPT phases protected by generic discrete physical symmetries $(G_b, \omega_2, s_1)$. This construction organizes the topological data into four layers: 
\begin{enumerate}
    \item[(1)] the $p+ip$ layer, characterized by $n_1 \in C^1(G_b, \mathbb{Z}_T)$;
    \item[(2)] the Kitaev chain (KC) layer, by $n_2 \in C^2(G_b, \mathbb{Z}_2)$;
    \item[(3)] the complex fermion layer, by $n_3 \in C^3(G_b, \mathbb{Z}_2)$;
    \item[(4)] the bosonic SPT layer, by $\nu_4 \in C^4(G_b, U_T(1))$.
\end{enumerate}
  The subscript $T$ denotes a twisted action by anti-unitary symmetries, which map $n \mapsto -n$ in $\mathbb{Z}_T$ and $a \mapsto a^*$ in $U_T(1)$. We now proceed to identify the specific subsets of this data that define distinct topological phases. First, the data must satisfy the following obstruction conditions:
\begin{align}
        &dn_1=0 \label{eq:p+ip_obs1}\\
        &dn_2=O_3 \quad \text{ mod }2\label{eq:p+ip_obs2}\\
        &dn_3=O_4 \quad \text{ mod }2 \label{eq:obs3}\\
        &d\nu_4=O_5.\label{eq:obs4}
    \end{align}
    We discuss two cases, depending on the presence or absence of the $p+ip$ layer labeled by $n_1$.

\begin{enumerate}
    \item \textbf{With the $p+ip$ layer:} Two obstruction conditions are necessary for the classification:
    \begin{align}
        &dn_1=0 \label{eq:p+ip_obs1}\\
        &dn_2=O_3[n_1] \text{ mod }2\label{eq:p+ip_obs2}
    \end{align}
    where 
    \begin{align}
        O_3[n_1]=(\omega_2+s_1\cup n_1)\cup n_1.
    \end{align}
    
    In general, we must also consider the additional two  obstruction conditions given by Eqs.\eqref{eq:obs3} and \eqref{eq:obs4}. However, we argue that these higher-order conditions are automatically satisfied provided that Eqs.\eqref{eq:p+ip_obs1} and \eqref{eq:p+ip_obs2} hold. 
    
    The condition in Eq.\eqref{eq:p+ip_obs1} implies $n_1\in H^1(G_b, \Z_T)$, which generally decomposes into two parts: a free part $\Z^r$ and a torsion part $\Z_2$. Let us denote their generators by $n_1^\mathrm{free}$ and $n_1^\mathrm{tor}$. For the free generator $n_1^\mathrm{free}$, there always exists a minimal integer $k$ such that for the $p+ip$ layer decoration specified by $k\cdot n_1^\mathrm{free}$, all obstruction functions $O_3, O_4$, and $O_5$ vanish (become trivial):
    \begin{align}
       & O_3=0\text{ mod }2,\quad  O_4=0\text{ mod }2,\quad  O_5=1, \nonumber
    \end{align}
    assuming the phase factor of $O_5$ can be a rational fraction of $2\pi$. Consequently, the corresponding $k\cdot n_1^\mathrm{free}$ will generate a $\Z^r$ classification. 
    
    For the torsion part $n_1^\mathrm{tor}$, we argue that provided $O_3$ vanishes, there always exist $n_2$ and $n_3$ such that the $O_4$ and $O_5$ obstructions vanish. Specifically:
    \begin{align}
  &\quad \textit{ If } O_3[n_1^\mathrm{tor}]\in B^3(G_b,\Z_2),\, \textit{there must exist $n_2$} \nonumber\\
  &\quad \textit{ and $n_3$ such that } O_i \in B^i(G_b,\Z_2),\text{ for }i=4,5. \nonumber
 \end{align}

    In summary, for the $p+ip$ layer, only the two obstruction conditions in Eqs.\eqref{eq:p+ip_obs1} and \eqref{eq:p+ip_obs2} are required for the classification. 
    
    More specifically, depending on whether $G_b$ is unitary or anti-unitary, we have the following two classifications:
    \begin{enumerate}
        \item $G_b$ is unitary: Since $H^1(G_b, \Z_T)=\Z^r$, we obtain a $\Z^r$ classification from the $p+ip$ layer.
        \item $G_b$ is anti-unitary: Since $H^1(G_b, \Z_T)=\Z^r\times \Z_2$, we obtain a $\Z^r\times \Z_2 ^s$ classification from the $p+ip$ layer. 
    \end{enumerate}

    Note that the anti-unitary structure of $G_b$ yields a non-trivial $\Z_2$ factor in $H^1(G_b, \Z_T)$. We can choose a representative $n_1(g)=s_1(g)$, taking values $0$ or $1$ when $g$ is unitary or anti-unitary, respectively. The exponent $s$ is defined as follows: $s=1$ if $\omega_2^{\mathrm{m}}$ is trivial in $H^2(G_b, \Z_T)$, and $s=0$ otherwise. Following the fermionic CEP in Eqs.\eqref{eq6.1} and \eqref{eq6.2}, $\omega_2^{\mathrm{m}}=\omega_2+s_1\cup s_1$. (For a detailed derivation, please refer to Sec.~\ref{sec:decoupling_pplusip}.)

    \item \textbf{Without the $p+ip$ layer:} In this case, the obstruction conditions reduce to the following three:
     \begin{align}
        &dn_2=0\text{ mod }2\label{eq:obs2}\\
        &dn_3=O_4[n_2]\text{ mod }2 \label{eq:nobs3}\\
        &d\nu_4=O_5[n_2,n_3],\label{eq:nobs4}
    \end{align}
    where the complete forms of $O_4[n_2]$ and $O_5[n_2,n_3]$ are explicitly given by:
    \begin{align}
        &\qquad O_4[n_2]=(\omega_2+n_2)\cup n_2+s_1\cup (n_2\cup_1n_2) \nonumber \\
        &\qquad O_5[n_2,n_3]=O_5^{\mathrm{sym}}[n_2,n_3]\; O_5^{c}[n_3]\; O_5^{c\gamma}[\mathrm{d}n_3]\; O_5^{\gamma}[n_2]. \nonumber
    \end{align}
    The explicit form of each piece of $O_5$ is provided in Sec.\ref{sec:O5_formula}. 

    More explicitly, for the KC layer, $n_2$ must be a 2-cocycle in $H^2(G_b, \Z_2)$. Those $n_2$ that contribute to the obstruction-free classification must satisfy two conditions: (1) $O_4[n_2]=(\omega_2+n_2)\cup n_2+s_1\cup (n_2\cup_1n_2)$ is trivial in $H^4(G_b, \Z_2)$, and (2) there exists at least one $n_3 \in C^3(G_b,\Z_2)$ satisfying $dn_3=O_4$ such that $O_5[n_2, n_3]$ is trivial in $H^5(G_b, U_T(1))$. 
        
    Next, passing to  the complex fermion layer, $n_3$ must be a 3-cocycle in $H^3(G_b, \Z_2)$. Only those $n_3$ for which $O_5[n_2=0, n_3]$ is trivial in $H^5(G_b, U_T(1))$ contribute to the obstruction-free classification. Finally, all bosonic SPT phases $\nu_4$, as 4-cocycles in $H^4(G_b, U_T(1))$, contribute to the obstruction-free classification.
\end{enumerate}

\begin{table*}[h]
    \centering
\resizebox{\textwidth}{!}{
\begin{tabular}{ccccccc|ccccccc|ccccccc}
\toprule
No.& Int. & $\mathrm{E}_{\mathrm{2D}}$ & $\mathrm{E}_{\mathrm{1D}}$ & $\mathrm{E}_{\mathrm{0D}}$ & $\mathrm{E}_{\mathrm{b}}$ & Total &No.& Int. & $\mathrm{E}_{\mathrm{2D}}$ & $\mathrm{E}_{\mathrm{1D}}$ & $\mathrm{E}_{\mathrm{0D}}$ & $\mathrm{E}_{\mathrm{b}}$ & Total &No.& Int. & $\mathrm{E}_{\mathrm{2D}}$ & $\mathrm{E}_{\mathrm{1D}}$ & $\mathrm{E}_{\mathrm{0D}}$ & $\mathrm{E}_{\mathrm{b}}$ & Total \\
\hline
1  & $P1$ & $\Z^3$ & $ \Z_1$ & $ \Z_2^3$ & $ \Z_2$   &  16$\times \Z^3$& 59  & $Pmmn$ & $\Z_1$ & $ \Z_1$ & $ \Z_1$ & $\Z_2^4$  & 16 & 117  & $P\overline{4}b2$ & $\Z_1$ & $\Z_2$ & $\Z_2^2$ & $ \Z_2^4\times \Z_4$  &  512\\
2  & $P\overline{1}$ &  $\Z^3$ & $ \Z_1 $ & $\Z_2^4$ & $ \Z_2$  & 32$\times\Z^3$& 60  & $Pbcn$ & $\Z_1$ & $\Z_2$ & $\Z_2^2$ & $\Z_2$  &16& 118  & $P\overline{4}n2$ & $\Z_1$ & $\Z_2$ & $ \Z_2^2$ & $ \Z_2^4\times \Z_4$  & 512\\
3  & $P2$ & $\Z$ & $ \Z_2^3$ & $ \Z_2^4$ & $ \Z_2^4$  & 2048$\times \Z$  & 61  & $Pbca$ & $\Z_1$ & $\Z_2^2$ & $\Z_2$ & $\Z_2$  &16& 119  & $I\overline{4}m2$ & $\Z_1$ & $\Z_1$ & $\Z_1$ & $\Z_2^6$  & 64\\
4  & $P2_1$ & $\Z$ & $ \Z_2^3$ & $ \Z_2 $ & $ \Z_1$  &16$\times\Z$& 62  & $Pnma$ & $\Z_1$ & $ \Z_1$ & $\Z_2^2 $ & $ \Z_1$  &4& 120  & $I\overline{4}c2$ & $\Z_1$ & $\Z_2$ & $ \Z_2^2$ & $\Z_2^4$ & 128\\
5  & $C2$ & $\Z$ & $ \Z_2^2 $ & $\Z_2^2 $ & $ \Z_2^2 $  & 64$\times\Z$& 63  & $Cmcm$ & $\Z_1$ & $ \Z_1$ & $\Z_2$ & $\Z_2^3$  &8& 121 & $I\overline{4}2m$ & $\Z_1$ & $\Z_1$ & $\Z_2$ & $ \Z_2^5$  &64 \\
6  & $Pm$ & $\Z$ & $ \Z_2 $ & $ \Z_2^2$ & $ \Z_2^2$  &32$\times\Z$& 64  & $Cmca$ & $\Z_1$ & $ \Z_2$ & $\Z_2$ & $\Z_2^2$  &16& 122 & $I\overline{4}2d$ & $\Z_1$ & $\Z_2$ & $ \Z_2^3$ & $ \Z_2\times \Z_4$  &128\\
7  & $Pc$ & $\Z$ & $ \Z_2^3$ & $ \Z_2 $ & $ \Z_1$  &16$\times\Z$& 65  & $Cmmm$ & $\Z_1$ & $ \Z_1$ & $ \Z_2$ & $ \Z_2^{13}$  &16384& 123 & $P4/mmm$ & $\Z_1$ & $\Z_1$ & $\Z_1$ & $\Z_2^{18}$  &262144\\
8  & $Cm$ & $\Z$ & $ \Z_2 $ & $ \Z_2$ & $ \Z_2$  &8$\times\Z$& 66  & $Cccm$ & $\Z_1$ & $ \Z_1$ & $ \Z_2^2$ & $\Z_2^6$  &256& 124 & $P4/mcc$ & $\Z_1$ & $\Z_1$ & $\Z_2^2$ & $\Z_2^7$  & 512\\
9  & $Cc$ & $\Z$ & $ \Z_2^2$ & $ \Z_2$ & $ \Z_1$   &8$\times\Z$& 67  & $Cmma$ & $\Z_1$ & $ \Z_1$ & $\Z_2$ & $\Z_2^6$  &128& 125 & $P4/nbm$ & $\Z_1$ & $\Z_1$ & $\Z_1$ & $\Z_2^7$  & 128\\
10 & $P2/m$ & $\Z $ & $ \Z_1$ & $ \Z_2^5$ & $ \Z_2^6$  &2048$\times\Z$& 68  & $Ccca$ & $\Z_1$ & $ \Z_2$ & $ Z_1$ & $ \Z_2^4$  &32& 126 & $P4/nnc$ & $\Z_1$ & $\Z_1$ & $\Z_2$ & $ \Z_2^{6}$  &128  \\
11 & $P2_1/m$ & $\Z$ & $ \Z_1$ & $ \Z_2^2$ & $ \Z_2$  &8$\times\Z$& 69  & $Fmmm$ & $\Z_1$ & $ \Z_1$ & $\Z_2$ & $\Z_2^8$  &512& 127 & $P4/mbm$ &  $\Z_1$ & $\Z_1$ & $\Z_2^2$ & $ \Z_2^7\times\Z_4$  &1024 \\
12 & $C2/m$ & $\Z $ & $ \Z_1$ & $ \Z_2^3 $ & $ \Z_2^3$  & 64$\times\Z$& 70  & $Fddd$ & $\Z_1$ & $\Z_1$ & $\Z_2$ & $\Z_2^4$  &32& 128 & $P4/mnc$ & $\Z_1$ & $\Z_1$ & $\Z_2^3$ & $\Z_2^3\times \Z_4$  &256\\
13 & $P2/c$ & $\Z$ & $ \Z_2$ & $ \Z_2^3$ & $ \Z_2^2$  &64$\times\Z$ & 71  & $Immm$ & $\Z_1$ & $\Z_1$ & $\Z_1$ & $\Z_2^{12}$  &4096& 129 & $P4/nmm$ & $\Z_1$ & $\Z_1$ & $\Z_1$ & $\Z_2^5$   &32\\
14 & $P2_1/c$ & $\Z$ & $ \Z_2$ & $ \Z_2^2$ & $ \Z_2$  &16$\times\Z$& 72  & $Ibam$ & $\Z_1$ & $\Z_1$ & $\Z_2$ & $\Z_2^4$  &32& 130 & $P4/ncc$ & $\Z_1$ & $\Z_2$ & $ \Z_2^2$ & $ \Z_2^2$   &32\\
15 & $C2/c$ & $\Z$ & $ \Z_1$ & $ \Z_2^3$ & $ \Z_2$  &16$\times\Z$& 73  & $Ibca$ & $\Z_1$ & $\Z_2^2$ & $\Z_2 $ & $ \Z_1$  &8& 131 & $P4_2/mmc$ & $\Z_1$ & $\Z_1$ & $\Z_1$ & $\Z_2^{14}$   &16384\\
16 & $P222$ & $\Z_1$ & $ \Z_1$ & $ \Z_1$ & $ \Z_2^{16}$  &65536& 74  & $Imma$ & $\Z_1$ & $\Z_1$ & $\Z_2^2$ & $\Z_2^4$  &64& 132 & $P4_2/mcm$ & $\Z_1$ & $\Z_1$ & $\Z_1$ & $\Z_2^{10}$  &1024\\
17 & $P222_1$ & $\Z_1$ & $ \Z_2$ & $\Z_2^4$ & $\Z_2^4$  &512 & 75  & $P4$ & $\Z$ & $ \Z_2^2$ & $ \Z_2^2 $ & $ \Z_2 \times\Z_4^2$  &512$\times\Z$& 133 & $P4_2/nbc$ & $\Z_1$ & $\Z_1$ & $\Z_2$ & $ \Z_2^6$  &64\\
18 & $P2_12_12$ & $\Z_1$ & $ \Z_2^2$ & $ \Z_2^2$ & $ \Z_2^2$  &64& 76  & $P4_1$ & $\Z$ & $\Z_2^2$ & $ \Z_2 $ & $ \Z_1$  &8$\times\Z$& 134 & $P4_2/nnm$ & $\Z_1$ & $\Z_1$ & $\Z_1$ & $\Z_2^7$  &128\\
19 & $P2_12_12_1$ & $\Z_1$ & $ \Z_2^2$ & $ \Z_2 $ & $ \Z_1$  &8& 77  & $P4_2$ & $\Z$ & $ \Z_2^2$ & $ \Z_2^3$ & $ \Z_2^3$  &256$\times\Z$& 135 & $P4_2/mbc$ & $\Z_1$ & $\Z_1$ & $\Z_2^3$ & $ \Z_2^3$  &64\\
20 & $C222_1$ & $\Z_1$ & $ \Z_2^2$ & $\Z_2^2$ & $\Z_2^2$  &64& 78  & $P4_3$ & $\Z$ & $ \Z_2^2$ & $ \Z_2 $ & $ \Z_1$  &8$\times\Z$& 136 & $P4_2/mnm$ & $\Z_1$ & $\Z_1$ & $\Z_2^2$ & $ \Z_2^7$   &512\\
21 & $C222$ & $\Z_1$ & $\Z_2$ & $ \Z_2$ & $ \Z_2^9$  &2048& 79  & $I4$ & $\Z$ & $\Z_2$ & $ \Z_2^2$ & $ \Z_2\times \Z_4$  &64$\times\Z$& 137 & $P4_2/nmc$ & $\Z_1$ & $\Z_1$ & $\Z_1$ & $\Z_2^4$   &16\\
22 & $F222$ & $\Z_1$ & $ \Z_2$ & $ \Z_1$ & $ \Z_2^8$  & 1024& 80  & $I4_1$ & $\Z$ & $ \Z_2$ & $ \Z_2$ & $ \Z_2$  &8$\times\Z$& 138 & $P4_2/ncm$ & $\Z_1$ & $\Z_1$ & $\Z_2^2$ & $\Z_2^3$   &32\\
23 & $I222$ & $\Z_1$ & $ \Z_1$ & $ \Z_1$ & $\Z_2^8$  &256& 81   & $P\overline{4}$ & $\Z_2\times \Z$ & $ \Z_2^2$ & $ \Z_2^5$ & $ \Z_2\times \Z_4^2$  &8192$\times\Z$& 139 & $I4/mmm$ & $\Z_1$ & $ \Z_1$ & $ \Z_1$ & $\Z_2^{10}$ &1024\\
24 & $I2_12_12_1$ & $\Z_1$ & $ \Z_2^2$ & $ \Z_2^3$ & $ \Z_2^3$  &256& 82   & $I\overline{4}$ & $\Z_2\times \Z$ & $ \Z_2$ & $ \Z_2^4$ & $ \Z_4^2$  & 256$\times\Z$& 140 & $I4/mcm$ & $\Z_1$ & $\Z_1$ & $\Z_2$ & $\Z_2^{6}$  &128\\
25 & $Pmm2$ & $\Z_1$ & $ \Z_1$ & $ \Z_1 $ & $\Z_2^8$  &256& 83   & $P4/m$ & $\Z$ & $  \Z_1$ & $ \Z_2^5 $ & $ \Z_2^4\times \Z_4^2$  &2048$\times\Z$& 141 & $I4_1/amd$ & $\Z_1$ & $\Z_1$ & $\Z_2$ & $\Z_2^4$   &32\\
26 & $Pmc2_1$ & $\Z_1$ & $ \Z_2$ & $\Z_2 $ & $ \Z_1$  &4& 84   & $P4_2/m$ & $\Z$ & $ \Z_1$ & $ \Z_2^5$ & $ \Z_2^4$   &512$\times\Z$& 142 & $I4_1/acd$ & $\Z_1$ & $\Z_2$ & $\Z_2$ & $ \Z_2^2$  &16$\times\Z$\\
27 & $Pcc2$ & $\Z_1$ & $ \Z_2^3$ & $\Z_2 $ & $ \Z_1$  &8& 85   & $P4/n$ & $\Z$ & $ \Z_2$ & $ \Z_2^3$ & $\Z_4^2$   &64$\times\Z$& 143 & $P3$ & $\Z$ & $ \Z_2$ & $ \Z_2$ & $ \Z_3^3$  &108$\times\Z$\\
28 & $Pma2$ & $\Z_1$ & $\Z_2^2$ & $ \Z_2^3$ & $\Z_2^3$  &256& 86   & $P4_2/n$ & $\Z$ & $ \Z_2$ & $ \Z_2^3$ & $ \Z_2\times \Z_4$   &128$\times\Z$& 144 &$P3_1$ & $\Z$ & $ \Z_2$ & $ \Z_2 $ & $ \Z_1$  &4$\times\Z$\\
29 & $Pca2_1$ & $\Z_1$ & $ \Z_2^3$ & $ \Z_2$ & $ \Z_1$  &16& 87   & $I4/m$ & $\Z$ & $ \Z_1$ & $ \Z_2^4$ & $ \Z_2^2\times \Z_4$  &256$\times\Z$& 145 & $P3_2$ & $\Z$ & $ \Z_2$ & $ \Z_2 $ & $ \Z_1$  &4$\times\Z$\\
30 & $Pnc2$ & $\Z_1$ & $\Z_2^2$ & $ \Z_2^2$ & $ \Z_2^2$  &64& 88   & $I4_1/a$ & $\Z$ & $ \Z_1$ & $ \Z_2^3$ & $ \Z_4$  &32$\times\Z$& 146 &  $R3$ & $\Z$ & $ \Z_2$ & $ \Z_2$ & $ \Z_3$   &12$\times\Z$\\
31 & $Pmn2_1$ & $\Z_1$ & $ \Z_2$ & $\Z_2$ & $ \Z_2$  &64& 89   & $P422$ & $\Z_1$ & $ \Z_1$ & $ \Z_1$ & $ \Z_2^{12}$  &4096& 147 & $P\overline{3}$ & $\Z$ & $\Z_1$ & $ \Z_2^2$ & $ \Z_3\times \Z_6$  &72$\times\Z$\\
32 & $Pba2$ & $\Z_1$ & $ \Z_2^2$ & $\Z_2^2 $ & $ \Z_2^2$  &64& 90   & $P42_12$ & $\Z_1$ & $\Z_2$ & $ \Z_2$ & $ \Z_2^4\times \Z_4$  &256& 148 & $R\overline{3}$ & $\Z$ & $ \Z_1$ & $\Z_2^2$ & $ \Z_6$   &24$\times\Z$\\
33 & $Pna2_1$ & $\Z_1$ & $ \Z_2^2$ & $ \Z_2 $ & $ \Z_1$  &8& 91   & $P4_122$ & $\Z_1$ & $\Z_2^2$ & $ \Z_2^3$ & $ \Z_2^3$  &256& 149 & $P312$ & $\Z_1$ & $\Z_2$ & $ \Z_2^2$ & $ \Z_2^2$   &32\\
34 & $Pnn2$ & $\Z_1$ & $\Z_2^2$ & $\Z_2^2 $ & $ \Z_2^2$  &64& 92   & $P4_12_12$ & $\Z_1$ & $\Z_2$ & $ \Z_2$ & $ \Z_2$   &8& 150 & $P321$ & $\Z_1$ & $\Z_2$ & $\Z_2^2$ & $ \Z_2\times \Z_6$  &96\\
35 & $Cmm2$ & $\Z_1$ & $ \Z_2$ & $ \Z_2$ & $ \Z_2^5$  &128& 93   & $P4_222$ & $ \Z_1$ & $ \Z_1$ & $ \Z_1$ & $\Z_2^{12}$   &4096& 151 & $P3_112$ & $\Z_1$ & $\Z_2$ & $ \Z_2^2$ & $ \Z_2^2$  &32\\
36 & $Cmc2_1$ & $\Z_1$ & $ \Z_2$ & $\Z_2 $ & $ \Z_1$  &4& 94   & $P4_22_12$ & $\Z_1$ & $\Z_2$ & $\Z_2$ & $ \Z_2^5$  &128& 152 & $P3_121$ & $\Z_1$ & $\Z_2$ & $ \Z_2^2$ & $ \Z_2^2$  &32\\
37 & $Ccc2$ & $\Z_1$ & $\Z_2^2$ & $\Z_2^2$ & $ \Z_2$  &16& 95   & $P4_322$ & $\Z_1$ & $\Z_2^2$ & $\Z_2^3$ & $ \Z_2^3$  &256& 153 & $P3_212$ & $\Z_1$ & $\Z_2$ & $ \Z_2^2$ & $ \Z_2^2$  &32\\
38 & $Amm2$ & $\Z_1$ & $ \Z_1$ & $ \Z_1$ & $\Z_2^4$  &16& 96   & $P4_32_12$ & $\Z_1$ & $\Z_2$ & $ \Z_2$ & $\Z_2$   &8& 154 & $P3_221$ & $\Z_1$ & $\Z_2$ & $ \Z_2^2$ & $ \Z_2^2$   &32\\
39 & $Abm2$ & $\Z_1$ & $\Z_2^2$ & $\Z_2 $ & $ \Z_1$  &8& 97   & $I422$ & $\Z_1$ & $\Z_1$ & $ \Z_1$ & $\Z_2^8$  &256& 155 & $R32$ & $\Z_1$ & $\Z_2$ & $ \Z_2^2$ & $ \Z_2^2$   &32\\
40 & $Ama2$ & $\Z_1$ & $\Z_2$ & $ \Z_2^2$ & $ \Z_2^2$  &32& 98   & $I4_122$ & $\Z_1$ & $ \Z_2$ & $\Z_2$ & $ \Z_2^5$  &128& 156 &  $P3m1$ & $\Z_1$ & $\Z_1$ & $\Z_2$ & $ \Z_2$  &4\\
41 & $Aba2$ & $\Z_1$ & $\Z_2^2$ & $ \Z_2 $ & $ \Z_2$    &16& 99   & $P4mm$ & $\Z_1$ & $\Z_1$ & $\Z_1$ & $\Z_2^6$  &64& 157 & $P31m$ & $\Z_1$ & $\Z_1$ & $\Z_2$ & $ \Z_6$  &12\\
42  & $Fmm2$ & $\Z_1$ & $\Z_2$ & $ \Z_1$ & $ \Z_2^2$   &8& 100  & $P4bm$ & $\Z_1$ & $\Z_2$ & $ \Z_2$ & $ \Z_2^2\times \Z_4$  &64& 158 & $P3c1$ & $\Z_1$ & $\Z_2$ & $\Z_2 $ & $ \Z_1$   &4\\
43  & $Fdd2$ & $\Z_1$ & $ \Z_2$ & $\Z_2$ & $ \Z_2$  &8& 101  & $P4_2cm$ & $\Z_1$ & $\Z_2$ & $ \Z_1$ & $\Z_2^2$   &8& 159 & $P31c$ & $\Z_1$ & $\Z_2$ & $\Z_2$ & $\Z_2$  &8\\
44  & $Imm2$ & $\Z_1$ & $ \Z_1$ & $ \Z_1$ & $ \Z_2^4$  &16& 102  & $P4_2nm$ & $\Z_1$ & $ \Z_2$ & $ \Z_2$ & $ \Z_2^3$  &32& 160 & $R3m$ & $\Z_1$ & $\Z_1$ & $\Z_2$ & $ \Z_2$  &4\\
45  & $Iba2$ & $\Z_1$ & $ \Z_2^2$ & $\Z_2 $ & $ \Z_1$  &8& 103  & $P4cc$ & $\Z_1$ & $ \Z_2^2$ & $ \Z_2^2 $ & $ \Z_1$  &16& 161 & $R3c$ & $\Z_1$ & $\Z_2$ & $\Z_2$ & $ \Z_1$  &4\\
46  & $Ima2$ & $\Z_1$ & $\Z_2$ & $\Z_2^2$ & $\Z_2$  &16& 104  & $P4nc$ & $\Z_1$ & $\Z_2$ & $\Z_2^2$ & $ \Z_4$   &32& 162 & $P\overline{3}1m$ & $\Z_1$ & $\Z_1$ & $\Z_2^2$ & $ \Z_2^3$  &32\\
47  & $Pmmm$ & $\Z_1$ & $ \Z_1$ & $ \Z_1$ & $\Z_2^{24}$  &16777216& 105  & $P4_2mc$ & $\Z_1$ & $\Z_1$ & $\Z_1$ & $\Z_2^4$   &16& 163 & $P\overline{3}1c$ & $\Z_1$ & $\Z_1$ & $\Z_2^2$ & $ \Z_2$   &8\\
48  & $Pnnn$ & $\Z_1$ & $ \Z_1$ & $ \Z_1$ & $ \Z_2^8$  &256& 106  & $P4_2bc$ & $\Z_1$ & $\Z_2$ & $ \Z_2^2$ & $ \Z_2$  &16& 164 & $P\overline{3}m1$ & $\Z_1$ & $\Z_1$ & $\Z_2^2$ & $ \Z_2^3$  &32\\
49  & $Pccm$ & $\Z_1$ & $ \Z_1$ & $\Z_2$ & $ \Z_2^9$  &1024& 107  & $I4mm$ & $\Z_1$ & $\Z_1$ & $\Z_1$ & $\Z_2^3$  &8& 165 & $P\overline{3}c1$ & $\Z_1$ & $\Z_1$ & $\Z_2^2$ & $\Z_2$   &8\\
50  & $Pban$ & $\Z_1$ & $ \Z_1$ & $ \Z_1$ & $ \Z_2^8$  &256& 108  & $I4cm$ & $\Z_1$ & $\Z_2$ & $ \Z_2$ & $ \Z_2$   &8& 166 & $R\overline{3}m$ & $\Z_1$ & $\Z_1$ & $\Z_2^2$ & $\Z_2^3$  &32\\
51  & $Pmma$ & $\Z_1$ & $ \Z_1$ & $\Z_2^2$ & $\Z_2^6$  &256& 109  & $I4_1md$ & $\Z_1$ & $\Z_1$ & $\Z_1$ & $\Z_2^2$  &4& 167 & $R\overline{3}c$ & $\Z_1$ & $\Z_1$ & $\Z_2^2$ & $\Z_2$  &8\\
52  & $Pnna$ & $\Z_1$ & $\Z_2$ & $\Z_2^3$ & $ \Z_2^2$  &64& 110  & $I4_1cd$ & $\Z_1$ & $\Z_2$ & $ \Z_2$ & $ \Z_1$  &4& 168 & $P6$ & $\Z$ & $ \Z_2$ & $ \Z_2^2$ & $ \Z_6^2$   &288$\times\Z$\\
53  & $Pmna$ & $\Z_1$ & $\Z_2$ & $\Z_2^3$ & $ \Z_2^4$  &256& 111  & $P\overline{4}2m$ & $\Z_1$ & $\Z_1$ & $\Z_1$ & $\Z_2^{10}$   &1024& 169 & $P6_1$ & $\Z$ & $ \Z_2$ & $ \Z_2 $ & $ \Z_1$   &4$\times\Z$\\
54  & $Pcca$ & $\Z_1$ & $\Z_2^2$ & $\Z_2^2$ & $\Z_2$  &32& 112  & $P\overline{4}2c$ & $\Z_1$ & $\Z_1$ & $\Z_2^2$ & $ \Z_2^8$  &1024& 170 & $P6_5$ & $\Z$ & $\Z_2$ & $ \Z_2 $ & $ \Z_1$  &4$\times\Z$\\
55  & $Pbam$ & $\Z_1$ & $ \Z_1$ & $\Z_2^3$ & $\Z_2^3$  &64& 113  & $P\overline{4}2_1m$ & $\Z_1$ & $\Z_2$ & $\Z_2^2$ & $ \Z_2^2\times \Z_4$  &128& 171 & $P6_2$ & $\Z$ & $ \Z_2$ & $ \Z_2^2$ & $\Z_2^2$  &32$\times\Z$\\
56  & $Pccn$ & $\Z_1$ & $ \Z_2$ & $\Z_2^2 $ & $ \Z_1$   &8& 114  & $P\overline{4}2_1c$ & $\Z_1$ & $\Z_2$ & $ \Z_2^3$ & $\Z_4$  &64& 172 & $P6_4$ & $\Z$ & $ \Z_2$ & $ \Z_2^2$ & $\Z_2^2$  &32$\times\Z$\\
57  & $Pbcm$ & $\Z_1$ & $\Z_2$ & $ \Z_2^2$ & $ \Z_2$  &16& 115  & $P\overline{4}m2$ & $\Z_1$ & $\Z_1$ & $\Z_1$ & $\Z_2^{8}$  &256& 173 & $P6_3$ & $\Z$ & $ \Z_2$ & $ \Z_2$ & $\Z_3^2$  &36$\times\Z$\\
58  & $Pnnm$ & $\Z_1$ & $ \Z_1$ & $ \Z_2^3$ & $\Z_2^3$  &16& 116  & $P\overline{4}c2$ & $\Z_1$ & $\Z_2$ & $ \Z_2^2$ & $ \Z_2^4$  &128& 174 & $P\overline{6}$ & $\Z$ & $ \Z_2$ & $ \Z_2^2$ & $\Z_3\times \Z_6^2$  &864$\times\Z$\\
 \bottomrule
\end{tabular}
}
    \caption{Classification of  topological crystalline superconductors protected by space groups in interacting electronic systems for space group from No.1-174 whose international short symbols are given by the row Int. The classification of  $p+ip$, Kitaev chain, complex fermion and bosonic SPT layers are separately counted denoted as $\mathrm{E}_{\mathrm{2D}}$, $\mathrm{E}_{\mathrm{1D}}$, $\mathrm{E}_{\mathrm{0D}}$ and $\mathrm{E}_{\mathrm{b}}$ respectively. Each layer is classified by an abelian group, such as  $\Z_2^n$ and $\Z_2^n\times \Z_4$ and so on. $\Z_2^n$ is $\Z_2\times \Z_2\times ... \Z_2$ with $n$ number of $\Z_2$. The total number of the classification are given in the row Total. Note that there might be some infinite factors $\Z$ from $p+ip$,  for  example, the No.12 space group, for which we will denote as $64\times \Z$ in the Total row. Specifically we note that $64\times\Z$ does not mean the set of the multiples of 64.}
    \label{tab:TCSc}
\end{table*}

\begin{table*}[t]
    \centering
   \resizebox{\textwidth}{!}{
\begin{tabular}{ccccccc|ccccccc|ccccccc}
\toprule
No.& Int. & $\mathrm{E}_{\mathrm{2D}}$ & $\mathrm{E}_{\mathrm{1D}}$ & $\mathrm{E}_{\mathrm{0D}}$ & $\mathrm{E}_{\mathrm{b}}$ & Total &No.& Int. & $\mathrm{E}_{\mathrm{2D}}$ & $\mathrm{E}_{\mathrm{1D}}$ & $\mathrm{E}_{\mathrm{0D}}$ & $\mathrm{E}_{\mathrm{b}}$ & Total &No.& Int. & $\mathrm{E}_{\mathrm{2D}}$ & $\mathrm{E}_{\mathrm{1D}}$ & $\mathrm{E}_{\mathrm{0D}}$ & $\mathrm{E}_{\mathrm{b}}$ & Total \\
\hline
175 & $P6/m$ & $\Z$ & $ \Z_1$ & $ \Z_2^3$ & $ \Z_2^2\times \Z_6^2$  &1152$\times\Z$& 194 & $P6_3/mmc$ & $\Z_1$ & $\Z_1$ & $\Z_2$ & $\Z_2^3$  &16& 213 & $P4_132$ & $\Z_1$ & $\Z_1$ & $\Z_2$ & $ \Z_2$  &4\\
176 & $P6_3/m$ &  $\Z$ & $ \Z_2^2$ & $ \Z_1$ & $\Z_3\times \Z_6$   &72$\times\Z$& 195 & $P23$ & $\Z_1$ & $\Z_1$ & $\Z_1$ & $\Z_2^3\times \Z_6$   &48& 214 & $I4_132$ & $\Z_1$ & $\Z_1$ & $\Z_1$ & $\Z_2^4$  &16\\
177 & $P622$ & $\Z_1$ & $\Z_1$ & $\Z_1$ & $\Z_2^8$  &256& 196 & $F23$ & $\Z_1$ & $\Z_2$ & $\Z_1$ & $\Z_3$  &6& 215 & $P\overline{4}3m$ & $\Z_1$ & $\Z_1$ & $\Z_1$ & $\Z_2^4$  &16\\
178 & $P6_122$ & $\Z_1$ & $\Z_2$ & $ \Z_2^2$ & $\Z_2^2$  &32& 197 & $I23$ & $\Z_1$ & $\Z_1$ & $\Z_1$ & $\Z_2\times \Z_6$ &12& 216 & $F\overline{4}3m$ & $\Z_1$ & $\Z_1$ & $\Z_1$ & $\Z_2^2$  &4\\
179 & $P6_522$ & $\Z_1$ & $\Z_2$ & $ \Z_2^2$ & $\Z_2^2$   &32& 198 & $P2_13$ & $\Z_1$ & $\Z_1$ & $\Z_2$ & $ \Z_3$ &6& 217 & $I\overline{4}3m$ & $\Z_1$ & $\Z_1$ & $\Z_2$ & $\Z_2^2$  &8\\
180 & $P6_222$ & $\Z_1$ & $\Z_1$ & $\Z_1$ & $\Z_2^8$  &256& 199 & $I2_13$ & $\Z_1$ & $\Z_1$ & $\Z_2$ & $\Z_6$   &12& 218 & $P\overline{4}3n$ & $\Z_1$ & $\Z_1$ & $\Z_2^2$ & $\Z_2^2$  &16\\
181 & $P6_422$ & $\Z_1$ & $\Z_1$ & $\Z_1$ & $\Z_2^8$  &256& 200 & $Pm\overline{3}$ & $\Z_1$ & $\Z_1$ & $\Z_1$ & $\Z_2^7\times \Z_6$   &768& 219 & $F\overline{4}3c$ & $\Z_1$ &&&\\
182 & $P6_322$ & $\Z_1$ & $\Z_2$ & $ \Z_2^2$ & $\Z_2^2$  &32& 201 & $Pn\overline{3}$ & $\Z_1$ & $\Z_1$ & $\Z_1$ & $\Z_2\times \Z_6$  &12& 220 & $I\overline{4}3d$ &  $\Z_1$ & $\Z_1$ & $\Z_2^2$ & $\Z_4$  &16\\
183 & $P6mm$ & $\Z_1$ & $ \Z_1$ & $\Z_1$ & $\Z_2^4$  &16& 202 & $Fm\overline{3}$ & $\Z_1$ & $\Z_1$ & $\Z_2$ & $ \Z_2\times \Z_6$  &24& 221 & $Pm\overline{3}m$ & $\Z_1$ & $\Z_1$ & $\Z_1$ & $\Z_2^{10}$   &1024\\
184 & $P6cc$ & $\Z_1$ & $\Z_2$ & $\Z_2 $ & $ \Z_1$   &4& 203 & $Fd\overline{3}$ & $\Z_1$ & $\Z_1$ & $\Z_2$ & $\Z_3$ &6& 222 & $Pn\overline{3}n$ & $\Z_1$ & $\Z_1$ & $\Z_2$ & $\Z_2^3$  &16\\
185 & $P6_3cm$ & $\Z_1$ & $\Z_1$ & $ \Z_2 $ & $ \Z_1$ &2& 204 & $Im\overline{3}$ & $\Z_1$ & $\Z_1$ & $\Z_1$ & $\Z_2^3\times \Z_6$  &48& 223 & $Pm\overline{3}n$ &  $\Z_1$ & $\Z_1$ & $\Z_1$ & $\Z_2^6$ &64\\
186 & $P6_3mc$ & $\Z_1$ & $\Z_1$ & $\Z_2 $ & $ \Z_1$  &2& 205 & $Pa\overline{3}$ & $\Z_1$ & $\Z_1$ & $\Z_2$ & $\Z_6$  &12& 224 & $Pn\overline{3}m$ & $\Z_1$ & $\Z_1$ & $\Z_1$ & $\Z_2^4$  &16\\
187 & $P\overline{6}m2$ & $\Z_1$ & $\Z_1$ & $\Z_1$ & $\Z_2^4$  &16& 206 & $Ia\overline{3}$ & $\Z_1$ & $\Z_1$ & $\Z_2$ & $\Z_3$ &6& 225 & $Fm\overline{3}m$ & $\Z_1$ & $\Z_1$ & $\Z_1$ & $\Z_2^7$ &128\\
188 & $P\overline{6}c2$ & $\Z_1$ & $\Z_2$ & $ \Z_2^2$ & $ \Z_2^2$  &32& 207 & $P432$ & $\Z_1$ & $\Z_1$ & $\Z_1$ & $\Z_2^6$  &64& 226 & $Fm\overline{3}c$ & $\Z_1$ & $\Z_1$ & $ \Z_2$ & $\Z_2^3$ &16\\
189 & $P\overline{6}2m$ & $\Z_1$ & $\Z_1$ & $\Z_1$ & $\Z_2^3\times \Z_6$   &48 & 208 & $P4_232$ & $\Z_1$ & $\Z_1$ & $\Z_1$ & $\Z_2^6$  &64 & 227 & $Fd\overline{3}m$ & $\Z_1$ & $\Z_1$ & $\Z_2$ & $\Z_2^2$ &8 \\
190 & $P\overline{6}2c$ & $\Z_1$ & $\Z_2$ & $\Z_2^2$ & $\Z_2\times \Z_6$  &96 & 209 & $F432$ & $\Z_1$ & $\Z_1$ & $\Z_1$ & $\Z_2^4$  & 16& 228 & $Fd\overline{3}c$ & $\Z_1$ &&& \\
191 & $P6/mmm$ & $\Z_1$ & $\Z_1$ & $\Z_1$ & $\Z_2^{12}$  &4096 & 210 & $F4_132$ & $\Z_1$ & && & & 229 & $Im\overline{3}m$ & $\Z_1$ & $\Z_1$ & $\Z_1$ & $\Z_2^6$ &64 \\
192 & $P6/mcc$ & $\Z_1$ & $\Z_1$ & $\Z_2$ & $\Z_2^5$   & 64& 211 & $I432$ & $\Z_1$ & $\Z_1$ & $\Z_1$ & $\Z_2^5$  & 32& 230 & $Ia\overline{3}d$ & $\Z_1$ & $\Z_1$ & $\Z_2$ & $\Z_2^2$ &8 \\
193 & $P6_3/mcm$ & $\Z_1$ & $\Z_1$ & $\Z_2$ & $\Z_2^3$  &16 & 212 & $P4_332$ & $\Z_1$ & $\Z_1$ & $\Z_2$ & $ \Z_2$  &4 & & & && \\
 \bottomrule
\end{tabular}
}
   \caption{Classification of  topological crystalline superconductors protected by space groups in interacting electronic systems for space group from No.175-230. \footnote{The three examples No.210,  219 and 228 are under computation, which will update soon.} }
    \label{tab:TCSc2}
\end{table*}

To obtain the final classification, we still need to quotient out those decorations that actually belong to trivial states, namely those decorations that can have symmetric short range entangled states on their surfaces \footnote{Here, we consider the symmetries to be internal, in contrast to spatial ones that may also have symmetric short range entangled states on their surfaces, such as the non-trivial 3D rotation SPT, which can have symmetric short range entangled states on their boundary that is parallel to the rotation axis.}.   Such states can correspond one-to-one-to the anomalous SPT states. In fact, such anomalous SPT are obtained in Ref.\cite{Wang2020}, which are labeled by $\Gamma_2,\Gamma_3$ and $\Gamma_4$. More explicitly, they are given by

    \begin{align}
\Gamma^2 &= \{\omega_2\cup n_0 \in H^2(G_b,\Z_2) | n_0\in H^0(G_b,\Z_T) \}, \nonumber \\
\Gamma^3 &= \{\omega_2 \cup n_1+s_1\cup n_1\cup n_1 + (\omega_2\cup \omega_2) \left\lfloor{n_0/2}\right\rfloor \in  \nonumber \\
&\quad H^3(G_b,\Z_2) | n_1\in H^1(G_b,\Z_2), n_0\in H^0(G_b,\Z_T) \}.\nonumber 
\end{align}
where $\left\lfloor{a}\right\rfloor$ is the floor function, namely it output the integral part of the number $a$.
The explicit form of $\Gamma_4$ can refer to Ref.\cite{Wang2020}. The $\Gamma_2$ will correspond to the trivial decoration layer KC $n_2$, $\Gamma_3$ corresponds to the trivialization of the decoration layer of complex fermions $n_3$ and $\Gamma_4$ will trivialize the bosonic SPT $\nu_4$. 
So, the final classification is given by the obstruction-free classification quotient out the trivialization, namely $(\{n_1\},\{n_2\}/\Gamma_2, \{n_3\}/\Gamma_3,\{\nu_4\}/\Gamma_4)$. We note that no trivialization for $n_1$.  More mathematical aspects of the trivialization can be refered to Sec.\ref{sec:math_ss}.

\subsection{3+1D  FSPT involving  spatial symmetries}
\label{sec:framework_for_spatial}

Building upon the framework for internal symmetries established above, we can now classify 3+1D TCS/FSPT phases involving spatial symmetries using the fermionic CEP. The key step is to determine the correspondence between a spatial (or spatial-internal) symmetry and an effective purely internal symmetry, which allows for classification under the fermionic CEP.

Following the notation in Sec.\ref{sec:fermion_symm}, we denote the spatial or spatial-internal symmetry $G_{f}$ by the triple $(G_b, \omega_2, s_1)$ and the effective purely internal symmetry $G_{f}^{\mathrm{eff}}$ by $(G_b, \omega_2^{\mathrm{eff}}, s_1^{\mathrm{eff}})$. We use the same symbol $G_b$ in both cases because, after quotienting out the fermion parity $\Z_2^f$, the bosonic symmetry groups are identical as abstract groups. The distinction lies in the anti-unitary structure, which is characterized by the group homomorphisms $s_1$ and $s_1^{\mathrm{eff}}$. 

To encode the geometric information of the spatial symmetries, we require additional quantities: $w_1$ and $w_2$. These are the pullbacks of the first and second Stiefel-Whitney classes, $\mathsf{w}_1\in H^1(BO(3),\Z_2)$ and $\mathsf{w}_2\in H^2(BSO(3),\Z_2)$, respectively. We adopt the conjecture proposed in Ref.\cite{Manjunath2023}, which is supported by the mathematical framework in Ref.\cite{Debray2021} and has been verified through various examples:
\begin{align}
s_{1} ^{\mathrm{eff}} &=s_{1} +w_{1} ,\label{eq6.1}\\
\omega _{2} ^{\mathrm{eff}} &=\omega _{2} + w_{2} + w_{1} \cup ( s_{1} + w_{1}) .\label{eq6.2}
\end{align}
Based on this correspondence principle, we can map the classification of topological crystalline phases protected by a space group (more generally, mixed spatial-internal symmetries) to the classification of phases protected by internal symmetries.
   
We note that the framework established here does not strictly depend on the specific mapping described in Eqs.\eqref{eq6.1} and \eqref{eq6.2}. In other words, even if the mapping requires adjustment for more intricate situations, our general framework remains applicable using the adjusted mapping. However, given that we have verified the validity of Eqs.\eqref{eq6.1} and \eqref{eq6.2} for 2D wallpaper groups and 3D point groups, we expect these relations to hold for 3D space groups as well.

\section{Decoration and classification from  $p+ip$ layer} 
\label{sec:decoupling_pplusip}
Here, we address the challenging layer of decoration: the $p+ip$ superconductor decoration on 2D domain walls. We first discuss the decoration procedure and rederive the $O_3$ obstruction~\cite{Wang2020}. However, the absence of a fixed-point wavefunction for $p+ip$ superconductors hinders the derivation of higher layers of obstruction. Consequently, we adopt an alternative approach to obtain the classification.

\subsection{Decoration of $p+ip$ layer}
We first review the decoration of the $p+ip$ layer. For a general description of the decoration procedure, refer to Sec.~\ref{sec:decoration}. Roughly speaking, given a spatial manifold $\mathcal{M}$, we require a triangulation $\mathcal{T}$ that admits a branching structure, achieved by labeling each vertex with a natural number.

\begin{figure}
     \centering
     \includegraphics[width=0.8\linewidth]{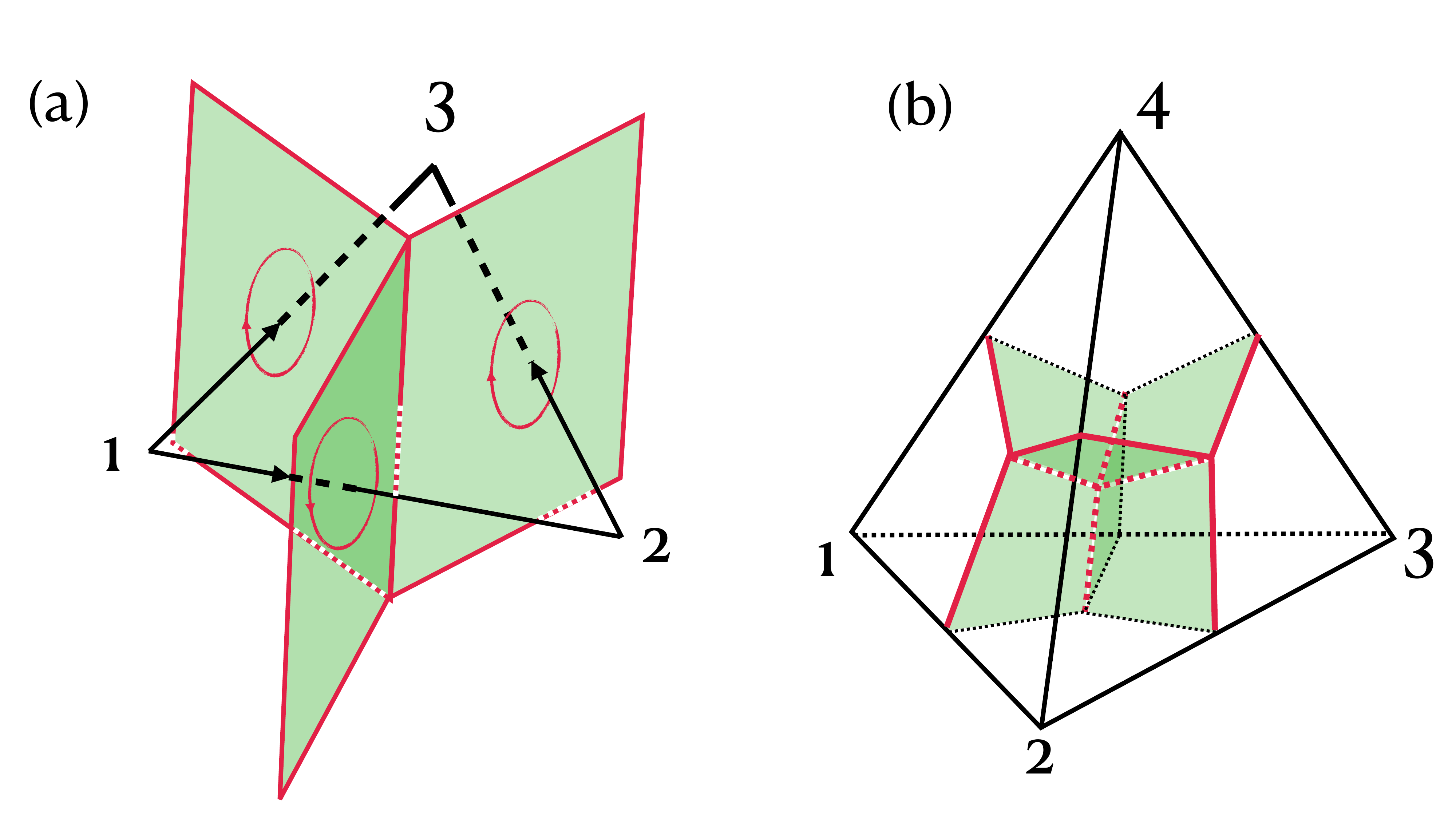}
     \caption{(a). The chiral or antichiral Majorana fermions of $p+ip$ ($p-ip$) superconductor decoration on the dual planes of the three links of the triangle $\langle 123\rangle$ meet at the link dual to the triangle. If $n_1(12)>0$ ($n_1(12)<0$), then there is $|n_1(12)|$ gapless Majorana fermions propagating into (out of) the triangle. (b) The chiral fermions meet in a tetrahedron. }
     \label{fig:p+ip_on_2cell}
 \end{figure}

As shown in Fig.\ref{fig:p+ip_on_2cell}, for each plane dual to a link $\langle ij\rangle$ of $\mathcal{T}$, there are $|G_{b}|$ species of fermions. Among these, $n_1(g_i,g_j)>0$ (or $<0$) species are assigned to $p+ip$ (or $p-ip$) chiral superconductors with boundary chiral Majorana fermions $\psi _{ij,R,\alpha}^{g_i }$ or $\psi _{ij,L,\alpha}^{g_i }$ ($\alpha=1,2...,n_1(g_i,g_j)$). Depending on the sign of $n_1(g_i,g_j)$, the chirality of the chiral Majorana fermions is determined by the right- or left-hand rule with respect to the orientation of the link $\langle ij \rangle$. For example, in Fig.~\ref{fig:p+ip_on_2cell}(a), where all $n_1(ij)$ are positive, the chirality of the boundary Majorana fermions is determined by the right-hand rule, as indicated by the directed circle around the links. We refer to these as chiral fermions hereafter.

The chiral fermions of $p \pm ip$ superconductors transform under symmetry as:
\begin{align}
&U( g) \psi _{ij,R,\alpha}^{\sigma } U^{\dagger }( g) =( -1)^{\omega _{2}( g,\sigma )} \psi _{ij,g(R),\alpha}^{g\sigma } , \label{eq:chi_ferm_sym1}\\
&U( g) \psi _{ij,L,\alpha}^{\sigma } U^{\dagger }( g) =( -1)^{\omega _{2}( g,\sigma ) +s_{1}( g)} \psi _{ij,g(L),\alpha}^{g\sigma } .\label{eq:chi_ferm_sym2}
\end{align}
Here, if $g$ is unitary, the chirality remains unchanged after the symmetry action ($g(R) = R$ and $g(L) = L$). Conversely, if $g$ is anti-unitary, the chirality is switched ($g(R) = L$ and $g(L) = R$). We note that while there is no fixed-point wavefunction for the chiral $p+ip$ superconductor, we can model them as free-fermion $p+ip$ superconductors with infinite mass terms. Consequently, the only degrees of freedom that remain relevant at low energies are their boundary chiral fermions.

The $p+ip$ states decorated on the three planes dual to the links of a triangle meet at the link dual to the triangle. This results in chiral fermions propagating along this link, either incoming or outgoing the triangle. As seen in Fig.~\ref{fig:p+ip_on_2cell}(a), for the plane dual to link $\langle 12\rangle$, there are $n_1(12)$ chiral fermions passing through the triangle; if $n_1(12)>0$ ($n_1(12)<0$), these fermions propagate into (out of) the triangle, according to the right (left) hand rule. Similarly, we have chiral fermions for the planes dual to links $\langle 13\rangle$ and $\langle 23\rangle$. To gap out these gapless fermions, the total number of fermions propagating into the triangle must equal the number propagating out. In other words, the chiral central charge along the link dual to the triangle must be zero. This leads to the following condition on $n_1$:
 \begin{align}
     n_1(g_1,g_2)-n_1(g_1,g_3)+n_1(g_2,g_3)=0.
 \end{align}
We define the inhomogeneous 1-cocycle $n_1(g):=n_1(e,g)$ (see Appendix \ref{ap:group cohomology} for group cohomology notation regarding homogeneous/inhomogeneous cochains). Using the relation $n_1(g_1,g_2)=(-1)^{s_1(g_1)}n_1(e,g_1^{-1}g_2)$, the condition above transforms into:
  \begin{align}
     n_1(g)-n_1(gh)+(-1)^{s_1(g)}n_1(h)=0,
     \label{eq:1_cocycle}
 \end{align}
 where we have defined $g=g_1^{-1}g_2$ and $h=g_2^{-1}g_3$. This condition implies that $n_1\in H^1(G_b, \Z_T)$. For a finite or compact group, $H^1(G_b, \Z_T)=\Z_2$, and the representative cocycle takes the form:
 \begin{align}
     n_1(g)=ks_1(g),
 \end{align}
where an even (odd) $k$ represents a trivial (non-trivial) class. For an infinite discrete group, such as the translation group, we have $H^1(\Z, \Z_T)=\Z$. These classes are labeled by integers $k\in\Z$, and the corresponding cocycle is $n_1(t^n)=nk$, with $t$ being the generator of the group and $n\in\Z$. For a generic discrete $G_b$, $H^1(G_b, \Z_T)$ may contain both a free part $\Z^r$ arising from the infinite structure and a torsion part $\Z_2$ arising from the anti-unitary structure of $G_b$.

Furthermore, if the condition in Eq.\eqref{eq:1_cocycle} is satisfied, we can gap out these fermions by adding symmetric mass terms. To obtain symmetric mass terms for all vertex configurations, we first fix the mass terms for the standard triangle, and then obtain terms for non-standard triangles by applying the symmetry action $U(g_1)$. We consider the mass terms for the standard triangle $\langle 123\rangle$, with group elements $e, g_1^{-1}g_2, g_1^{-1}g_3$ respectively, taking the form (assuming $m>0$):
 \begin{align}
     im \psi_\mathrm{out}^e \psi_\mathrm{in}^{g_1^{-1}g_i} \quad \mathrm{or} \quad im \psi_\mathrm{out}^{g_1^{-1}g_i} \psi_\mathrm{in}^e. \label{eq:mass_rule}
 \end{align}
Since mass terms with different signs result in gapped states differing by a Kitaev chain (corresponding to negative mass), we adopt the rule of always placing the outgoing chiral fermion at the front of the mass term to consistently count the effective Kitaev chain along the link dual to the triangle. Whether the chiral fermions are outgoing or incoming depends not only on the sign of $n_1$ but also on the orientation of the triangle's links. For example, in Fig.\ref{fig:p+ip_on_2cell}(a), assuming $n_1(e,g_1^{-1}g_2)$, $n_1(e,g_1^{-1}g_3)$, and $n_1(e,g_2^{-1}g_3)$ are all positive, we can add the mass terms as follows (assuming $m>0$):
\begin{align}
    H_{123}^\mathrm{std}= \int dx &\sum_{\alpha=1}^{n_1(12)} i m  \psi_{12,R;\alpha}^{e} \psi_{13,R;\alpha}^{e} \nonumber \\
    +& \sum_{\alpha=1}^{n_1(23)} i m  \psi_{23,R;\alpha}^{g_2^{-1}g_3} \psi_{13,R;\alpha}^{e},
\end{align}
where we denote $n_1(12)=n_1(e,g_1^{-1}g_2)$ and $n_1(23)=n_1(e,g_2^{-1}g_3)$.
Using the right-hand rule, the chiral fermions $\psi_{12,R;\alpha}^{e}$ and $\psi_{23,R;\alpha}^{g_2^{-1}g_3}$ in Fig.\ref{fig:p+ip_on_2cell}(a) are outgoing from the triangle, so these mass terms satisfy the rule in Eq.\eqref{eq:mass_rule}. For other patterns of $n_1$, we can explicitly write down the mass terms respecting Eq.\eqref{eq:mass_rule}. The mass terms for standard triangles are chosen to be positive, ensuring no effective Kitaev chain passes through them. 

Mass terms for non-standard triangles are obtained by acting with the symmetry transformation $U(g_1)$, which transforms the vertices from $(e,g_1^{-1}g_2,g_1^{-1}g_3)$ to $(g_1,g_2,g_3)$. (By a standard simplex, we mean that the group element associated with its first vertex is the identity.) According to the symmetry properties of chiral fermions in Eqs.\eqref{eq:chi_ferm_sym1} and \eqref{eq:chi_ferm_sym2}, some mass terms will change sign under this symmetry, leading to the presence of an effective Kitaev chain across the non-standard triangle. Evaluating all possible configurations, the number (mod 2) of effective Kitaev chains across the triangle with vertices $(g_1,g_2, g_3)$ is given by:
\begin{align}
    [\omega_2(g_1,g_1^{-1}g_2)+s_1(g_1)n_1(g_1^{-1}g_2)]n_1(g_2^{-1}g_3).
\end{align}
Thus, for a tetrahedron $\langle 1234\rangle$, as in Fig.\ref{fig:p+ip_on_2cell}(b), the total number of Kitaev chains (including both effective and decorated ones) can be calculated by summing the contributions from each triangle, resulting in (mod 2):
    \begin{align}
     O_3(g_1,g_2,g_3,g_4)&= (\omega_2\cup n_1 + s_1\cup n_1\cup n_1)(g_1,g_2,g_3,g_4)\nonumber \\
     &=\omega_2(g_1^{-1}g_2,g_2^{-1}g_3) n_1(g_3^{-1}g_4) +\nonumber \\
     & \,\,\quad  s_1(g_1^{-1}g_2) n_1(g_2^{-1}g_3) n_1(g_3^{-1}g_4).
     \label{eq:O3_main}
     \end{align}
If $O_3$ equals one, the effective Kitaev chain would naively leave a single (effective) Majorana zero mode inside the tetrahedron, violating the gapped condition. However, we can still decorate a Kitaev chain on each link dual to the triangle, specified by $n_2(g_1,g_2,g_3)$ for each triangle (we discuss this in more detail in Sec.\ref{sec:ddw_3Danti}), which will interplay with the effective Kitaev chain. The total number of decorated Kitaev chains is given by (mod 2):
   \begin{align}
       dn_2(g_1,g_2,g_3,g_4) &=n_2(g_1,g_2,g_3)+n_2(g_1,g_2,g_4)\nonumber \\ & \,\,\,+n_2(g_1,g_3,g_4)+n_2(g_2,g_3,g_4).
   \end{align}
Therefore, to avoid dangling Majorana zero modes in the tetrahedron, the total number of all Kitaev chains, both effective and decorated, must equal zero modulo 2. This leads to the following condition:
 \begin{align}
     dn_2(g_1,g_2,g_3,g_4) &= O_3(g_1,g_2,g_3,g_4).
     \label{eq:O3}
     \end{align}
 If $n_1=0$, this condition is satisfied as long as $n_2$ is a 2-cocycle. For nonzero $n_1$, if $O_3$ is a 3-coboundary in $B^3(G_b,\Z_2)$, the resolution of $O_3$ can still avoid a single Majorana zero mode in the tetrahedron (mod 2) by choosing $n_2$ as a torsor over $H^2(G_b, \Z_2)$. However, if $O_3$ is not a coboundary, this condition cannot be satisfied. This implies that the non-trivial decoration of the $p+ip$ superconductor specified by nonzero $n_1$ cannot result in a well-defined FSPT state. In this case, we call this $p+ip$ superconductor decoration ``obstructed", meaning it will not contribute to the classification of $G_f$ FSPT.

For finite symmetry, we can have a non-trivial $p+ip$  decoration layer only if $[n_1]=[s_1]$; therefore, when $\omega_2+s_1\cup s_1$ is a 2-coboundary in $H^2(G_b, \Z_2)$, $O_3$ is a coboundary. Then there exists a 1-cochain $u_1\in C^1(G_b,\Z_2)$ such that $\omega_2+s_1\cup s_1=du_1$. We can add the coboundary $d u_1$ to $\omega_2$ to define a new $\tilde \omega_2$ for a new fermionic symmetry $\tilde G_f$ that has the same $G_b$ and is isomorphic to $G_f$. Since $G_f$ and $\tilde G_f$ must have the same FSPT classification, we work with $\tilde G_f$. In this frame, $O_3=0$, implying that $n_1$ is decoupled from $n_2$ and potentially from other layers of decoration. This observation motivates the claim that for $[n_1]=[s_1]$, whether a valid FSPT can be formed is determined purely by whether $\omega_2+s_1\cup s_1$ is a coboundary. In the following, we provide a rigorous argument to support this claim.


\subsection{No higher obstruction of $p+ip$ layer from anti-unitary structure}
\label{sec:crystalZ2T}
Now we address the question of how the decoration of the $p+ip$ layer contributes to the classification of FSPT phases protected by the internal symmetry $G_\mathrm{f}=(G_\mathrm{b},\omega_{2}, s_{1})$. As established above, the decoration of the $p+ip$ layer is characterized by a 1-cocycle $n_1\in H^1(G_{\mathrm{b}}, \Z_T)$. As mentioned in Sec.~\ref{sec:framework_for_internal}, this cohomology group is isomorphic to $\Z^r$ for unitary symmetry or $\Z^r\times \Z_2$ for anti-unitary symmetry. The free part ($\Z^r$) originates from discrete elements of infinite order, while the torsion part ($\Z_2$) arises from the anti-unitary structure. Below, we discuss these two parts separately.

First, we focus on the $\Z_2$ contribution. For finite or compact groups, a non-trivial contribution arises only when $G_\mathrm{f}$ contains anti-unitary elements. We can decompose $G_\mathrm{f}$ into the disjoint union $G_\mathrm{f}=G_\mathrm{f,1} \cup G_\mathrm{f,2}$, where $G_\mathrm{f,1}$ comprises all unitary operators in $G_\mathrm{f}$, and $G_\mathrm{f,2}$ contains all anti-unitary operators. Since the product of two unitary elements is unitary, the set $G_\mathrm{f,1}$ forms a subgroup of $G_\mathrm{f}$. Furthermore, since the product of two anti-unitary elements is unitary, $G_\mathrm{f,1}$ is a normal subgroup of $G_\mathrm{f}$. Consequently, the quotient group is $G_\mathrm{f}/G_\mathrm{f,1} \cong \Z_2^T=\{1,T\}$, yielding the following short exact sequence:
\begin{align}
    1\rightarrow G_\mathrm{f,1}\rightarrow G_\mathrm{f}\rightarrow \Z_2^T\rightarrow 1.
\end{align}
In other words, $G_\mathrm{f}$ is a group extension of $\Z_2^T$ by $G_\mathrm{f,1}$.

With this understanding of the structure of $G_\mathrm{f}$, we now invoke the fermionic CEP. We map the time-reversal symmetry $\Z_2^T$ to a crystalline mirror symmetry $\Z_2^M$. According to the fermionic CEP, the classification of $G_\mathrm{f}$-FSPT phases is in one-to-one correspondence with the classification of crystalline SPT phases protected by a group $G_\mathrm{f}^{\mathrm{m}}$, which contains both internal symmetry and crystalline (mirror) symmetry.
Using the notation of Sec.~\ref{sec:fermion_symm}, $G_\mathrm{f}^{\mathrm{m}}$ is denoted as $(G_\mathrm{b}, \omega_{2}^{\mathrm{m}}, s_1^{\mathrm{m}} = 0)$. The condition $s_1^\mathrm{m} = 0$ reflects the fact that all elements in $G_\mathrm{f}^{\mathrm{m}}$ are unitary.

\begin{figure}
    \centering
    \includegraphics[width=0.7\linewidth]{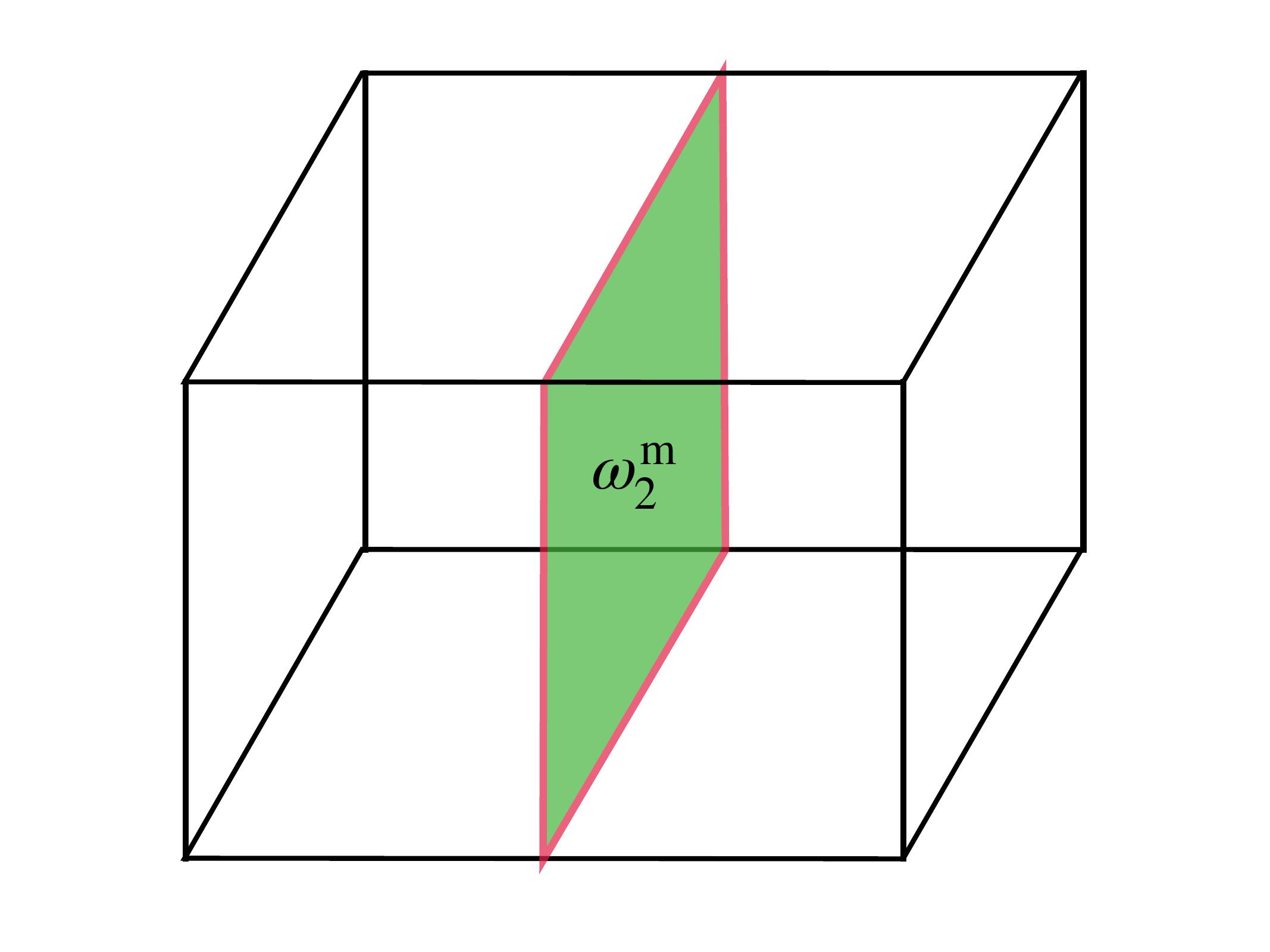}
    \caption{3D $G^{\mathrm{m}}_\mathrm{f}$ FSPT from $p+ip$ superconductor state decorated on the mirror plane. On the mirror plane, the total symmetry is $G^{\mathrm{m}}_\mathrm{f}$, whether the $p+ip$ state is compatible with $G^{\mathrm{m}}_\mathrm{f}$  is fully determined by the fact whether $\omega_2^{\mathrm{m}}=\omega_2+s_1\cup s_1$ is coboundary or not. } 
    \label{fig:mirror}
\end{figure}

Among the classification of $G_\mathrm{f}^{\mathrm{m}}$ FSPT, we are interested in the phases 
that assign the $p+ip$ state to the mirror plane (see Fig.\ref{fig:mirror}), which  correspond to the $p+ip$ layer decorated states of $G_\mathrm{f}$.  On the mirror plane, the total symmetry now is  $G_\mathrm{f}^{\mathrm{m}}$ which now is internal since  $\Z_2^M$ now is  internal symmetry on the mirror plane.   So,  the key that whether the assigned $p+ip$ states on the mirror plane is a valid  $G_\mathrm{f}^{\mathrm{m}}$ FSPT is determined by    whether it is compatible with $G_\mathrm{f}^{\mathrm{m}}$ or not, namely whether there is  enforced symmetry breaking of  $G_\mathrm{f}^{\mathrm{m}}$ symmetry  by the $p+ip$ states \cite{Ning2023enforced}. According to Ref.\cite{Ning2023enforced},   $p+ip$ state is compatible with the unitary fermionic symmetry $ G_\mathrm{f}^{\mathrm{m}}=(G_b,\omega_{2}^\mathrm{m})$ as long as  $[\omega_{2}^\mathrm{m}]=[0]$.    
According to Sec.\ref{sec:fermion_symm}, we have 
\begin{align}
\omega_{2}=\omega_{2}^\mathrm{m}+w_1\cup w_1
\end{align}
As $w_1$ is identical to $s_1$ in $G_{\mathrm{b}}$, we can have
\begin{align}
\omega_{2}^\mathrm{m}=\omega_{2}+s_1\cup s_1.
\end{align}
Therefore, we conclude that if $[\omega_{2}+s_1\cup s_1]$ is trivial (non-trivial), the $p+ip$ state on the mirror plane is (is not) a valid mirror FSPT with $G_{\mathrm{f}}^\mathrm{m}$ symmetry. In other words, the triviality of $[\omega_{2}+s_1\cup s_1]$ is the necessary and sufficient condition for the obstruction-free existence of such a $G_{\mathrm{f}}^\mathrm{m}$ FSPT, and by CEP, for its crystalline counterpart—the 3D $G_{\mathrm{f}}$ FSPT derived from the $p+ip$ layer specified by $n_1=s_1$.

For example, the well-known 3D topological superconductor with $T^2=P_f$ is classified by $\Z_{16}$, where the root state is the $p+ip$ decoration given by $n_1=s_1$. For the symmetry group $\Z_4^{Tf}$ (where $T^2=P_f$), it is established that $\omega_2=s_1\cup s_1$. Consequently, $\omega_{2}^\mathrm{m}=\omega_{2}+s_1\cup s_1 \equiv 0 \pmod 2$. Thus, the $p+ip$ layer decoration yields a valid 3D $T^2=P_f$ topological superconducting state. In contrast, the 3D $T^2=1$ superconductor has a trivial classification with $\omega_2=0$; this implies $\omega_{2}^\mathrm{m}=s_1\cup s_1$ is non-trivial, rendering the $p+ip$ layer decoration obstructed.

A remarkable observation follows from the definition of $O_3$ in Eq.(\ref{eq:O3_main}): when the class $[\omega_{2}+s_1\cup s_1]$ is trivial, the obstruction $O_3$ is also trivial. Recalling the obstruction structure in Eqs.(\ref{eq:obs3})–(\ref{eq:obs4}), the necessary and sufficient condition for the decoration $n_1=s_1$ to be obstruction-free implies the following:
\begin{align}
       &\textit{For a discrete group, the $O_4$ and $O_5$ obstructions for} \nonumber \\
       &\textit{the $p+ip$ layer decoration specified by $n_1=s_1$ }\nonumber \\
       &\textit{vanish provided that $O_3$ is a coboundary.}\nonumber
\end{align}
In other words, the vanishing of higher obstruction functions beyond $O_3$ is guaranteed (i.e., they are coboundaries) as long as $O_3[n_1=s_1]$ is a coboundary.\footnote{More precisely, $O_4$ and $O_5$ may explicitly depend on the choice of lower-order data. Thus, the statement should be interpreted as follows: If $O_3[n_1=s_1]\in B^3(G_b,\mathbb{Z}_2)$, there must exist $n_2$ and $n_3$ such that $O_4[n_1=s_1,n_2] \in B^4(G_b,\mathbb{Z}_2)$ and $O_5[n_1=s_1,n_2,n_3]\in B^5(G_b, U(1))$.}

The following remark is in order. Since the compatibility condition depends solely on the group structure, it is independent of the specific $G_\mathrm{f}^\mathrm{m}$ symmetry enrichment of the $p+ip$ states. Consequently, whether this construction yields a valid 3D $G_\mathrm{f}^\mathrm{m}$ FSPT state depends only on the group structure of $G_\mathrm{f}^\mathrm{m}$ and is decoupled from the topological quantities characterizing other 3D $G_\mathrm{f}^\mathrm{m}$ FSPT phases. Accordingly, whether the $p+ip$ layer decoration leads to a valid 3D $G_\mathrm{f}$ FSPT state is determined exclusively by its group structure and remains decoupled from the decorations of other layers.

\subsection{$p+ip$ layer from discrete infinite structure}
\label{sec:p+ip_infinite}

Next, we consider the free part, $\mathbb{Z}^r$. The rank $r$ is a non-negative integer determined by the free component of the group structure of $G_{\mathrm{b}}$. A physical origin for this free component in the context of the $p+ip$ layer is typically the translation group. We argue that the $\mathbb{Z}^r$ subgroup of $H^1(G_{\mathrm{b}}, \mathbb{Z}_T)$ will always contribute a $\mathbb{Z}^r$ factor to the FSPT classification.

This can be justified as follows. Consider a decoration labeled by the 1-cocycle $t n_1$, where $t$ is an integer constant and $n_1$ is a generator of the free part of $H^1(G_{\mathrm{b}}, \mathbb{Z}_T)$. To determine whether this decoration yields an obstruction-free FSPT state, we must, in principle, check three obstruction conditions characterized by $O_3$, $O_4$, and $O_5$. These obstruction classes $O_i$ are functions of $t$. Specifically, for the trivial case $t=0$, we have $O_i[t=0](g_1, \dots, g_i) = 0$  for $i=3,4$ and $O_5[t=0](g_1, \dots, g_5) = 1$ for any set of group elements $(g_1, \dots, g_i)$.\footnote{Here we use the inhomogeneous expression of the $i$-cocycle $O_i$, so they depend on only $i$ group elements.} This holds for at least one choice of higher-level decorations, specifically by setting $n_2=n_3=0$ and $\nu_4=1$ (see Sec.\ref{sec:antiuni_SPT} for the precise definition).

We can therefore express these functions in the form:
$$O_3[t](g_1,g_2,g_3)=F_3[t](g_1,g_2,g_3) t \,\,\text{ mod }2$$
$$O_4[t](g_1,g_2,g_3,g_4)=F_4[t](g_1,g_2,g_3,g_4) t\,\,\text{ mod }2$$
$$O_5[t](g_1,g_2,\dots,g_5)=e^{i \frac{2\pi}{p} F_5[t](g_1,g_2,\dots,g_5) t}$$
where $F_i[t](g_1, \dots, g_i)$ is an integer-valued function of the group elements (and possibly $t$), and $p$ is an integer. The integer nature of $p$ follows from the assumption that the phase factors of $O_5$ (derived from the super-coherence condition) are rational multiples of $2\pi$ (up to a coboundary transformation).\footnote{Since it has been observed that the most fractional values of $O_5$ arising from the complex fermion layer and Kitaev chain layer are $\pi/2$ and $\pi/4$, one may conjecture that the most fractional value arising from the $p+ip$ layer is $\pi/8$.} We leave a rigorous proof of this for future work.

Consequently, we can choose $t$ to be even, ensuring that the $O_3$ and $O_4$ obstructions vanish. Furthermore, by choosing $t = 2 p \tilde{t} \in \mathbb{Z}$, we ensure that $O_5$ also vanishes. To conclude the classification argument, we observe that any decoration $2 p \tilde{t} n_1$ with non-zero $\tilde{t}$ generates a distinct state, thereby establishing a $\mathbb{Z}$ classification. We note that analogous integer ($\mathbb{Z}$) classifications appear in bosonic systems~\cite{chen13} and free fermion systems~\cite{Kitaev_2009}.

We now focus on the case where the translation group forms a normal subgroup, as is typical for spatial crystalline groups. The resulting phases are commonly referred to as weak SPT phases. We analyze the following scenarios:

\begin{enumerate}
\item Symmetry $Q=K\times \Z_2^f$ where $K$ is unitary. 
The root weak SPT state is constructed by stacking 2D $p+ip$ states along a translation direction, separated by a unit length. Such a state is characterized by a 1-cocycle $n_1$ satisfying $n_1(a_g)=n_1(a)=1$, where $a$ is the generator of the translation group $\Z$ and $g\in K$. This 1-cocycle can be understood as the result of lifting the generator of the translation cohomology group $\mathcal{H}^1(\Z, \Z)$ to the total group cohomology $\mathcal{H}^1(\Z\times K, \Z)$.

\item Symmetry $Q=K\times_{\omega_2} \Z_2^f$ where $K$ is unitary and $\omega_2$ is non-trivial.  
In this case, the root weak SPT is not a stack of single $p+ip$ states. Instead, it is a stack of a distinct invertible topological order (TO) $[\nu]$, which corresponds to $\nu$ copies of $p+ip$ states.

\item Symmetry $Q$ contains anti-unitary elements.  
If the internal symmetry group is anti-unitary, no weak SPT based on stacking invertible chiral TOs is allowed. This is because the anti-unitary symmetry maps the state $[\nu]$ to $[-\nu]$, which forbids a non-trivial stacking classification unless the symmetry acts on the translation direction in a specific way (discussed below).

\item General case: $Q$ acts on translation symmetry.  
For a 3D space group, we can decompose the group $G$ via the following extension:
\begin{align}
    1\rightarrow \Z^3\rightarrow G \rightarrow Q\rightarrow 1. 
\end{align}
This extension is characterized by the action $\rho: Q \to \text{Aut}(\Z^3)$ and the second cohomology class $\mathbf{w}_2 \in \mathcal{H}_\rho^2(Q,\Z^3)$. A generic 1-cocycle in $\mathcal{H}^1(\Z^3,\Z)$ (with trivial action on the coefficients) takes the form:
\begin{align}
    n_1(a)=t_1 a_1 + t_2 a_2+t_3 a_3,
\end{align}
where $t_i, a_i\in \Z$, and $a=(a_1,a_2,a_3)$ denotes a generic element of the translation group $\Z^3$. 

To lift this 1-cocycle from $\Z^3$ to the extended group $G$ in the presence of anti-unitary symmetries (which act as $-1$ on the cohomology coefficients), the action of $Q$ on the translation sector must satisfy the invariance condition:
\begin{align}
    \rho_h(a)=(-1)^{s_1(h)} a.
\end{align}
Here, $s_1(h)=1$ if $h$ is anti-unitary and $0$ otherwise. This implies that only anti-unitary elements of $Q$ can act non-trivially on the specific translation direction $a$, and they must invert it. The action $\rho$ is diagonal (acting separately on each component of $\Z^3$), then any $h \in Q$ maps $a_i$ to $(-1)^{s_1(h)} a_i$.

Under this condition, the obstruction to lifting the 1-cocycle is determined by the class $\mathbf{w}_2=(w_1,w_2,w_3)\in (\mathcal{H}_\rho^2(Q,\Z))^3$. For a valid lifting of the vector $t=(t_1, t_2, t_3)$, we require the product $t_i  w_i$ to be trivial. Since $Q$ is finite, $\mathcal{H}_\rho^2(Q,\Z)$ is a torsion group. Consequently, there always exists an integer scalar for $t_i$ such that the obstruction vanishes. Therefore, as long as the action satisfies $\rho_h=(-1)^{s_1(h)}$, a valid 1-cocycle for $G$ exists, generating a $\Z$ classification.

We conclude that this is the only non-trivial condition for counting the obstruction-free $p+ip$ classification derived from translation symmetry. Specifically, a translation direction contributes a $\Z$ classification if and only if the $Q$ action on that direction satisfies $\rho_h=(-1)^{s_1(h)}$; otherwise, it does not contribute.
\end{enumerate}

\subsection{Classification}

In summary, the classification of the $p+ip$ layer is generally given by $\mathbb{Z}_2^s \times \mathbb{Z}^r$, where the integers $s$ and $r$ depend on whether the symmetry group $G_b$ is unitary or anti-unitary, as well as on the free part of the cohomology of $G_b$. Explicitly, we distinguish between the following two cases:

\begin{enumerate}
\item $G_b$ is unitary. In this case, $H^1(G_b, \mathbb{Z}_T) \cong \mathbb{Z}^r$, leading to a $\mathbb{Z}^r$ classification for the $p+ip$ layer.
\item $G_b$ contains anti-unitary symmetries. In this case, $H^1(G_b, \mathbb{Z}_T) \cong \mathbb{Z}^r \times \mathbb{Z}_2$, which results in a classification of $\mathbb{Z}^r \times \mathbb{Z}_2^s$.
\end{enumerate}

For the anti-unitary case, the non-trivial $\mathbb{Z}_2$ factor of $H^1(G_b, \mathbb{Z}_T)$ is generated by the representative cocycle $n_1(g) = s_1(g)$, which takes values $0$ or $1$ depending on whether $g$ is unitary or anti-unitary, respectively. The parameter $s$ determines whether this phase survives: $s=1$ if the obstruction class $\omega_2^{\mathrm{m}}$ is trivial in $H^2(G_b, \mathbb{Z}_2)$, and $s=0$ if it is non-trivial. Following the fermionic CEP in Eqs.\eqref{eq6.1} and \eqref{eq6.2}, the obstruction class is given by $\omega_2^{\mathrm{m}} = \omega_2 + s_1 \cup s_1$.

\section{Domain wall decoration and classification of 3+1D FSPT  beyond $p+ip$ layer  } 
\label{sec:antiuni_SPT}
In this section, we develop the domain wall decoration (DWD) framework for (3+1)D fermionic symmetry-protected topological (FSPT) phases protected by discrete internal symmetries, with a particular focus on generic anti-unitary symmetries. We detail the decoration procedure involving KC layers and complex fermion layers, as well as the stacking of bosonic SPT phases. Specifically, we derive the necessary formulas for anti-unitary symmetries, including the explicit form of the $O_5$ obstruction function. Combining these results allows us to establish the complete classification of (3+1)D FSPT phases. The corresponding mathematical framework is elaborated upon in Sec.~\ref{sec:math_ss}.

\subsection{Domain wall decoration of Kitaev chain and complex fermion  layers} 
\label{sec:ddw_3Danti}

Below we will first review the basic setup of two fermionic decoration layers, which aligns with those discussed in Ref.\cite{Wang2020}. With this preparation, we will discuss the  $F$ move generally for both unitary and anti-unitary symmetries, in which anti-unitary symmetry presents essential new aspects. The new $F$ move should satisfy the so-called super hexagon equation, a condition for the fixed point wavefunction, and will lead to new $O_5$ obstruction function, which will reduce to the previous results when restricting to unitary symmetries.    A remarkable new feature of the new $O_5$ function is that it may be inevitable to involve the $e^{i\pi/4}$ phase factors when considering the KC decoration, as compared to that $e^{i\pi/2}$ is sufficient  for unitary symmetries. In this way, it is more ``fractionalized'' for the KC decoration of anti-unitary symmetries than unitary ones.

\subsubsection{Decoration of two fermionic layers} \label{sec:decoration}

\begin{figure}
    \centering
    \includegraphics[width=\linewidth]{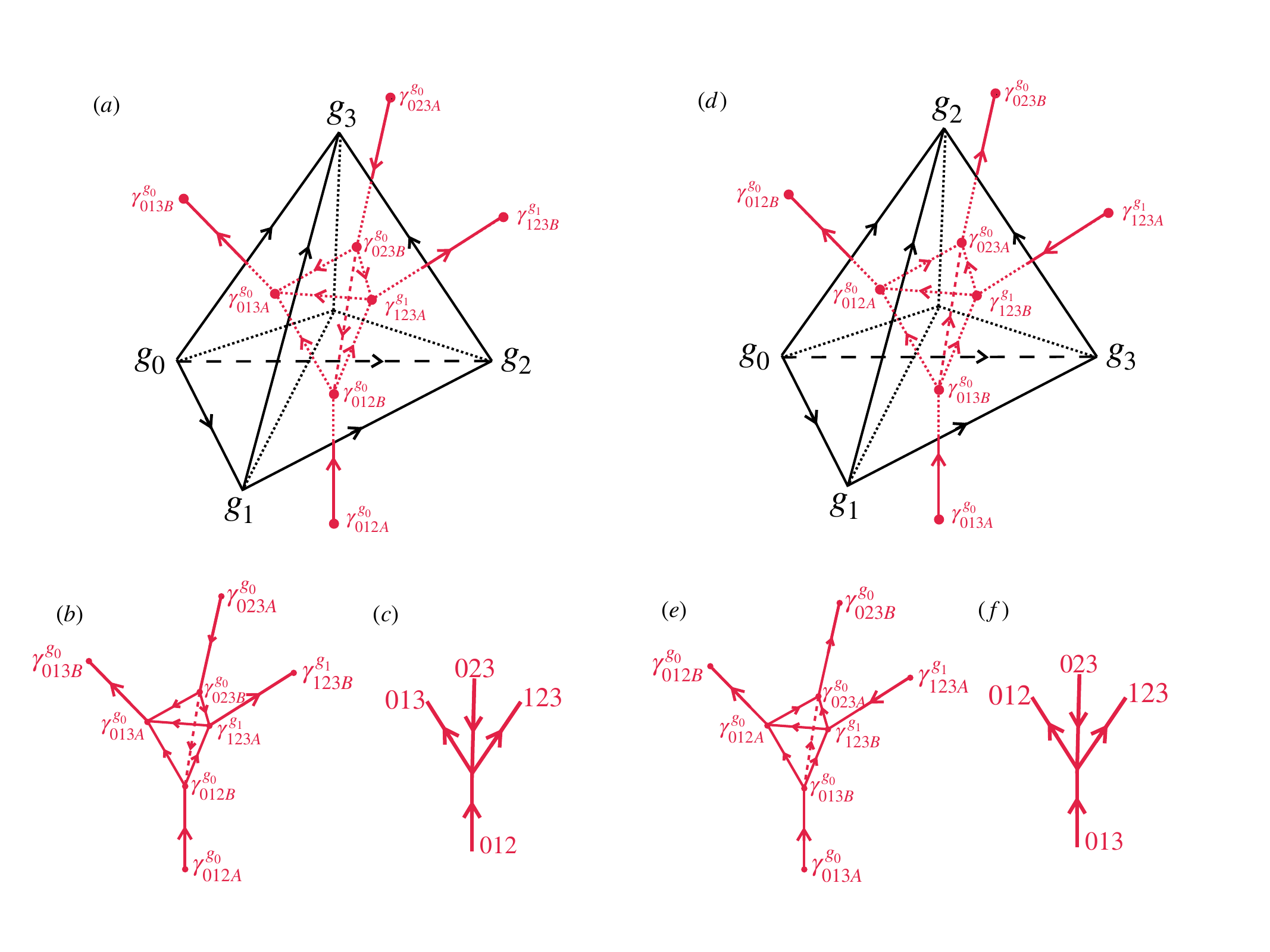}
    \caption{The resolved dual lattice. The black tetrahedra in (a) is the positive oriented with the red in (a) sketch the resolved dual lattice.  (b) only present the resolved dual lattice corresponding to the (black) tetrahedra in (a), while (c) is the simplified version of (b) for convenience. (d)-(f)  is for the negative oriented.}
    \label{fig:3D_resolved_dual_positive_negative}
\end{figure}

 The FSPT state can be defined in any 
 {triangulation} with a branching structure of a spatial manifold $\mathcal{M}$.
A convenient branching structure is to label every vertex of the triangulation $\mathcal{T}$ (or lattice site) with a natural number $n \in \mathbb N$ and
then  assign every link $\langle ij \rangle$ ($i<j$) with a direction from $i$ to $j$.
This choice of link orientation guarantees that there is no oriented loop in any triangle (2-simplex), so a branching structure is obtained.  For our decoration purpose, we  define a resolved dual lattice $\tilde{\mathcal{P}}$ from this oriented triangulation $\mathcal{T}$. First, we construct a polyhedral decomposition $\mathcal{P}$ of $\mathcal{T}$, for which we add a new vertex in the center of each tetrahedron of $\mathcal{T}$ and then link it with the four vertices around it. Secondly,  the dual lattice of $\mathcal{P}$ is the so-called resolved dual lattice $\tilde{\mathcal{P}}$, illustrated by the red vertices and directed links of a tetrahedron (3-simplex) in Fig.\ref{fig:3D_resolved_dual_positive_negative}(a) and (d). 
This specific direction of the links of $\tilde{\mathcal{P}}$ is promised to    satisfy the \textit{local Kasteleyn orientation} property: the smallest loops in the dual lattice $\tilde{\mathcal{P}}$ are always Kasteleyn oriented. As crucial for KC  decoration,  the construction of the local Kasteleyn orientation is related to the discrete spin structure. For readers interested in this construction, please refer to Ref.\cite{Wang2018}.

  \textit{{Degree of freedom}} For each vertex $i$ of $\mathcal{T}$, we assign $|G_b|$ bosonic states, denoted as $|g_i\rangle$ ($g_i\in G_b$). Then we assign $|G_b|$ species of complex fermions $a_{ijk}^{\sigma }$  $( \sigma \in G_{b})$  to each 2-simplex of $\mathcal{T}$. Each of these complex fermions can be decomposed into two Majorana fermions $a_{ijk}^{\sigma }=( \gamma _{ijk,A}^{\sigma } +i\gamma _{ijk,B}^{\sigma })/2$. The positions of the two types of fermions are assigned on the two sides of each triangle $\langle ijk\rangle $ by the rule that the A(B) type is at the start (end) of the directed link in $\tilde{\mathcal{P}}$ intersecting the triangle.  We also assign  $|G_b|$ species of complex fermions $c_{ijkl}^{\sigma }$  to each 3-simplex $\langle ijkl \rangle$, whose position is just at the center of the 3-simplex. 

\textit{Symmetry transformation} These degrees of freedom should transform suitably under the symmetry action $U( g)$  as follows: For any $g, \sigma \in G_{b}$, 
\begin{align}
U( g)\ket{g_{i}} &=\ket{gg_{i}} \\
U( g) c_{ijkl}^{\sigma } U^{\dagger }( g) &=( -1)^{\omega _{2}( g,\sigma )} c_{ijkl}^{g\sigma }\\
U( g) a_{ijk}^{\sigma } U^{\dagger }( g) &=( -1)^{\omega _{2}( g,\sigma )} a_{ijk}^{g\sigma }.
\end{align}
Equivalently,  the Majorana fermions transform as 
\begin{align}
U( g) \gamma _{ijk,A}^{\sigma } U^{\dagger }( g) &=( -1)^{\omega _{2}( g,\sigma )} \gamma _{ijk,A}^{g\sigma } ,\label{eq:sym trans Majorana A}\\
U( g) \gamma _{ijk,B}^{\sigma } U^{\dagger }( g) &=( -1)^{\omega _{2}( g,\sigma ) +s_{1}( g)} \gamma _{ijk,B}^{g\sigma } .\label{eq:sym trans Majorana B}
\end{align}
Under these transformations, the bosonic degrees of freedom always form a linear representation of $G_{b}$ (and $G_{f}$).
On the other hand, the fermionic degrees of freedom always form projective representations of $G_{b}$ with coefficient $(-1)^{\omega_{2}}$ in order to get linear representations of $G_{f}$.

\textit{Decoration rules}
Now we discuss how to decorate the KC and complex fermions to construct an FSPT state. For a given $G_b$ pattern (i.e., a given $G_b$ configuration on the vertices of $\mathcal{T}$), we define the vacuum state for fermions on all the 2-simplex and 3-simplex as a reference. By fermion in its vacuum state, we mean that $c^\sigma_{ijkl}|\psi\rangle=0$ or $a^\sigma_{ijk}|\psi\rangle =0$, while the latter equivalently implies that $-i\gamma_{ijk,A}^\sigma \gamma_{ijk,B}^\sigma|\psi\rangle=|\psi\rangle$. The decoration rules are nothing but the rules for telling us how to construct the possible non-trivial fermionic states for each $G_b$ pattern. 

For the KC layer, the decoration rules are specified by a function $n_2(g_i,g_j,g_k)=0, 1$. On a  triangle $\langle ijk\rangle$, if $n_2(g_i,g_j,g_k)=0$, there is no non-trivial KC passing through the triangle, so all the Majorana fermions are in the vacuum pairing, namely $-i\gamma_{ijk,A}^\sigma \gamma_{ijk,B}^\sigma=1$ for all $\sigma\in G_b$ when acting on the state. On the other hand, when $n_2(g_i,g_j,g_k)=1$, there is a non-trivial KC passing through, formed by Majorana fermions $\gamma_{ijk,A}^{g_i}$ and  $\gamma_{ijk,B}^{g_i}$ and together with fermions of other triangle(s), while all other species (i.e., $\sigma\neq g_i$) of Majorana fermions are in the vacuum pairing. More explicitly, the concrete decoration rules for an arbitrary tetrahedron are discussed in the following.

\begin{figure}
    \centering
    \includegraphics[width=0.5\textwidth]{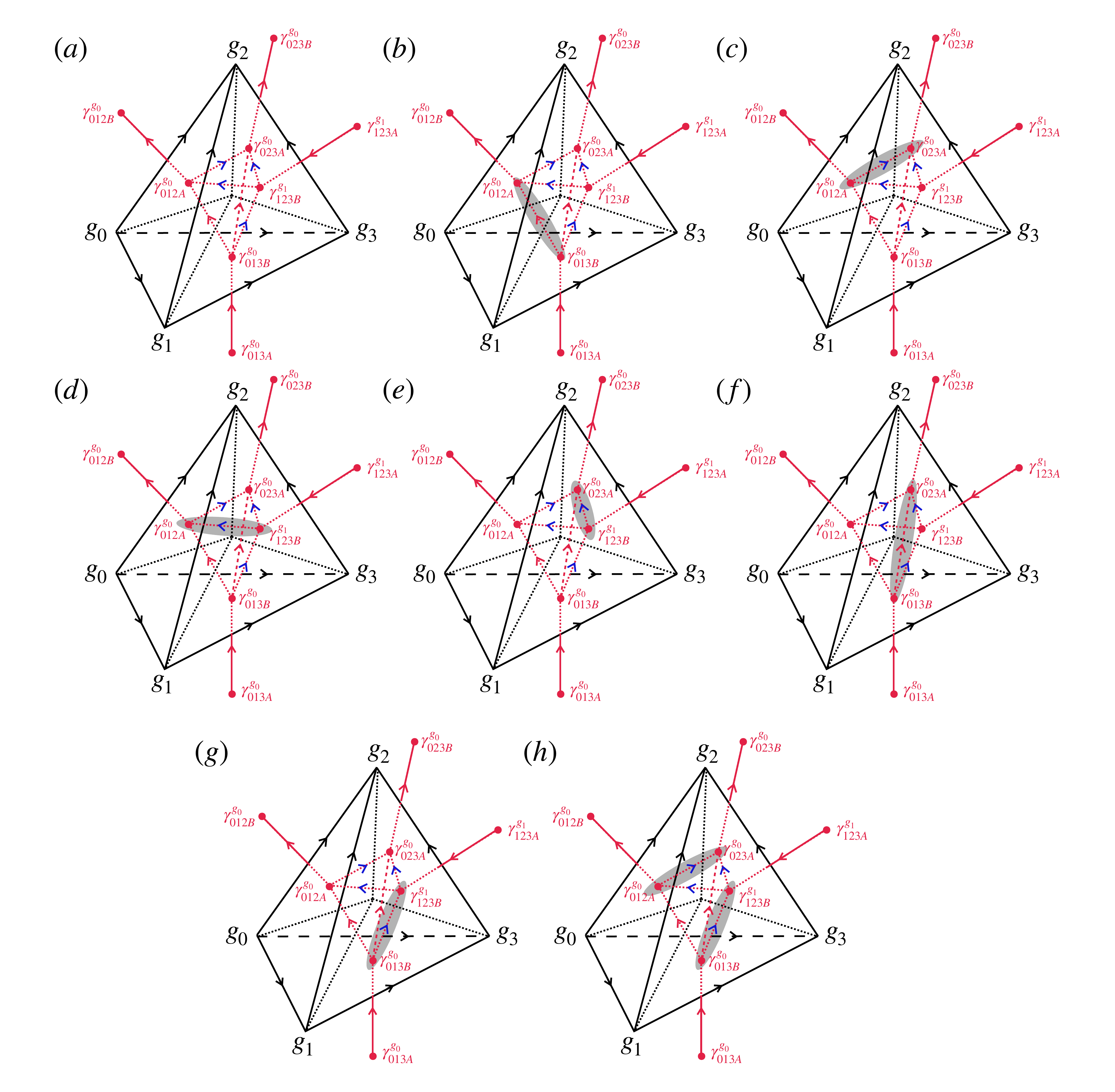}
    \caption{All possibilities of Kitaev chain decoration within a single 3-simplex.
    The solid red line with arrow indicates there is a Majorana pairing, while the dotted red line with arrow indicates no Majorana pairing (just drawn as reference). 
    (a) Vacuum state, also called reference state, in which all $n_2 = 0$.
    (b) $n_2(012) = n_2(013) = 1$, while all other $n_2 = 0$.
    (c-g) Similar as (b), only one non-trivial Majorana pairing within the 3-simplex. 
    (h) All $n_2 = 1$, two non-trivial Majorana pairing within the 3-simplex.
    We take the convention that $\gamma_{012A}$ is paired with $\gamma_{023A}$ while $\gamma_{013B}$ is paired with $\gamma_{123B}$.}
    \label{fig:Kitaev chain decoration}
\end{figure}

In a tetrahedron,  there are four triangles; if there are an odd number of triangles having $n_2(g_i,g_j,g_k)=1$, there are  odd Kitaev chain(s) ending in the tetrahedron and leaving odd Majorana zero mode(s) there, which is incompatible with the construction of a gapped state. So there must be an even number of $n_2(g_i,g_j,g_k)=1$ for a tetrahedron, which leads to the following condition: 
\begin{align}
\mathrm d n_{2} (0123) = 0 \mod 2
\label{eq:2-cocyle}
\end{align}
which means $n_2 \in \mathcal{Z}_2(G_b,\Z_2)$. So in general among the $16$ cases,  there are  $8$ cases for a tetrahedron, depending whether $n_2(ijk)=0, 1$ as shown in Fig.\ref{fig:Kitaev chain decoration}.  

To obtain the symmetric pairing of Majorana fermions, the strategy is given as follows:   (1) construct the pairing rules for the standard tetrahedron and  (2) apply the symmetry action on the standard tetrahedron to obtain those for the non-standard. For the standard tetrahedron, denote its vertices group elements  by $(e, g_0^{-1}g_1, g_0^{-1}g_2, g_0^{-1}g_3)$.  Among the eight configurations  in Fig.\ref{fig:Kitaev chain decoration}, the case (a) has all $n_2(ijk)=0$, so no non-trivial pairing exists inside the tetrahedron.  
For the six cases (b)-(g), there are two   nonzero $n_2(ijk)$, corresponding to the six choices choosing any two from the four: $n_2(012)$, $n_2(013)$, $n_2(023)$ and $n_2(123)$.  Now two Majorana fermions are non-trivial pairing in the tetrahedron, namely $-i\gamma_{ijk,C}^{g_0^{-1}g_i}\gamma_{lkn,D}^{g_0^{-1}g_l}=1$ when acting on the states, with the pairing convention that puts the bottom Majorana $\gamma_{ijk,C}^{g_0^{-1}g_i}$ of their direction links in front. 
For the case $(h)$, all the four $n_2$ are nonzero.  In this case, all these four Majorana fermions in the tetrahedron are non-trivially paired. Although there are several ways of pairing, we choose the convention $-i \gamma_{012A}^e\gamma_{023A}^e=-i \gamma_{013B}^e\gamma_{123B}^{g_0^{-1}g_1}=1$ when acting on the state.
Simply speaking,  the rule of non-trivial pairings in standard tetrahedron is to pair them up according to direction of their links. These pairings can be characterized by the following  Majorana projection operators
\begin{align}
P_{013B, 012A}^{\mathrm{std}}&=(1-i \gamma_{013B}^e\gamma_{012A}^e)/2 \nonumber\\
P_{012A, 023A}^{\mathrm{std}}&=(1-i \gamma_{012A}^e\gamma_{023A}^e)/2\nonumber\\
P_{123B, 012A}^{\mathrm{std}}&=(1-i \gamma_{123B}^{g_0^{-1}g_1}\gamma_{012A}^{e})/2\nonumber\\
P_{013B, 123B}^{\mathrm{std}}&=(1-i \gamma_{013B}^{e}\gamma_{123B}^{g_0^{-1}g_1})/2\nonumber\\
P_{013B, 023A}^{\mathrm{std}}&=(1-i \gamma_{013B}^{e}\gamma_{023A}^{e})/2\nonumber\\
P_{023A, 123B}^{\mathrm{std}}&=(1-i \gamma_{023A}^{e}\gamma_{123B}^{g_0^{-1}g_1})/2.
\end{align}

The projectors $U(g_0)P_{ijk C,lmnD}^{\mathrm{std}} U(g_0)^\dagger$ of the nonstandard tetrahedron obtained by applying symmetry transformation have possible minus sign in the pairing following Eqs.\eqref{eq:sym trans Majorana A} and \eqref{eq:sym trans Majorana B}.  To count the explicit signs, we carry out the transformation for all the six projectors using Eqs.\eqref{eq:sym trans Majorana A} and \eqref{eq:sym trans Majorana B}:
\begin{align*}
&U(g_0)P_{013B, 012A}^{\mathrm{std}}U(g_0)^\dagger
=(1- i \gamma_{013B}^{g_0}\gamma_{012A}^{g_0})/2\nonumber\\
&U(g_0)P_{012A, 023A}^{\mathrm{std}}U(g_0)^\dagger
=(1-(-)^{s_1(0)}i \gamma_{012A}^{g_0}\gamma_{023A}^{g_0})/2\nonumber\\
&U(g_0)P_{123B, 012A}^{\mathrm{std}}U(g_0)^\dagger
=(1-(-)^{\omega_2(0,\bar 01)}i \gamma_{123B}^{g_1}\gamma_{012A}^{g_0})/2\nonumber\\
&U(g_0)P_{013B, 123B}^{\mathrm{std}}U(g_0)^\dagger \nonumber \\
&\qquad \qquad \qquad =(1-(-)^{s_1(0)+\omega_2(0,\bar 0 1)} i \gamma_{013B}^{g_0}\gamma_{123B}^{g_1})/2 \nonumber\\
&U(g_0)P_{013B, 023A}^{\mathrm{std}}U(g_0)^\dagger
=(1-i \gamma_{013B}^{g_0}\gamma_{023A}^{g_0})/2\nonumber\\
&U(g_0)P_{023A, 123B}^{\mathrm{std}}U(g_0)^\dagger
=(1-(-)^{\omega_2(0,\bar 01)}i \gamma_{023A}^{g_0}\gamma_{123B}^{g_1})/2
\end{align*}
where $s_1(0)=s_1(g_0)$ and $\omega_2(0,\bar 01)=\omega_2(g_0,g_0^{-1}g_1)$. These sign changes are labeled in blue of the arrow in Fig.\ref{fig:Kitaev chain decoration}. 
We can see that the sign from $\omega_2$ will be present as long as $n_2(123)$ is nonzero, given by $\omega_2(g_0,g_0^{-1}g_1) n_2(g_1, g_2, g_3)$; the sign from $s_1$ have two contribution: $s_1(g_0)n_2(g_0, g_1, g_2)n_2(g_0,g_2,g_3)$ and $s_1(g_0)n_2(g_0, g_1, g_3)n_2(g_1,g_2,g_3)$. So the sign-induced parity change can be summarized into
\begin{align}
    \Delta P_f^\gamma(0123)=(-1)^{[w_2\cup n_2+s_1\cup (n_2\cup_1 n_2)](g_0,g_0^{-1}g_1, g_0^{-1}g_2,g_0^{-1}g_3)}. 
    \label{eq:Maj_parity_change}
\end{align}
This is related to the change of fermion parity of Majorana fermions in the $F$ move as shown below.

For the complex fermion layer,  we use another function $n_3(g_i,g_j,g_k,g_l)$ taking values $0$ or $1$. If $n_3(g_i,g_j,g_k,g_l)=0$, all the complex fermions $c_{ijkl}^\sigma$ ($\sigma\in G_b$) are in vacuum; if $n_3(g_i,g_j,g_k,g_l)=1$, the fermion $c_{ijkl}^{g_i}$ is occupied while other species are in vacuum. 

In summary, using the two function $n_2$ and $n_3$, we can decorate one and only one 1D KC chain and complex fermions on the 1D and 0D domain wall/defect, respectively.

\subsubsection{F move}

Above we have discussed how to ensure the gapness of the fermionic decoration on a specific triangulation of the closed manifold.  As a topological fixed-point state, it should be invariant under wave-function renormalization, which is finite-depth fermionic symmetric local unitary (FSLU) transformation induced by 3D Pachner moves for retriangulation.   For the FSPT state, we  only need to consider one of the 3D Pachner moves since other Pachner moves can be derived from this one \cite{Wang2020}. We again first consider the \textit{standard} $F$ moves while the other nonstandard ones can be obtained by applying symmetry actions. On the resolved dual lattice, the standard $F$ move is given by 
  \begin{align}
    \psi\left(\raisebox{-0.5\height}{\includegraphics[width=0.10\textwidth]{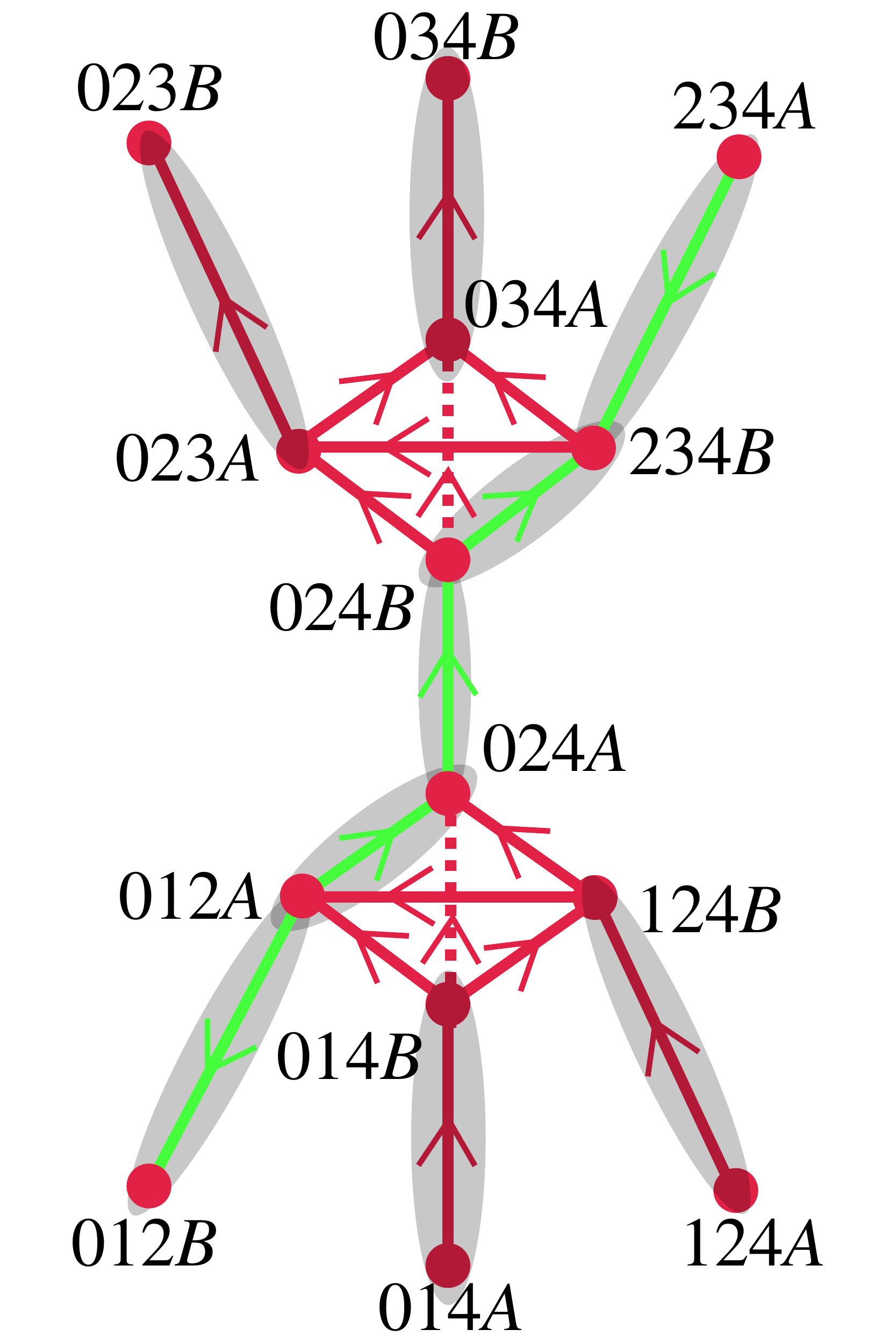}}\right)=F(e,\bar 0 1, \bar 0 2, \bar 0 3, \bar 0 4) \psi\left(\raisebox{-0.5\height}{\includegraphics[width=0.10\textwidth]{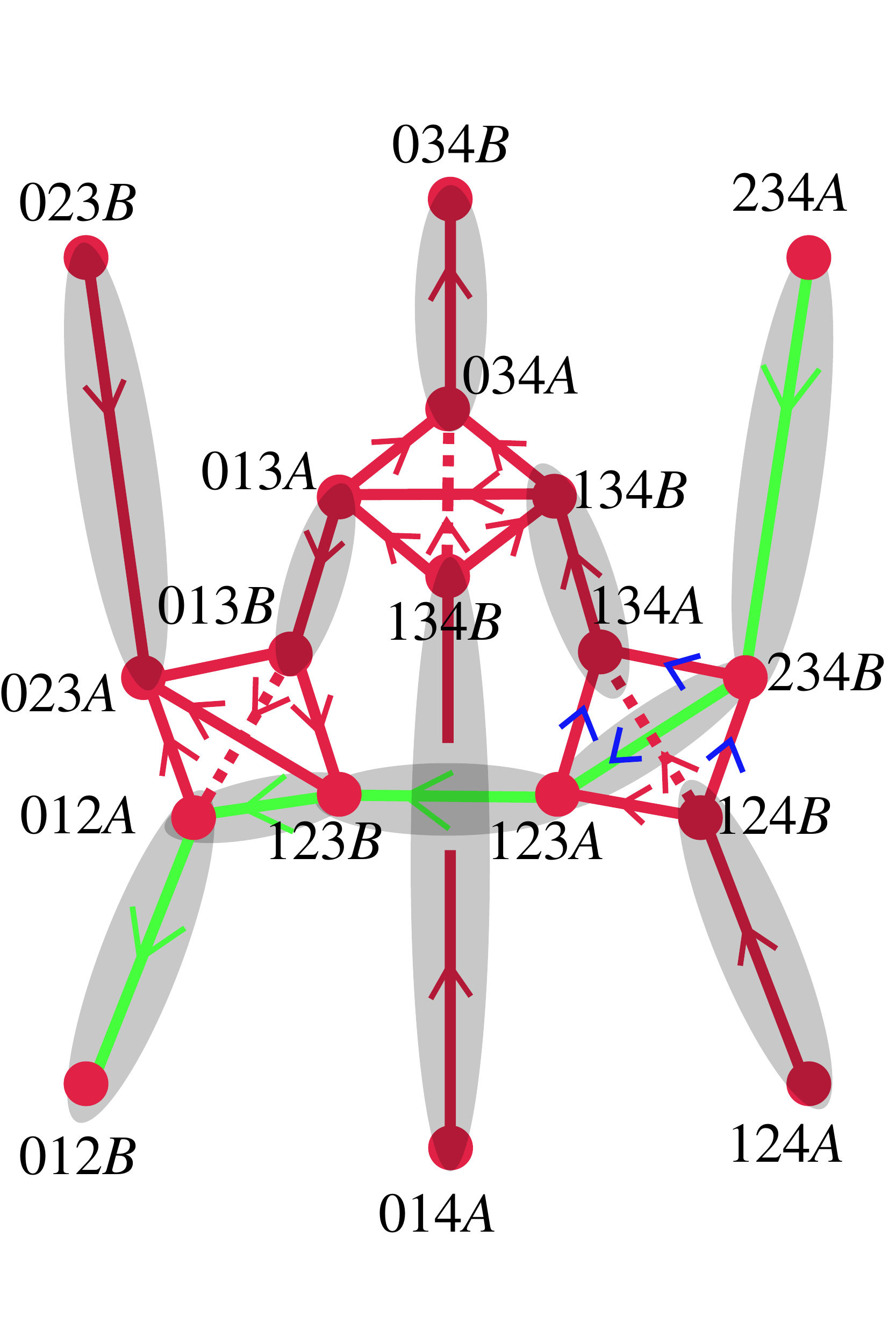}}\right)
    \label{eq:Fmove}
\end{align}
where 
{\small 
\begin{align}
    F(e,\bar 0 1, \bar 0 2, \bar 0 3, \bar 0 4)=\nu_4(\bar 0 1, \bar 1 2, \bar 2 3, \bar 3 4) X_{01234}^c[n_3]X_{01234}^\gamma[n_2].
\end{align}
}
The FSLU transformation mainly consists of three parts $\nu_4, X^c$ and $X^\gamma$.  The phase factor $\nu_4(\bar 0 1, \bar 1 2, \bar 2 3, \bar 3 4) \equiv \nu_4(e,\bar 0 1, \bar 0 2, \bar 0 3, \bar 0 4)$ is due to that wavefunction is defined up to an $U(1)$ phase. The operator $X_{01234}^c[n_3]$ and $X_{01234}^\gamma[n_2]$ maps the complex fermion and Kitaev chain decoration configuration from the right to the left respectively.  Apparently, depending on $n_3$ we may annihilate   the corresponding complex fermions for tetrahedra $\langle0123 \rangle$, $\langle0134 \rangle$ and $\langle1234 \rangle$ on the right side  of Eq.(\ref{eq:Fmove})  and then create the complex fermion for $\langle0124 \rangle$ and $\langle023 4\rangle$: 
\begin{align}
X_{01234}^c[n_3]&=(c_{0124}^{e\dagger})^{n_3(0124)}(c_{0234}^{e\dagger})^{n_3(0234)}\nonumber \\
& (c_{0123}^{e})^{n_3(0123)}(c_{0134}^{e})^{n_3(0134)} (c_{1234}^{g_0^{-1}g_1})^{n_3(1234)}
\end{align}
The definition of $X_{01234}^\gamma[n_2]$ is more complicated which we will discuss soon.

Importantly, as a FSLU transformation, it should  conserve the fermion parity, namely the wavefunction before and after this transformation should have the same fermion parity.  First of all, the fermion parity change due to the complex fermion decoration can be calculated straightforwardly, which is the number of the presence of either creation or annihilate fermionic operators, given by
\begin{align}
\Delta P_f^c=(-1)^{dn_3(01234)}.
\end{align}
On the other hand, the fermion parity change due to the Kitaev chain decoration is more complicated. We can calculate them using the concept of transition loop of two Majorana dimer states. 
\begin{table*}
    \centering
    \begin{tabular}{|c|c|c|c|c|c|c|c|c|c|c|}\hline
 & \multicolumn{6}{c}{$n_2(0ij)$}& \multicolumn{2}{|c|}{Loop 1} & \multicolumn{2}{c|}{Loop 2}\\\hline
         Case&  $n_2(012)$&  $n_2(013)$&  $n_2(014)$&  $n_2(023)$& $n_2(024)$ &$n_2(034)$ & Length&Kasteleyn & Length&Kasteleyn\\\hline
         1&  0&  1&  0&  0&  1&0 & 6& $(-1)^{s_1(01)+\omega_2(012)}$& 4& $(-1)^{s_1(01)}$\\\hline
         2&  0&  1&  0&  1&  1&1 & 4&$(-1)^{s_1(01)+\omega_2(012)}$ & 4&1\\\hline
         3&  0&  0&  0&  1&  1&1 & 4& $(-1)^{s_1(01)+\omega_2(012)}$& 6&$(-1)^{s_1(01)}$\\\hline
         4&  0&  1&  1&  1&  0&1 & 4& 1& 6&1\\\hline
         5&  1&  1&  0&  1&  0&0 & 6& $(-1)^{1+s_1(01)+\omega_2(012)}$& 4&$(-1)^{s_1(01)}$ \\\hline
         6&  1&  0&  1&  1&  1&0 & 4& 1& 6&1\\\hline
         7&  1&  1&  1&  1&  1&0 & 4& 1& 6&1\\\hline
         8&  0&  1&  1&  1&  1&1 & 4& 1& 6&1\\\hline
         9&  1&  1&  1&  1&  1&1 & 6& $(-1)^{\omega_2(012)}$& 4&1\\ \hline
    \end{tabular}
    \caption{All cases with 2 transition loops with length $2N\ge 4$.}
    \label{tab:2_loop}
\end{table*}

\begin{figure}[h]
   \centering
   \includegraphics[width=0.45\textwidth]{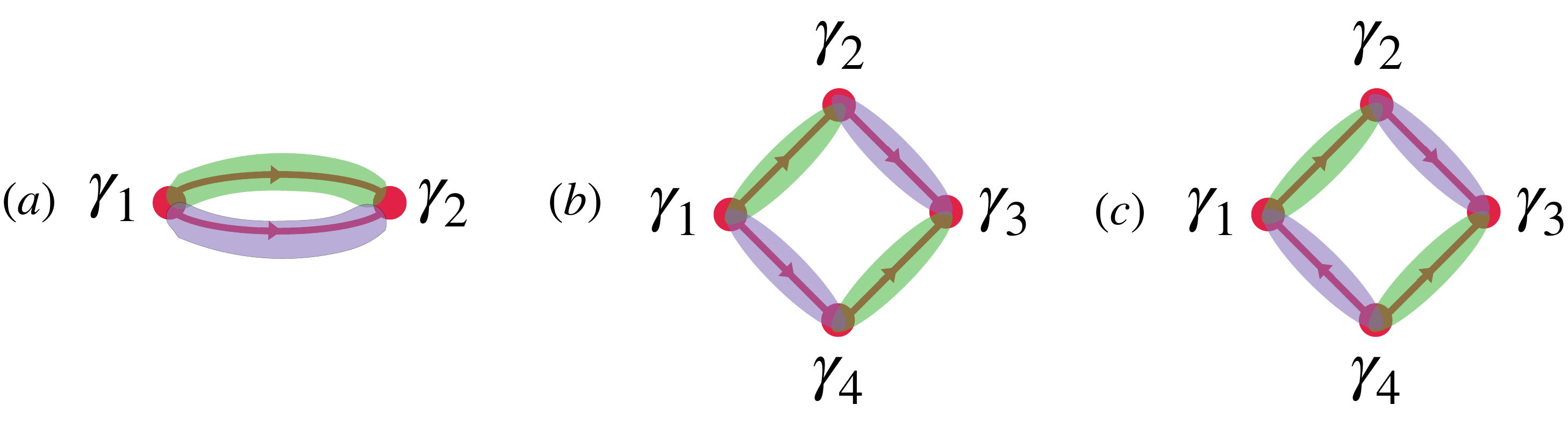} 
   \caption{Transition loop with 2 and 4 sites}
   \label{fig:2site}
\end{figure}

\textit{Transition loop}.  For given  $2N$  Majorana fermions, we line up at site $1,2,..., 2N$ of a ring, with an oriented direction---any two nearest neighbor sites are given a directed arrow.  We consider two ways to pair them up into Majorana dimer states: one  is to pair them up at $(2i-1, 2i)$ for all $i=1,...N$  while the other  is to  pair them up at $(2i, 2i+1)$ for all $i=1,...N$ (assuming $2N+1=1$), both putting them into states according to the arrow linking them. The transition loop of the two dimer states is the underlying oriented loop (ring) of sites. A very crucial property of the transition loop of the Majorana dimer states is that \textit{if the transition loop is Kasteleyn oriented, then the two Majorana dimer states share the same fermion parity; otherwise they have different fermion parity.} The definition of Kasteleyn orientation is that if  the number of counter-clockwise directed arrows of the loop is odd, then the loop is Kasteleyn oriented. For the loop with even number of sites, the parity of numbers of counter-clockwise and clockwise directed arrows are the same.  The simplest example in Fig.\ref{fig:2site}(a) is $N=1$ so that we can have only one Majorana pair.
Putting the two sites in a loop, if $1\rightarrow 2$ and $2\rightarrow1$ links have inversion direction, then the loop is Kasteleyn oriented, so the two pairings from pair $(2i-1,2i)$ and from $(2i,2i+1)$ have the same fermion parity. This is apparent since these two pairs are the same  $-i\gamma_1\gamma_2=1$, which means there is no change in the dimer state.
We will usually not view this case as a transition loop becauase nothing changes here.
Another example is the one with $N=2$. As shown in Fig.\ref{fig:2site}(b),  
we consider the two dimer states: (1) $-i\gamma_1\gamma_2=-i\gamma_4\gamma_3=1$,  and (2) $-i\gamma_2\gamma_3=-i\gamma_1\gamma_4=1$. The transition loop of the two dimer states is non-Kasteleyn oriented, so they have different fermion parity. One straightforward way to see this point is to detect the  eigenvalues of fermion parity operator $P_f=i^2 \gamma_1\gamma_2\gamma_3\gamma_4$, which take $-1$ and $1$, respectively. On the other hand, if we reverse the direction of the arrow between $4$ and $1$ as in Fig.\ref{fig:2site}(c), the second state (2) has $-i\gamma_2\gamma_3=-i\gamma_4\gamma_1=1$. Now the transition loop of the dimer states (1) and (2) is Kasteleyn oriented and they have the same fermion parity, which can be directly checked.

Using the transition loop, we can count the fermion parity change from the right Majorana dimer states to the left dimer states in the $F$ move. The Majorana dimer states on these tetrahedron are determined  explicitly by $n_2$, with some dimer pairings affected by $\omega_2$ and $s_1$. In the $F$ move, there are 10 different $n_2(ijk)$ involved, among which only 6 are independent due to the 2-cocycle conditions Eq.(\ref{eq:2-cocyle}), namely $n_2(0ij)$ for $1\le i < j\le 4$. So there are $2^6=64$ different $n_2$ configurations. Remarkably, there are cases with more than one transition loop with length $2N\ge 4$. It turns out there are in total 9 such cases that will have two transition loops with length $2N\ge 4$. We summarize them in Table \ref{tab:2_loop} where we present the length of both loops and whether they are Kasteleyn oriented or not---if the number in the row named Kasteleyn is plus (minus) one, then it is (non-)Kasteleyn oriented. 

Below we discuss four examples of the transition loop. 
\begin{enumerate}
    \item 
\begin{figure}[h]
    \centering
    \includegraphics[width=\linewidth]{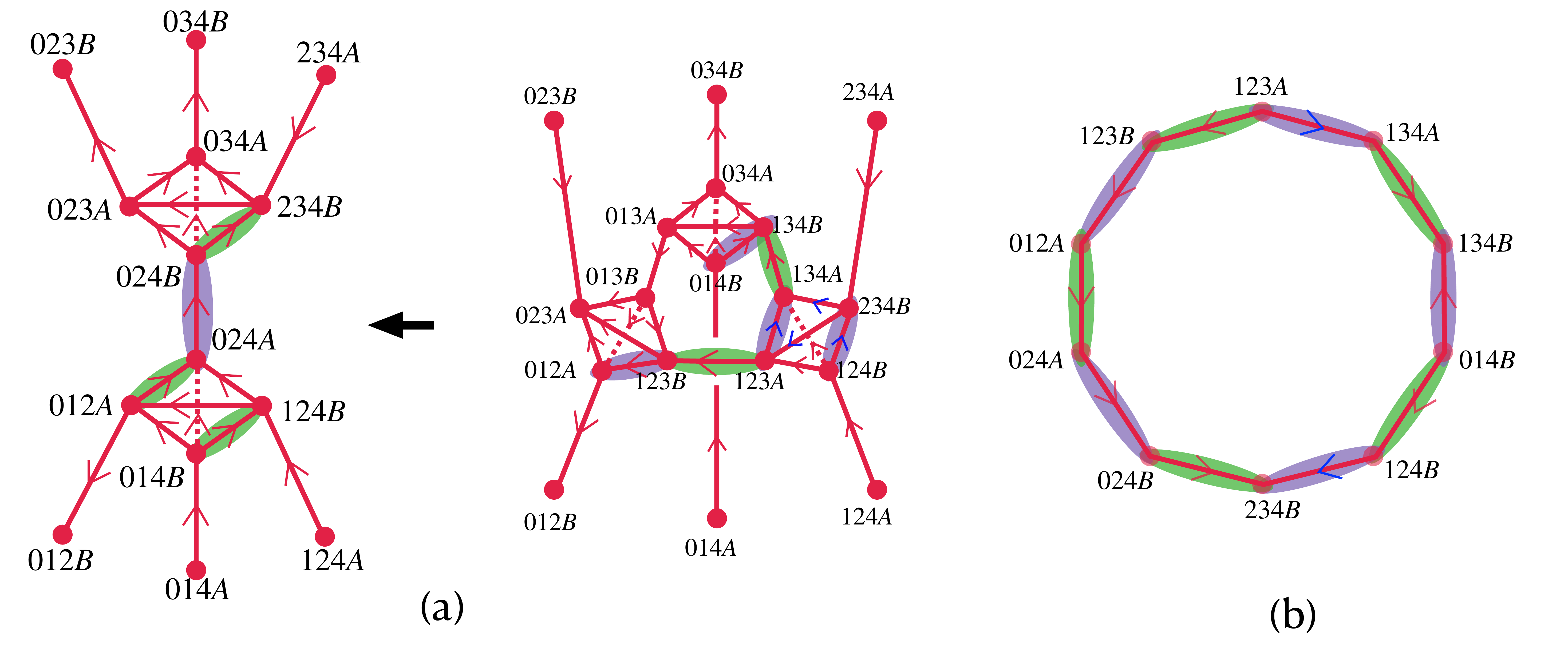}
    \caption{The 1st Example of $F$ move and its transition loop. The pairing in violet (green) is the pairing before (after) the $F$ move.}
    \label{fig:F_move_example}
\end{figure}
The first example is shown in Fig.\,\ref{fig:F_move_example} where the Kitaev chain decoration are specified by nonzero  $n_2(012)$, $n_2(014)$, $n_2(024)$, $n_2(123)$, $n_2(124)$, $n_2(134)$ and $n_2(234)$. In Fig.\ref{fig:F_move_example}(a), before (after) the $F$ move, the Majorana pairings are under violet (green) shadows  where we ignore the vacuum pairings that do not change before and after the $F$ move. Putting the two Majorana dimers before and after the $F$ move, we can construct the transition loop of the Majorana dimers as depicted in Fig.\ref{fig:F_move_example}(b). The underlying transition loop is non-Kasteleyn oriented, so it will result in parity change. Meanwhile the blue arrows may have additional signs due to the symmetry transformation, which are precisely given by $(-1)^{s_1(g_0^{-1}g_1)}$ and $(-1)^{s_1(g_0^{-1}g_1)+\omega_2(g_0^{-1}g_1, g_1^{-1}g_2)}$. Considering these two parts of possible change, the final fermion parity change for this case is given by $\Delta P_f^\gamma|_{\{n_2\}}=(-1)^{1+\omega_2(g_0^{-1}g_1, g_1^{-1}g_2)}$. 



\item The second example is drawn in Fig.\ref{fig:Majorana parity eg1}.
The Kitaev chain decoration is specified by setting $n_2(014)$, $n_2(124)$, $n_2(023)$, $n_2(034)$, $n_2(013)$, $n_2(134)$ to $1$, while others $0$.
In Fig.\ref{fig:Majorana parity eg1}(a), before (after) the $F$ move, the Majorana pairings are under violate (green) shadows  where we ignore the vacuum pairings that do not change before and after the $F$ move.
Before the $F$ move, we have four non-trivial pairings  (marked by violet shadows) with eight different majorana fermions:  $-i \gamma_{013B}^{e} \gamma_{023A}^{e}$, $-i \gamma_{013A}^{e} \gamma_{034A}^{e}$, $-i \gamma_{014B}^{e} \gamma_{134B}^{g_0^{-1} g_1}$ and $-i \gamma_{124B}^{g_0^{-1} g_1} \gamma_{134A}^{g_0^{-1} g_1}$.
After the $F$ move, the eight majorana four majorana are paired in four different pairing (marked by green shadows):  $-i \gamma_{023A}^{e} \gamma_{034A}^{e}$, $-i \gamma_{014B}^{e} \gamma_{124B}^{g_0^{-1} g_1}$, $-\gamma_{013A}^e\gamma_{013B}^e$ and $-\gamma_{134A}^{g_0^{-1}g_1}\gamma_{134B}^{g_0^{-1}g_1}$. 
Then connecting these Majorana pairings before and after the $F$ move as Fig.\ref{fig:Majorana parity eg1}(b), we will get two transition loops, which are all Kasteleyn oriented.
This means there is no fermion parity change for this Kitaev chain decoration. 

\begin{figure}
    \centering
    \includegraphics[width=1.0\linewidth]{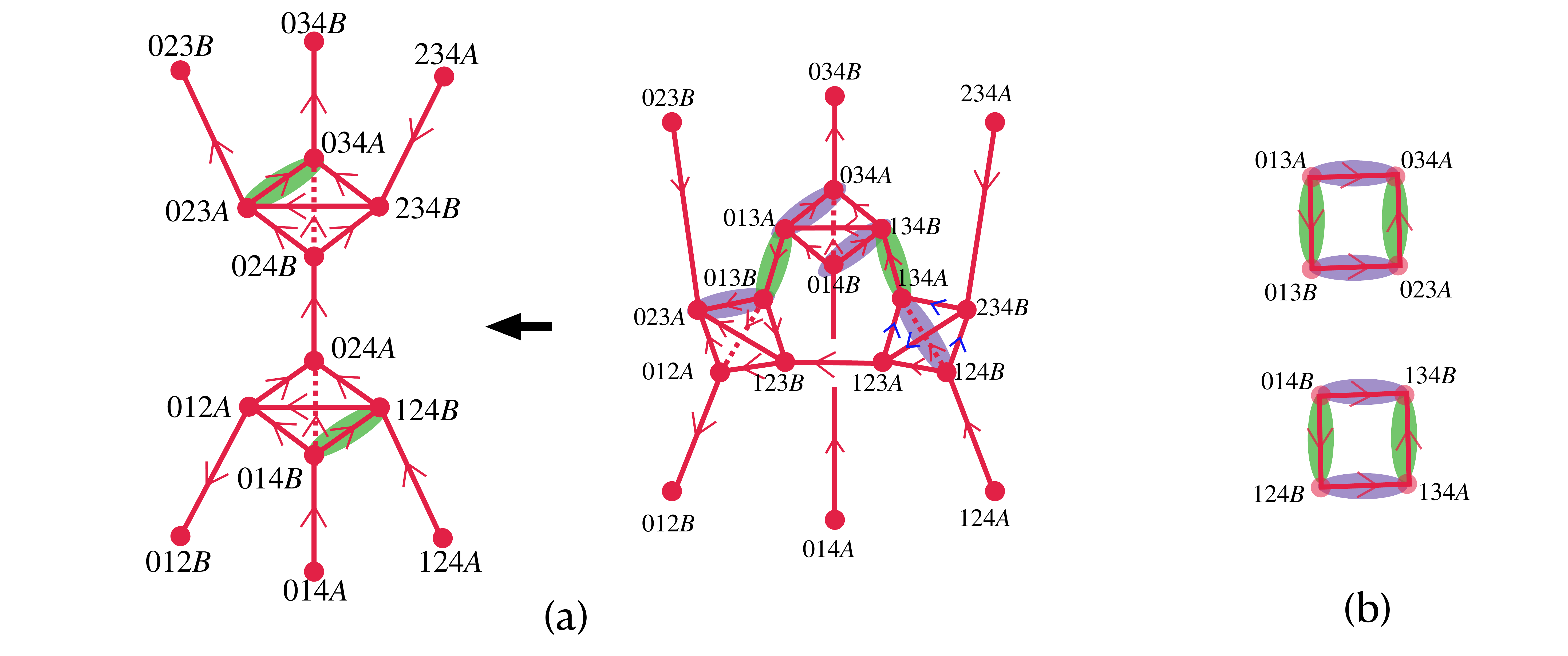}
    \caption{The 2nd Example of Kitaev chain fermion parity change during $F$ move.}
    \label{fig:Majorana parity eg1}
\end{figure}

\item The third example is drawn in Fig.\ref{fig:Majorana parity eg3}.
The Kitaev chain decoration is specified by setting $n_2(024)$, $n_2(124)$, $n_2(234)$, $n_2(013)$, $n_2(123)$, $n_2(134)$ to $1$, while others $0$.
In Fig.\ref{fig:Majorana parity eg3}(a), before (after) the $F$ move, the Majorana pairings are under violate (green) shadows, ignoring the vacuum pairings that do not change in the $F$ move,  where  ten majorana fermions are involved in the move.  We note that the $\gamma_{024A/B}^e$ are not drawn on the right but on the left of Fig.\,\ref{fig:Majorana parity eg3}(a). 
Before the $F$ move, the ten majorana fermions are in five pairings (marked by violet shadows): $-i \gamma_{013B}^{e} \gamma_{123B}^{g_0^{-1} g_1}$,  $-i \gamma_{134B}^{g_0^{-1} g_1} \gamma_{013A}^{e}$, $-i\gamma_{024A}^e\gamma_{024B}^e$,  $- (-1)^{s_1 (g_0^{-1} g_1)} i \gamma_{123A}^{g_0^{-1} g_1} \gamma_{134A}^{g_0^{-1} g_1}$ and $- (-1)^{\omega_2(g_0^{-1} g_1, g_1^{-1} g_2) + s_1 (g_0^{-1} g_1)} i \gamma_{124B}^{g_0^{-1} g_1} \gamma_{234B}^{g_0^{-1} g_2}$.  After the $F$ move, they are paired up in different dimers:  $-i \gamma_{124B}^{g_0^{-1} g_1} \gamma_{024A}^{e}$, $-i \gamma_{024B}^{e} \gamma_{234B}^{g_0^{-1} g_1}$, and three vacuum pairing $-i\gamma_{013A}^e\gamma_{013B}^e$, $-i\gamma_{123A}^{g_0^{-1}g_1}\gamma_{123B}^{g_0^{-1}g_1}$ and $-i\gamma_{134A}^{g_0^{-1}g_1}\gamma_{134B}^{g_0^{-1}g_1}$. 

\begin{figure}
    \centering
    \includegraphics[width=1.0\linewidth]{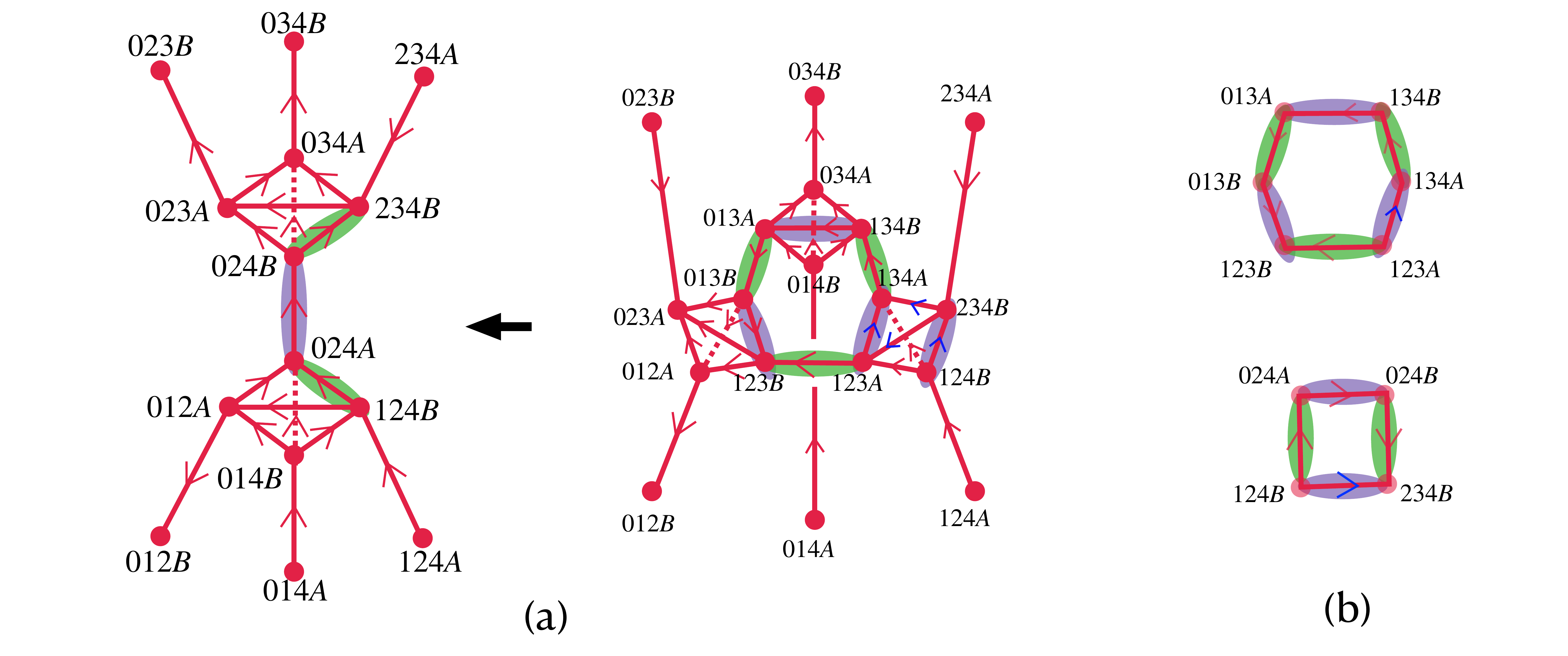}
    \caption{The 3rd example of Kitaev chain fermion parity change during $F$ move.}
    \label{fig:Majorana parity eg3}
\end{figure}

Then connecting these Majorana pairings before and after the $F$ move as Fig.\ref{fig:Majorana parity eg3}(b), we will get two transition loops, which with the underlying orientation are  Kasteleyn oriented but the blue arrows indicate they may become non-Kasteleyn since the blue arrows  reverse the directions of pairing.
The fermion parity change of the lower transition loop in Fig.\ref{fig:Majorana parity eg3}(b) is $(-1)^{s_1(g_0^{-1} g_1)}$; while the fermion parity change of the upper transition loop is $(-1)^{\omega_2(g_0^{-1} g_1, g_1^{-1} g_2) + s_1(g_0^{-1} g_1)}$. In total, the fermion parity changes in this $F$ move is given by $(-1)^{\omega_2(g_0^{-1} g_1, g_1^{-1} g_2)}$ since the two $s_1(g_0^{-1} g_1)$s add to even parity.

\end{enumerate}

Similarly, after checking the transition loops for all the 64 cases, the fermion parity change in the $F$ move can be summarized into 
\begin{align}
     \Delta P_f^\gamma=(-1)^{[(\omega_2+n_2)\cup n_2+s_1\cup (n_2\cup_1 n_2)](01234)}.
     \end{align}
One way to understand this result can be divide it into two parts.
On the one hand, we can see that the tetrahedron $\langle 1234\rangle$ is non-standard, so according to Eq.(\ref{eq:Maj_parity_change}), this will have a contribution to  fermion parity change
\begin{align}
    \Delta P_{f}^\gamma(1234)=(-1)^{[\omega_2\cup n_2+s_1\cup (n_2\cup_1 n_2)](\bar 0 1, \bar 1 2,\bar 23,\bar 34)}. 
\end{align}
On the other hand, according to Ref.\cite{Wang2018}, there is a contribution purely from the retriangulation, independent of $w_2$ and $s_1$, to fermion parity change that is
 \begin{align}
      \Delta P_{f}^\gamma(01234)=(-1)^{n_2\cup n_2 (01234)}. 
 \end{align}
 Therefore, total fermion parity change is given by 
 \begin{align}
     \Delta P_f^\mathrm{tot}=(-)^{dn_3(01234)+[(\omega_2+n_2)\cup n_2+s_1\cup (n_2\cup_1 n_2)](01234)},
 \end{align}
 which should be parity even. 
So we should have the condition on $n_3$ and $n_2$ as follows
    \begin{align}
    dn_3(01234)=O_4(01234) \quad (\text{mod}\ 2).
\end{align}
where 
\begin{align}
   O_4 := (\omega_2+n_2)\cup n_2+s_1\cup (n_2\cup_1 n_2).
\end{align}
The physical meaning of above equation is clear: the fermion parity change due to the Kitaev chain decoration should be canceled by that from the complex fermion decoration. Whether this cancellation can be satisfied or not depends on whether $O_4$ is a coboundary in $B^4(G_b, \Z_2)$ or not. If it is not a coboundary, then the corresponding Kitaev chain decoration specified $n_2$ is obstructed, so it will not contribute to the final classification. 

With the understanding of fermion parity change, now we can explicitly construct the operator $X_{01234}[n_2]$ as  
\begin{align}
    X_{01234}[n_2]=P_{01234}[n_2] (\gamma_{234B}^{g_0^{-1}g_2})^{ \alpha (01234)}(\gamma_{123A}^{g_0^{-1}g_1})^{ \beta (01234)}.
    \label{eq:X}
\end{align}
 Here  $P_{01234}[n_2]$ is the projector (up to a normalization factor) on the Majorana fermion configuration of the left side of the Eq.(\ref{eq:Fmove}).   More specifically, we can decompose the  $P_{01234}[n_2]$ into five sectors:
\begin{align}
    P_{01234}[n_2]=&\mathcal{N}P_{\mathrm{tet}}(0124)P_{\mathrm{tet}}(0234)\nonumber \\
    \qquad \qquad & P_{\mathrm{tri}}(013)P_{\mathrm{tri}}(123) P_{\mathrm{tri}}(134).
    \label{eq:P01234def}
\end{align}
where $\mathcal{N}=\prod_{loop\,i}2^{(L_i-1)/2}$ is the normalization factor and $2L_i$ is the length of the $i$th transition loop.
We now define the projectors $P_{\mathrm{tet}}$ and $P_{\mathrm{tri}}$.  First, the definition of $P_{\mathrm{tri}}(ijk)$ depends on $n_2(ijk)$, namely, when $n_2(ijk)=0$, it is the identity operator while when $n_2(ijk)=1$, it is the local projector of the pair Majorana on triangle $\langle ijk\rangle$, i.e.,  
\begin{align}
    P_{\mathrm{tri}}(ijk)=\bigg(\frac{1-i\gamma_{ijkA}\gamma_{ijkB}}{2}\bigg)^{n_2(ijk)} .
\end{align}
One can easily show that 
\begin{align}
     P_{\mathrm{tri}}(ijk)^2= P_{\mathrm{tri}}(ijk).\label{eq:Ptri_prop}
\end{align}
The reason of such form of $P_{\mathrm{tri}}(013)$, $P_{\mathrm{tri}}(123)$ and $P_{\mathrm{tri}}(134)$ is due to the fact that if  $n_2(ijk)$ of the corresponding triangle is nonzero, then the corresponding Majorana fermions attend in the non-trivial pairing for implementing the KC layer of decoration; if the corresponding $n_2(ijk)$ is zero, then the Majorana fermions will always be in vacuum pairing and not affected in the $F$ move transformation. 
Second,  $P_{\mathrm{tet}}(ijkl)$ is the projector of Majorana fermion in the tetrahedron $\langle ijkl \rangle$. Depending on the configuration of $n_2$, we can write the corresponding projectors to project on the Majorana dimer configuration, corresponding to one choice in Fig.\ref{fig:Kitaev chain decoration}.
It is easy to see that all four terms in $ P_{01234}[n_2]$ commute with each other, so the position of these operators can be freely moved in $ P_{01234}[n_2]$.

Now let us discuss the two inserted Majorana fermions $\gamma_{234B}^{g_0^{-1}g_2}$  and  $\gamma_{123A}^{g_0^{-1}g_1}$. With a  non-Kasteleyn oriented transition loop, the Majorana fermion states before and after $F$ move are in different fermion parity sector so that  direct action of the projector will result in a null state because the projector preserves the fermion parity. So the insertion of these dangling Majorana fermions is necessary for cases with non-Kasteleyn oriented  transition loop since the fermion operators will first flip the fermion parity sectors, then following projectors can map on the correct Majorana states.  The reason we need two Majorana fermions is that there may be two non-Kasteleyn oriented transition loops for one $n_2$ configuration. As shown in Table \ref{tab:2_loop}, the case 1 and 3 can have two non-Kasteleyn transition loops simultaneously when $s_1(01)=1, \omega_2(012)=0$ while the case 5 have two when $s_1(01)=\omega_2(012)=1$.  In these cases, the loop 1 always contain the fermion $\gamma_{234B}^{g_0^{-1}g_2}$ while the loop 2 always contains $\gamma_{123A}^{g_0^{-1}g_1}$. Furthermore, for other cases that contain only one non-Kasteleyn oriented transition loop, either $\gamma_{234B}^{g_0^{-1}g_2}$ or $\gamma_{123A}^{g_0^{-1}g_1}$ will be contained. Therefore, we can choose these two Majorana fermions $\gamma_{234B}^{g_0^{-1}g_2}$ and $\gamma_{123A}^{g_0^{-1}g_1}$ inserted to properly flip the fermion parity sector in the $F$ move. Taking Fig.\ref{fig:Majorana parity eg3} as an example, corresponding to case 1 in Table \ref{tab:2_loop}, we insert $ \gamma_{123A}^{g_0^{-1} g_1}$ and $ \gamma_{234B}^{g_0^{-1} g_1} $ for the lower and upper transition loop, respectively, to the $F$ move if $s_1(g_0^{-1} g_1)=\omega_2(g_0^{-1} g_1, g_1^{-1} g_2)=1$.

To precisely control when the dangling Majorana fermion insertion is needed, we use two quantities $\alpha(01234)$ and $\beta(01234)$ defined as 
 \begin{align}
 \alpha(01234)=\tilde \alpha(01234)+N_{1234}(1-L_{01234})\tilde \beta(01234) ,\label{eq:def_tal}\\
 \beta(01234)=\tilde \beta(01234)+N_{1234}(L_{01234}-1)\tilde \beta(01234).\label{eq:def_tbe}
\end{align}
We used the notations $\tilde \alpha$ and $\tilde \beta$ defined as
 \begin{align}
\tilde \alpha (01234)& = \big[n_2(012) + \omega_2(012) \nonumber \\
&\qquad + s_1(01)n_2(124)\big]n_2(234) ,\label{eq:alpha}\\
\tilde \beta (01234) &= s_1(01)n_2(134)n_2(123) \label{eq:beta} ,
\end{align}
 while $N$ and $L$ are 
    \begin{align}
N_{1234} &=n_2(123)n_2(124)n_2(134)n_2(234) ,\\
L_{01234} &= n_2(012)+n_2(024)+ n_2(012)n_2(024)\times \nonumber \\&\qquad \qquad \qquad [n_2(034) + n_2(023) n_2(034)] .
\end{align}
Several explanations. First of all, $\alpha$ and $\beta$ add to $O_4(01234)$ modulo 2, namely they provide a finer resolution to the fermion parity change in the $F$ move. Secondly, when considering only unitary symmetry, $s_1=0$ and then $\alpha$  and $\beta$ become exactly $\tilde\alpha$  and $\tilde \beta$ that were used in Ref.\cite{Wang2020}.  

Thirdly, we discuss the explicit implication from $\alpha$ and $\beta$.  If $\alpha$ and $\beta$ both are zero modulo 2, the fermionic states spanned by the Majorana fermions before and after the $F$ move have the same fermion parity sector. In this case, the  operator $P_{01234}[n_2]$ will directly map the Majorana fermions configuration from the right side to the left side in Eq.(\ref{eq:Fmove}).  However,  with $\alpha$ or $\beta$ taking 1 mod 2, as there will be non-Kasteleyn oriented transition loop,  the insertion of $\gamma_{234B}^{g_0^{-1}g_2}$  or $\gamma_{123A}^{g_0^{-1}g_1}$ will change the fermion parity sector and then  can  the (normalized) projector  $P_{01234}[n_2]$  yield the proper Majorana fermion configuration.  

Lastly, there is a key difference compared to Ref.\cite{Wang2020}  where  both the two Majorana fermions are inserted when $\tilde \alpha=\tilde \beta=1$. This is suitable for cases that the two Majorana fermions sit in different transition loops. However, when there is only one transition loop, both Majorana fermions are in the same transition loop, and the fermion parity sector are not changed for their insertion. So, the proper way is not to  insert any of them, which is exactly realized by adding the term $N(1-L)\tilde \beta$.  In fact,  $\tilde \alpha=\tilde \beta=1$ implies $N=1$, under which the quantity  $L$  reflects the number of transition loops. Namely,  under $N=1$, $L=0 (1)$ means there is only one (two) transition loop(s) in the F move. (Note that we ignore the smallest transition loop that only cover one local link.) 

The $F$ move should respect the symmetry. So the nonstandard $F$ move can be obtained by applying the symmetry actions
\begin{widetext}
 \begin{align}
        F(g_0,g_1,g_2,g_3,g_4)=U(g_0) F(e,g_0^{-1}g_1, g_0^{-1}g_2, g_0^{-1}g_3, g_0^{-1}g_4) U(g_0)^\dagger  
    \end{align}
    which  leads to the following condition on $n_2,n_3$ and $\nu_4$ 
\begin{align}
n_i(g_0,g_1,...,g_i)&=n_2(e,g_0^{-1}g_1,...,g_0^{-1}g_i)=n_2(g_0^{-1}g_1,...g_{i-1}^{-1}g_i)\quad \mathrm{for}\,\, i=2,3,\\
    \nu_4(g_0,g_1,g_2,g_3,g_4)&={}^{g_0}\nu_4(e,g_0^{-1}g_1,g_0^{-1}g_2,g_0^{-1}g_3,g_0^{-1}g_4)\nonumber \\
    &=
    \nu_4(g_0^{-1}g_1,g_1^{-1}g_2,g_2^{-1}g_3,g_3^{-1}g_4)^{1-s_1(g_0)} O_5^\mathrm{sym}(g_0,g_1,g_2,g_3,g_4)
\end{align}
The phase factor $O_5^\mathrm{sym}(g_0,g_1,g_2,g_3,g_4)$ comes from the symmetry transformation of $c_{1234}^{g_0^{-1}g_1}$, $\gamma_{234B}^{g_0^{-1}g_2}$ and $\gamma_{123A}^{g_0^{-1} g_1}$, and the explicit form is 
\begin{align}
    O_5^\mathrm{sym}(g_0,g_1,g_2,g_3,g_4)=(-1)^{(\omega_2\cup n_3+s_1\cup \alpha)(0,\bar 01,\bar 12,\bar 23,\bar 34)+\omega_2(0,\bar 02)\alpha(01234)+w_2(0,\bar 01)\beta_4(01234)}.
    \label{eq:O5sym}
\end{align}
\end{widetext}
When the symmetry is purely unitary, this reduces to the $O_5^\mathrm{sym}$ in Ref.\cite{Wang2020}. For generic anti-unitary symmetry, this term has different meaning due to the difference definition of $\alpha$ and $\beta$, even though it has the similar form.
\begin{figure*}[t]
    \centering
    \includegraphics[width=0.8\linewidth]{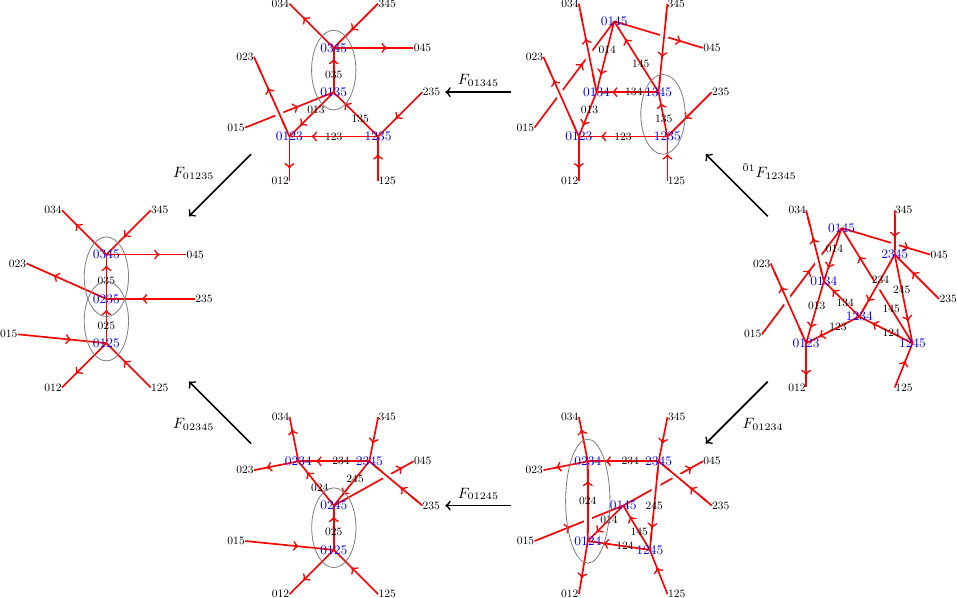}
    \caption{Superhaxagon equation on resolved dual lattice. See Fig.\ref{fig:3D_resolved_dual_positive_negative} for the simplified notation.}
    \label{fig:hexagon_dual}
\end{figure*}
\subsubsection{Consistent conditions of F move }
\label{sec:O5_formula}
The $F$ move should satisfy the superhexagon equation, which equates two different paths to transform the wavefunction with vertices $\langle 012345\rangle$ in Fig.\ref{fig:hexagon_dual}. Again, we only need to consider the standard case where the first vertex 0 has group element $e$, while other nonstandard ones can be obtained by applying the symmetry actions. Note that in the superhexagon equation, there is one nonstandard $F$ move on  $\langle \bar 01,\bar 02,\bar 03,\bar 04,\bar 05\rangle$.  From this equation, we  have a condition on  $\nu_4$ together with other decorated quantities, which is 
\begin{align}d\nu_4(\bar 01,\bar 12,\bar 23, \bar 34,\bar 45)=O_5(\bar 01,\bar 12,\bar 23, \bar 34,\bar 45). 
\end{align}
More explicitly, the left-hand side is 
    \begin{align*}
    &d\nu_4(\bar 01,\bar 12,\bar 23, \bar 34,\bar 45)\nonumber \\
    =&\frac{\nu_4(\bar 12,\bar23,\bar34,\bar45)^{1-2s_1(\bar 01)}\nu_4(\bar 01,\bar 23,\bar34,\bar45)\nu_4(\bar 01,\bar 12,\bar23,\bar45)}{\nu_4(\bar 01,\bar 12,\bar23,\bar 34)\nu_4(\bar 01,\bar 12,\bar34,\bar45)\nu_4(\bar 02,\bar 23,\bar34,\bar45)}
\end{align*}
%
while the right-hand side can be decomposed into 
    \begin{align}
        O_5=O_5^\mathrm{sym}[n_2,n_3] O_5^c[n_3] O_5^{c\gamma}[dn_3] O_5^{\gamma}[n_2] 
    \end{align}
    In the above decomposition, the first phase factor $O_5^\mathrm{sym}$, explicitly depending on $n_2$ and  $ n_3$,  is given by Eq.(\ref{eq:O5sym}), which comes from the nonstandard $F$ move on $\langle \bar 01,\bar 02,\bar 03,\bar 04,\bar 05\rangle$.  The second one purely comes from the complex fermion decoration layer,  which only  explicitly depends on $n_3$,  and is given by
    \begin{align}
         O_5^c[n_3]=(-)^{n_3\cup_1 n_3+dn_3\cup n_3}. 
    \end{align}
   The third term $O_5^{c\gamma}[dn_3]$ comes from the anticommutation between the complex fermions and Majorana fermions, which is 
    \begin{align}
        O_5^{c\gamma}[dn_3]=(-1)^{dn_3\cup_2dn_3}.
    \end{align}
The two terms $O_5^c[n_3]$ and $O_5^{c\gamma}[dn_3]$ have been discovered  in Ref.\cite{Wang2018} and \cite{Wang2020}. The last term $O_5^{\gamma}[n_2] $ we discover here is new and is generic for both unitary and anti-unitary discrete symmetries.    We will discuss the derivation of this new term below.

\subsubsection{Derivation of $O_5^{\gamma}$}
\label{sec:derivationO5}

The $O_5^{\gamma}$ only depends on the states spanned by the Majorana fermions involving in the superhexagon equation. 
It is given by the following expectation value on a reference state, which we will just call it ‘ground state' 
\begin{align}
O_5^\gamma=&{}_\gamma\langle \text{GS}|X |\text{GS}\rangle_\gamma ,
\end{align}
where $|\text{GS}\rangle_\gamma$ is the reference state spanned by the Majorana fermions for the most right configuration in Fig. \ref{fig:hexagon_dual} and 
\begin{align}
X=& (\gamma_{234A})^{ \beta_{\hat 0}} (\gamma_{345B})^{\alpha_{\hat 0}} P_3(\gamma_{134A})^{\beta_{\hat 2}} (\gamma_{345B})^{\alpha_{\hat 2}}P_2 \nonumber \\
&(\gamma_{123A})^{\beta_{\hat 4}} (\gamma_{235B})^{\alpha_{\hat 4}} P_0 
P_1 (\gamma_{345B})^{\beta_{\hat 1}} (\gamma_{234A})^{\alpha_{\hat 1}} P_4 \nonumber \\ 
&(\gamma_{245B})^{\beta_{\hat 3}} (\gamma_{124A})^{\alpha_{\hat 3}}P_5(\gamma_{234B})^{\beta_{\hat 5}} (\gamma_{123A})^{\alpha_{\hat 5}}.
\label{eqn:O5}
\end{align}
where $ \alpha_{\hat i}=\alpha(01..\hat i...45)$ where $i$ are removed and similarly for $\beta_{\hat i}$. Here we will ignore the group label for the inserted Majorana fermions which can be recovered easily from the definition of $F$ move.
Given the necessary values of $n_2, w_2$  and $s_1$, three aspects in calculation of $O_5^\gamma$ can be determined:
\begin{enumerate}
\item ${\beta}_{\hat i}$ and ${\alpha}_{\hat i}$ can be calculated according to their definitions in Eqs.(\ref{eq:def_tal})  and (\ref{eq:def_tbe}) and then the presence of the corresponding Majorana fermions can be determined. 
\item $P_i$ can also be determined according to the above definition of the $F$ move since we explicitly define the $F$  for each case of $n_2$-$w_2$-$s_1$ configurations.
\item The Majorana dimers in the $|GS\rangle_\gamma$ are also determined, namely there are some Majorana fermions that are in non-trivial pairing in $|GS\rangle_\gamma$.
\end{enumerate}
So given a $n_2$-$w_2$-$s_1$ pattern, we can evaluate the phase factor for the given string operator $X$ on the state $|GS\rangle_\gamma$.  As in general these projection operators and the state $|GS\rangle_\gamma$ are complicated, we illustrate the idea of our strategy to compute it by a minimal example.

\begin{figure}[h]
    \centering    \includegraphics[width=0.5\linewidth]{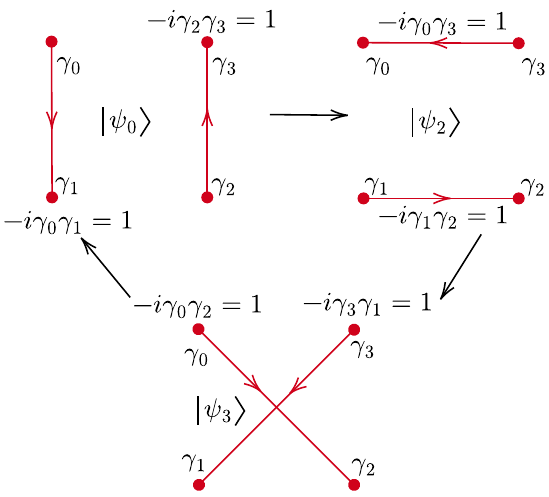}
    \caption{The simplest example with $\mathbb{Z}_8$ Majorana phase.}
    \label{fig:Majorana phase demo}
\end{figure}

To illustrate it, let us consider a simple example (see Fig.\ref{fig:Majorana phase demo}) that can lead to $\theta_\gamma=\frac{\pi}{4}$ in Majorana dimer state expectation value. 
Consider the initial state $|\psi_0\rangle$ with $-i\gamma_0\gamma_1=-i\gamma_2\gamma_3=1$. Applying three transformation as in Fig.\ref{fig:Majorana phase demo}, we can go back to the state $|\psi_0\rangle$ but up to a phase. 
This Majorana phase $e^{ i \theta_{\gamma}}$ in the above transition processes can be extracted by the following expectation value 
\begin{align}
e^{ i \theta_{\gamma}} = \bra{\psi_0} \left[ \sqrt{2} P_{0,1} P_{2,3} \right] \left[ \sqrt{2} P_{0,2} P_{3,1} \gamma_2 \right]\nonumber \\
 \left[ \sqrt{2} P_{3,0} P_{1,2} \gamma_1 \right] \ket{\psi_0} ,\label{eq:Majorana phase eg.1}
\end{align}
where the first two Majorana projectors project the $\ket{\psi_0}$ to the state $\ket{\psi_2}$, the middle two Majorana projectors project to the state $\ket{\psi_3}$, the last two Majorana projectors project back to the state $\ket{\psi_0}$, and the two dangling Majorana modes are inserted for fermion parity sector flipping. The factor $\sqrt{2}$ is due to the transition loops between them have length $4$.

\begin{table*}
    \centering
    \begin{tabular}{|c|c|c|c|c|c|c|c|c|c|c|}\hline
         {\footnotesize $n$} &  $P_{01}$&  $P_{23}$&  $P_{02}$&  $P_{31}$&  $\gamma_{2}$&  $P_{30}$&  $P_{12}$&  $\gamma_{1}$&  Simplified Term $R_n$ &{\footnotesize  $\langle \psi_0|R_n|\psi_0\rangle$}\\\hline
         1&  &  &  &  &  $\gamma_{2}$&  &  &  $\gamma_{1}$&  $\gamma_{2}\gamma_1$&0 \\\hline
         2&  &  &  &  &  $\gamma_{2}$&  &  $(-i\gamma_{1}\gamma_{2})$&  $\gamma_{1}$&  $i$&$i$ \\\hline
         3&  &  &  &  &  $\gamma_{2}$&  $(-i\gamma_{3}\gamma_{0})$&  &  $\gamma_{1}$&  $(-i\gamma_2\gamma_3)(\gamma_0\gamma_1)$&$i$ \\\hline
         4&  &  &  &  $(-i\gamma_{3}\gamma_{1})$&  $\gamma_{2}$&  &  &  $\gamma_{1}$&  $-i\gamma_{2}\gamma_{3}$&1 \\\hline
         5&  &  &  $(-i\gamma_{0}\gamma_{2})$&  &  $\gamma_{2}$&  &  &  $\gamma_{1}$&  $-i\gamma_{0}\gamma_{1}$&1 \\\hline
         6&  &  $(-i\gamma_{2}\gamma_{3})$&  &  &  $\gamma_{2}$&  &  &  $\gamma_{1}$&  $i\gamma_{3}\gamma_{1}$&0 \\\hline
         7&  $(-i\gamma_{0}\gamma_{1})$&  &  &  &  $\gamma_{2}$&  &  &  $\gamma_{1}$&  $i\gamma_{0}\gamma_{2}$&0 \\\hline
         8&  &  &  &  &  $\gamma_{2}$&  $(-i\gamma_{3}\gamma_{0})$&  $(-i\gamma_{1}\gamma_{2})$&  $\gamma_{1}$&   $\gamma_{3}\gamma_{0}$&0\\\hline
         9&  &  &  &  $(-i\gamma_{3}\gamma_{1})$&  $\gamma_{2}$&  &  $(-i\gamma_{1}\gamma_{2})$&  $\gamma_{1}$&   $\gamma_{3}\gamma_{1}$&0\\\hline
 10& & & $(-i\gamma_{0}\gamma_{2})$& & $\gamma_{2}$& & $(-i\gamma_{1}\gamma_{2})$& $\gamma_{1}$&  $\gamma_{0}\gamma_{2}$&0\\\hline
 11& & $(-i\gamma_{2}\gamma_{3})$& & & $\gamma_{2}$& & $(-i\gamma_{1}\gamma_{2})$& $\gamma_{1}$&  $\gamma_{2}\gamma_{3}$&$i$\\\hline
 12& $(-i\gamma_{0}\gamma_{1})$& & & & $\gamma_{2}$& & $(-i\gamma_{1}\gamma_{2})$& $\gamma_{1}$&  $\gamma_{0}\gamma_{1}$&$i$\\\hline
 13& & & & $(-i\gamma_{3}\gamma_{1})$& $\gamma_{2}$& $(-i\gamma_{3}\gamma_{0})$& & $\gamma_{1}$&  $\gamma_{0}\gamma_{2}$&0\\\hline
 14& & & $(-i\gamma_{0}\gamma_{2})$& & $\gamma_{2}$& $(-i\gamma_{3}\gamma_{0})$& & $\gamma_{1}$&  $\gamma_{3}\gamma_{1}$&0\\\hline
 15& & $(-i\gamma_{2}\gamma_{3})$& & & $\gamma_{2}$& $(-i\gamma_{3}\gamma_{0})$& & $\gamma_{1}$&  $\gamma_{0}\gamma_{1}$&$i$\\\hline
 16& $(-i\gamma_{0}\gamma_{1})$& & & & $\gamma_{2}$& $(-i\gamma_{3}\gamma_{0})$& & $\gamma_{1}$&  $\gamma_{2}\gamma_{3}$&$i$\\\hline
 17& & & $(-i\gamma_{0}\gamma_{2})$& $(-i\gamma_{3}\gamma_{1})$& $\gamma_{2}$& & & $\gamma_{1}$&  $\gamma_{3}\gamma_{0}$&0\\\hline
 18& & $(-i\gamma_{2}\gamma_{3})$& & $(-i\gamma_{3}\gamma_{1})$& $\gamma_{2}$& & & $\gamma_{1}$&  1&1\\\hline
 19& $(-i\gamma_{0}\gamma_{1})$& & & $(-i\gamma_{3}\gamma_{1})$& $\gamma_{2}$& & & $\gamma_{1}$&  $(-i\gamma_0\gamma_1)(-i\gamma_2\gamma_3)$&1\\\hline
 20& & $(-i\gamma_{2}\gamma_{3})$& $(-i\gamma_{0}\gamma_{2})$& & $\gamma_{2}$& & & $\gamma_{1}$&  $(-i\gamma_0\gamma_1)(-i\gamma_2\gamma_3)$&1\\\hline
 21& $(-i\gamma_{0}\gamma_{1})$& & $(-i\gamma_{0}\gamma_{2})$& & $\gamma_{2}$& & & $\gamma_{1}$&  1&1\\\hline
 22& $(-i\gamma_{0}\gamma_{1})$& $(-i\gamma_{2}\gamma_{3})$& & & $\gamma_{2}$& & & $\gamma_{1}$&  $\gamma_{3}\gamma_{0}$&0\\\hline
 23& & & & $(-i\gamma_{3}\gamma_{1})$& $\gamma_{2}$& $(-i\gamma_{3}\gamma_{0})$& $(-i\gamma_{1}\gamma_{2})$& $\gamma_{1}$&  $(-i\gamma_{0}\gamma_{1})$&1\\\hline
 24& & & $(-i\gamma_{0}\gamma_{2})$& & $\gamma_{2}$& $(-i\gamma_{3}\gamma_{0})$& $(-i\gamma_{1}\gamma_{2})$& $\gamma_{1}$&  $(-i\gamma_{2}\gamma_{3})$&1\\\hline
 25& & $(-i\gamma_{2}\gamma_{3})$& & & $\gamma_{2}$& $(-i\gamma_{3}\gamma_{0})$& $(-i\gamma_{1}\gamma_{2})$& $\gamma_{1}$&  $i\gamma_{0}\gamma_{2}$&0\\\hline
 26& $(-i\gamma_{0}\gamma_{1})$& & & & $\gamma_{2}$& $(-i\gamma_{3}\gamma_{0})$& $(-i\gamma_{1}\gamma_{2})$& $\gamma_{1}$&  $-i\gamma_{3}\gamma_{1}$&0\\\hline
 27& & & $(-i\gamma_{0}\gamma_{2})$& $(-i\gamma_{3}\gamma_{1})$& $\gamma_{2}$& & $(-i\gamma_{1}\gamma_{2})$& $\gamma_{1}$&  $(-i\gamma_0\gamma_1)(\gamma_2\gamma_3)$&$i$\\\hline 
 28& & $(-i\gamma_{2}\gamma_{3})$& & $(-i\gamma_{3}\gamma_{1})$& $\gamma_{2}$& & $(-i\gamma_{1}\gamma_{2})$& $\gamma_{1}$& $i\gamma_{1}\gamma_2$&0\\\hline
 29& $(-i\gamma_{0}\gamma_{1})$& & & $(-i\gamma_{3}\gamma_{1})$& $\gamma_{2}$& & $(-i\gamma_{1}\gamma_{2})$& $\gamma_{1}$& $i\gamma_{0}\gamma_3$&0\\\hline
 30& & $(-i\gamma_{2}\gamma_{3})$& $(-i\gamma_{0}\gamma_{2})$& & $\gamma_{2}$& & $(-i\gamma_{1}\gamma_{2})$& $\gamma_{1}$& $i\gamma_{0}\gamma_3$&0\\\hline
 31& $(-i\gamma_{0}\gamma_{1})$& & $(-i\gamma_{0}\gamma_{2})$& & $\gamma_{2}$& & $(-i\gamma_{1}\gamma_{2})$& $\gamma_{1}$& $i\gamma_{1}\gamma_2$&0\\\hline
 32& $(-i\gamma_{0}\gamma_{1})$& $(-i\gamma_{2}\gamma_{3})$& & & $\gamma_{2}$& & $(-i\gamma_{1}\gamma_{2})$& $\gamma_{1}$& $(-i\gamma_0\gamma_1)(\gamma_2\gamma_3)$&$i$\\\hline
 \end{tabular}
    \caption{Expansion of Eq.(\ref{eq:Majorana phase eg.1}) including terms 1-32 and evaluate each term in the expectation of state $|\psi_0\rangle$. }
    \label{tab:derivation1}
\end{table*}
 \begin{table*}
    \centering
    \begin{tabular}{|c|c|c|c|c|c|c|c|c|c|c|}\hline
         {\footnotesize $n$} &  $P_{01}$&  $P_{23}$&  $P_{02}$&  $P_{31}$&  $\gamma_{2}$&  $P_{30}$&  $P_{12}$&  $\gamma_{1}$&  Simplified Term $R_n$ &{\footnotesize  $\langle \psi_0|R_n|\psi_0\rangle$}\\\hline
 33& & & $(-i\gamma_{0}\gamma_{2})$& $(-i\gamma_{3}\gamma_{1})$& $\gamma_{2}$& $(-i\gamma_{3}\gamma_{0})$& & $\gamma_{1}$& $i$&$i$\\\hline
 34& & $(-i\gamma_{2}\gamma_{3})$& & $(-i\gamma_{3}\gamma_{1})$& $\gamma_{2}$& $(-i\gamma_{3}\gamma_{0})$& & $\gamma_{1}$& $i\gamma_{0}\gamma_3$&0\\\hline
 35& $(-i\gamma_{0}\gamma_{1})$& & & $(-i\gamma_{3}\gamma_{1})$& $\gamma_{2}$& $(-i\gamma_{3}\gamma_{0})$& & $\gamma_{1}$& $i\gamma_{1}\gamma_2$&0\\\hline
 36& & $(-i\gamma_{2}\gamma_{3})$& $(-i\gamma_{0}\gamma_{2})$& & $\gamma_{2}$& $(-i\gamma_{3}\gamma_{0})$& & $\gamma_{1}$& $i\gamma_{1}\gamma_2$&0\\\hline
 37& $(-i\gamma_{0}\gamma_{1})$& & $(-i\gamma_{0}\gamma_{2})$& & $\gamma_{2}$& $(-i\gamma_{3}\gamma_{0})$& & $\gamma_{1}$& $i\gamma_{0}\gamma_3$&0\\\hline
 38& $(-i\gamma_{0}\gamma_{1})$& $(-i\gamma_{2}\gamma_{3})$& & & $\gamma_{2}$& $(-i\gamma_{3}\gamma_{0})$& & $\gamma_{1}$& $i$&$i$\\\hline
 39& & $(-i\gamma_{2}\gamma_{3})$& $(-i\gamma_{0}\gamma_{2})$& $(-i\gamma_{3}\gamma_{1})$& $\gamma_{2}$& & & $\gamma_{1}$& $i\gamma_{0}\gamma_2$&0\\\hline
 40& $(-i\gamma_{0}\gamma_{1})$& & $(-i\gamma_{0}\gamma_{2})$& $(-i\gamma_{3}\gamma_{1})$& $\gamma_{2}$& & & $\gamma_{1}$& $-i\gamma_{1}\gamma_3$&0\\\hline
 41& $(-i\gamma_{0}\gamma_{1})$& $(-i\gamma_{2}\gamma_{3})$& & $(-i\gamma_{3}\gamma_{1})$& $\gamma_{2}$& & & $\gamma_{1}$& $-i\gamma_{0}\gamma_{1}$&1\\\hline
 42& $(-i\gamma_{0}\gamma_{1})$& $(-i\gamma_{2}\gamma_{3})$& $(-i\gamma_{0}\gamma_{2})$& & $\gamma_{2}$& & & $\gamma_{1}$& $-i\gamma_{2}\gamma_3$&1\\\hline 
 43& & & $(-i\gamma_{0}\gamma_{2})$& $(-i\gamma_{3}\gamma_{1})$& $\gamma_{2}$& $(-i\gamma_{3}\gamma_{0})$& $(-i\gamma_{1}\gamma_{2})$& $\gamma_{1}$& $\gamma_{1}\gamma_2$&0\\\hline
 44& & $(-i\gamma_{2}\gamma_{3})$& & $(-i\gamma_{3}\gamma_{1})$& $\gamma_{2}$& $(-i\gamma_{3}\gamma_{0})$& $(-i\gamma_{1}\gamma_{2})$& $\gamma_{1}$& $(-i\gamma_0\gamma_1)(-i\gamma_2\gamma_3)$&1\\\hline
 45& $(-i\gamma_{0}\gamma_{1})$& & & $(-i\gamma_{3}\gamma_{1})$& $\gamma_{2}$& $(-i\gamma_{3}\gamma_{0})$& $(-i\gamma_{1}\gamma_{2})$& $\gamma_{1}$& 1&1\\\hline
 46& & $(-i\gamma_{2}\gamma_{3})$& $(-i\gamma_{0}\gamma_{2})$& & $\gamma_{2}$& $(-i\gamma_{3}\gamma_{0})$& $(-i\gamma_{1}\gamma_{2})$& $\gamma_{1}$& 1&1\\\hline
 47& $(-i\gamma_{0}\gamma_{1})$& & $(-i\gamma_{0}\gamma_{2})$& & $\gamma_{2}$& $(-i\gamma_{3}\gamma_{0})$& $(-i\gamma_{1}\gamma_{2})$& $\gamma_{1}$& $(-i\gamma_0\gamma_1)(-i\gamma_2\gamma_3)$&1\\\hline
 48& $(-i\gamma_{0}\gamma_{1})$& $(-i\gamma_{2}\gamma_{3})$& & & $\gamma_{2}$& $(-i\gamma_{3}\gamma_{0})$& $(-i\gamma_{1}\gamma_{2})$& $\gamma_{1}$& $-\gamma_{1}\gamma_2$&0\\\hline
 49& & $(-i\gamma_{2}\gamma_{3})$& $(-i\gamma_{0}\gamma_{2})$& $(-i\gamma_{3}\gamma_{1})$& $\gamma_{2}$& & $(-i\gamma_{1}\gamma_{2})$& $\gamma_{1}$& $\gamma_{0}\gamma_{1}$ &$i$\\\hline
 50& $(-i\gamma_{0}\gamma_{1})$& & $(-i\gamma_{0}\gamma_{2})$& $(-i\gamma_{3}\gamma_{1})$& $\gamma_{2}$& & $(-i\gamma_{1}\gamma_{2})$& $\gamma_{1}$& $\gamma_{2}\gamma_{3}$&$i$\\\hline
 51& $(-i\gamma_{0}\gamma_{1})$& $(-i\gamma_{2}\gamma_{3})$& & $(-i\gamma_{3}\gamma_{1})$& $\gamma_{2}$& & $(-i\gamma_{1}\gamma_{2})$& $\gamma_{1}$& $\gamma_{0}\gamma_{2}$&0\\\hline
 52& $(-i\gamma_{0}\gamma_{1})$& $(-i\gamma_{2}\gamma_{3})$& $(-i\gamma_{0}\gamma_{2})$& & $\gamma_{2}$& & $(-i\gamma_{1}\gamma_{2})$& $\gamma_{1}$& $-\gamma_{1}\gamma_3$&0\\\hline
 53& & $(-i\gamma_{2}\gamma_{3})$& $(-i\gamma_{0}\gamma_{2})$& $(-i\gamma_{3}\gamma_{1})$& $\gamma_{2}$& $(-i\gamma_{3}\gamma_{0})$& & $\gamma_{1}$& $\gamma_{2}\gamma_3$&$i$\\\hline
 54& $(-i\gamma_{0}\gamma_{1})$& & $(-i\gamma_{0}\gamma_{2})$& $(-i\gamma_{3}\gamma_{1})$& $\gamma_{2}$& $(-i\gamma_{3}\gamma_{0})$& & $\gamma_{1}$& $\gamma_{0}\gamma_{1}$&$i$\\\hline
 55& $(-i\gamma_{0}\gamma_{1})$& $(-i\gamma_{2}\gamma_{3})$& & $(-i\gamma_{3}\gamma_{1})$& $\gamma_{2}$& $(-i\gamma_{3}\gamma_{0})$& & $\gamma_{1}$& $-\gamma_{1}\gamma_3$&0\\\hline 
 56& $(-i\gamma_{0}\gamma_{1})$& $(-i\gamma_{2}\gamma_{3})$& $(-i\gamma_{0}\gamma_{2})$& & $\gamma_{2}$& $(-i\gamma_{3}\gamma_{0})$& & $\gamma_{1}$& $\gamma_{0}\gamma_{2}$&0\\\hline
 57& $(-i\gamma_{0}\gamma_{1})$& $(-i\gamma_{2}\gamma_{3})$& $(-i\gamma_{0}\gamma_{2})$& $(-i\gamma_{3}\gamma_{1})$& $\gamma_{2}$& & & $\gamma_{1}$& $-\gamma_{1}\gamma_2$&0\\\hline
 58& & $(-i\gamma_{2}\gamma_{3})$& $(-i\gamma_{0}\gamma_{2})$& $(-i\gamma_{3}\gamma_{1})$& $\gamma_{2}$& $(-i\gamma_{3}\gamma_{0})$& $(-i\gamma_{1}\gamma_{2})$& $\gamma_{1}$& $-i\gamma_{1}\gamma_3$&0\\\hline
 59& $(-i\gamma_{0}\gamma_{1})$& & $(-i\gamma_{0}\gamma_{2})$& $(-i\gamma_{3}\gamma_{1})$& $\gamma_{2}$& $(-i\gamma_{3}\gamma_{0})$& $(-i\gamma_{1}\gamma_{2})$& $\gamma_{1}$& $i\gamma_{0}\gamma_{2}$&0\\\hline
 60& $(-i\gamma_{0}\gamma_{1})$& $(-i\gamma_{2}\gamma_{3})$& & $(-i\gamma_{3}\gamma_{1})$& $\gamma_{2}$& $(-i\gamma_{3}\gamma_{0})$& $(-i\gamma_{1}\gamma_{2})$& $\gamma_{1}$& $-i\gamma_{2}\gamma_3$&1\\\hline
 61& $(-i\gamma_{0}\gamma_{1})$& $(-i\gamma_{2}\gamma_{3})$& $(-i\gamma_{0}\gamma_{2})$& & $\gamma_{2}$& $(-i\gamma_{3}\gamma_{0})$& $(-i\gamma_{1}\gamma_{2})$& $\gamma_{1}$& $-i\gamma_{0}\gamma_{1}$&1\\\hline
 62& $(-i\gamma_{0}\gamma_{1})$& $(-i\gamma_{2}\gamma_{3})$& $(-i\gamma_{0}\gamma_{2})$& $(-i\gamma_{3}\gamma_{1})$& $\gamma_{2}$& & $(-i\gamma_{1}\gamma_{2})$& $\gamma_{1}$& $i$&$i$\\\hline
 63& $(-i\gamma_{0}\gamma_{1})$& $(-i\gamma_{2}\gamma_{3})$& $(-i\gamma_{0}\gamma_{2})$& $(-i\gamma_{3}\gamma_{1})$& $\gamma_{2}$& $(-i\gamma_{3}\gamma_{0})$& & $\gamma_{1}$& $i(-i\gamma_0\gamma_1)(-i\gamma_2\gamma_3)$&$i$\\\hline
 64& $(-i\gamma_{0}\gamma_{1})$& $(-i\gamma_{2}\gamma_{3})$& $(-i\gamma_{0}\gamma_{2})$& $(-i\gamma_{3}\gamma_{1})$& $\gamma_{2}$& $(-i\gamma_{3}\gamma_{0})$& $(-i\gamma_{1}\gamma_{2})$& $\gamma_{1}$& $-\gamma_{0}\gamma_3$&0\\ \hline
    \end{tabular}
    \caption{Expansion of Eq.(\ref{eq:Majorana phase eg.1}) including terms 33-64 and evaluate each term in the expectation of state $|\psi_0\rangle$. }
    \label{tab:derivation2}
\end{table*}

The most straightforward strategy to compute the expectation value above is to expand all the Majorana projectors directly and evaluate every term in the polynomial-like summation of the expansion one by one using the above properties.
There are 6 Majorana projectors in the expectation value Eq.\eqref{eq:Majorana phase eg.1}.
Since every Majorana projector has two terms $1$ and $-i\gamma_i\gamma_j$, we will get $2^6 = 64$ terms in total after expanding all Majorana projectors.   Namely, 
\begin{align}
   & \bra{\psi_0} \left[ \sqrt{2} P_{0,1} P_{2,3} \right] \left[ \sqrt{2} P_{0,2} P_{3,1} \gamma_2 \right]
 \left[ \sqrt{2} P_{3,0} P_{1,2} \gamma_1 \right] \ket{\psi_0} \nonumber \\
 =&\frac{(\sqrt{2})^3}{2^6} \sum_{\{ n_i\}} \bra{\psi_0} T_{01}^{n_1}T_{23}^{n_2}  T_{02}^{n_3} T_{31}^{n_4}\gamma_2 T_{30}^{n_5} T_{12}^{n_6}\gamma_1\ket{\psi_0}\nonumber \\
 =& \frac{(\sqrt{2})^3}{2^6} \sum_n \bra{\psi_0} R_n \ket{\psi_0}
\end{align}
where we denote the projector as $P_{ij}=\frac{1}{2} \sum_{n=0,1}(T_{ij})^n$ with $T_{ij}=-i\gamma_i\gamma_j$.  To evaluate the eigenvalue of $R_n$ for each $n$, we can first simplify $R_n$ using the anticommutation relation of the Majorana fermions so that the simplified $ R_n$ does not contain replicated Majorana fermions. Further using the fact that the state $|\psi_0\rangle$ is a Majorana dimer state with pairing $-i\gamma_0\gamma_1=-i\gamma_2\gamma_3=1$ when acting $|\psi_0\rangle$, the expectation of $ R_n$ will be nonzero (zero) if the Majorana fermion of $ R_n$ can (not) pair up into $\gamma_0\gamma_1$ and/or $\gamma_2\gamma_3$.   We present the concrete expansion and calculation of the 64 terms in Table \ref{tab:derivation1} and \ref{tab:derivation2}. 
For example, the term $R_n$ with $n=12$, the $R_{12}=(-i\gamma_0\gamma_1)\gamma_2(-i\gamma_1\gamma_2)\gamma_1$, where the term $(-i\gamma_0\gamma_1)$ comes from the projector $P_{01}$ and $(-i\gamma_1\gamma_2)$ from $P_{12}$. Using the anticommutation relations of majorana fermions, $R_{12}$ simplifies to $R_{12}=\gamma_0\gamma_1$. The expectation $\langle \psi_0 |R_{12} |\psi_0\rangle=i$ since we have $-i\gamma_0\gamma_1|\psi_0\rangle= |\psi_0\rangle$. 
     Another example is $n=8$ whose simplified $R_8=\gamma_3\gamma_0$ and its action on state $|\psi_0\rangle$ will map to anther state $|\psi_0'\rangle$ with eigenvalues of  $-i\gamma_0\gamma_1|\psi_0'\rangle=-i\gamma_2\gamma_3|\psi_0'\rangle=-|\psi_0'\rangle$. So $|\psi_0\rangle$ and $|\psi_0'\rangle$ are orthogonal since they have different eigenvalues of the two parity operators.  So we have $\langle \psi_0|R_8|\psi_0\rangle=0$. From the Table \ref{tab:derivation1} and \ref{tab:derivation2}, we  count the nonzero expectation values---sixteen $R_n$ have $1$ expectation and also sixteen $R_n$ have $i$ expectation, the Eq.(\ref{eq:Majorana phase eg.1}) are evaluated into 
\begin{align}
    e^{i\theta_\gamma}=  \frac{(\sqrt{2})^3}{2^6}\times \big[16\times (1+i)\big]=\frac{1}{\sqrt{2}}(1+i)=e^{i\frac{\pi}{4}}
    \label{eq:Majorana phase eg.2}
\end{align}
where $(\sqrt{2})^3$ comes from the normalization factors of the projectors and $2^6$ comes from the $1/2$ in each projector. 

We take a similar strategy to compute $O_5^\gamma$. 
To calculate $O_5^\gamma$, we  expand all $P_i=2^{\sum_j (L_j-1)/2}\prod_{\langle ij\rangle}\frac{1}{2}(1-i\gamma_{iC}\gamma_{jD})$ into a polynomial-like summation of exponentially many terms, and each term can be calculated independently and then summed as shown in the illustrated minimal example above. 
For simplicity, we can use the property Eq.\eqref{eq:Ptri_prop} and the commutativity in Eq.\eqref{eq:P01234def} to simplify the string operator $X$ in Eq.\eqref{eqn:O5}. Then the simplified form of $X$ becomes $\tilde X$ with $P_i$ replaced by $\tilde P_i$ given by 
 \begin{align}
 &\tilde P_0=\mathcal{N}_0P_{0125} P_{0235} P_{013} P_{123}\nonumber \\
 &\tilde P_1=\mathcal{N}_1P_{0345} P_{024}\nonumber \\
 &\tilde P_2=\mathcal{N}_2P_{0135}P_{0345}P_{014} P_{134} P_{145} \nonumber \\
 &\tilde P_3=\mathcal{N}_3P_{1235}P_{1345}P_{124}P_{234}P_{245}  \nonumber \\
 &\tilde P_4=\mathcal{N}_4P_{0245}\nonumber\\
 &\tilde P_5=\mathcal{N}_5P_{0124}P_{0234}
 \end{align}
%
where $\mathcal{N}_i=2^{\sum_j (L_j-1)/2}$ where $2L_j$ is the length of  $j$-th transition loop in the $i$-th $F$ move.
Note that we have removed one more projector $P_{135}$ from the string operator as $P_{135}$ acts on the $|\text{GS}\rangle_{\gamma}$ as identity.   
Expansion of $ X$ will involve many terms,  among which many, however, will contribute to zero since the mismatch of Majorana fermion action on the ground state. For those which will result in nonzero values in the ground state expectation, we call each term as a ‘path'. In fact, different paths may result in different values, among which there are four possibilities: $\pm 1, \pm i$.

Depending on various configurations of $n_2$ in the superhexagon equation, we can calculate them case by case. 
With the 2-cocycle condition on $n_2$ in Eq.(\ref{eq:2-cocyle}),  there are 10 among them are independent:
\begin{align}
n_2(012), n_2(013), n_2(014), n_2(015), n_2(023), \nonumber \\
n_2(024), n_2(025), n_2(034),
    n_2(035), n_2(045).
\end{align}
Furthermore, $O_5^\gamma$ also depends explicitly on $\omega_2$ and $s_1$, but not all of them will be involved. In fact, we only need to consider the following independent ones:
\begin{align}
\omega_2(012), \omega_2(013), \omega_2(023), s_2(01), s_1(02).
\end{align}
So, the $O_5^\gamma$ is actually a function (or more generally functional) on these fifteen quantities, which can be combined into a vector $\vec N$
 \begin{align}
\vec N =&[s_1(02), s_2(01), \omega_2(023), \omega_2(013),   \omega_2(012), \nonumber \\
&n_2(045),  n_2(035),n_2(034),n_2(025),n_2(024),\nonumber \\
& n_2(023), n_2(015), n_2(014), n_2(013), n_2(012) ]  \nonumber \\ 
\end{align}
whose elements  can all take $0$ or $1$. Alternatively, we can use the following case number to encode all the $n_2$-$\omega_2$-$s_1$ patterns
\begin{align}
\mathrm{Case}=\sum_{i=0}^{14} 2^{i} \times N_i.
\end{align} 
Finally, the function $O_5^\gamma$ can be denoted as a function depending on $\vec N$ or $\mathrm{Case}$, i,e, $O_5^\gamma[\vec N]$ or $O_5^\gamma[Case]$. As there are many cases (up to $2^{15}=32768$), we do this using a computer algorithm on a laptop with a CPU and the results for all the cases are given in open source project \cite{YangQi_github}. 

Let us illustrate it with a simple example, which has $n_2(012)=1$ while all other $n_2$ and $\omega_2$ and $s_1$ are zero. To carry out the corresponding $O_5^\gamma$ we take the following steps:
\begin{enumerate}
\item Determine other $n_2$ and $\omega_2$ and $s_1$: According to the cocycle conditions, we have $n_2(123)=n_2(124)=n_2(125)=1$ while all other $\omega_2,s_1$ equal to zero. 
\item Determine the state $|GS\rangle_\gamma$ corresponding to the right hand side of Fig.\ref{fig:hexagon_dual}. This state has non-trivial pairing $-i\gamma_{123B}^{g_0^{-1}g_1}\gamma_{012A}^e=-i\gamma_{124B}^{g_0^{-1}g_1}\gamma_{123A}^{g_0^{-1}g_1}=-i\gamma_{125B}^{g_0^{-1}g_1}\gamma_{124A}^{g_0^{-1}g_1}=1$ while all other Majorana dimers are in vacuum states. The configuration of decorated KC is shown below (indicated by the green line) in the resolved dual  lattice
\vspace{0pt}
\begin{center}
\includegraphics[width=0.25\textwidth]{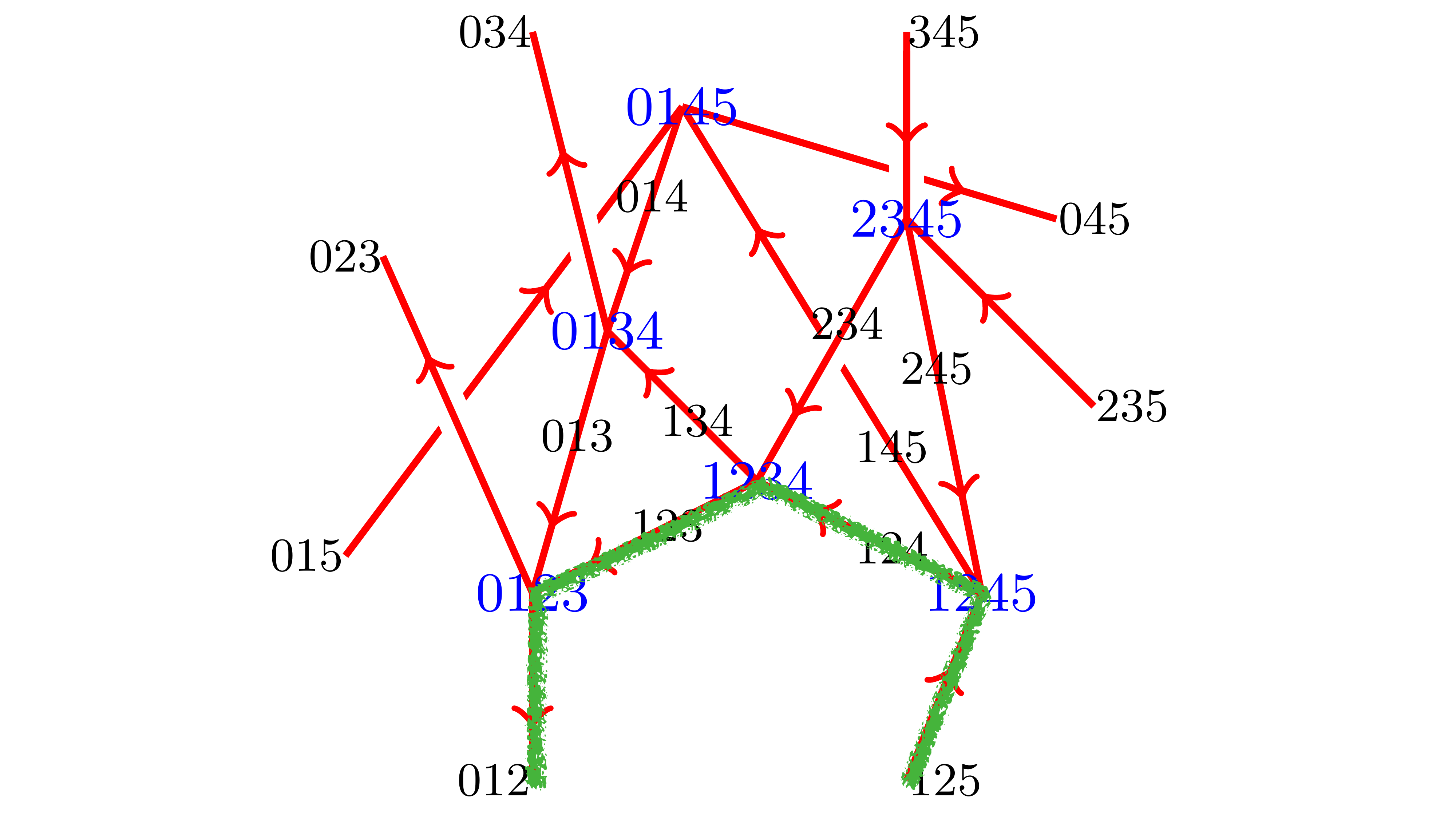}
\end{center}
\vspace{0pt}

\item Calculate all $\alpha_{\hat i}$ and $\beta_{\hat i}$ which can determine which Majorana fermions are inserted. According to Eqs.(\ref{eq:def_tal}) and (\ref{eq:def_tbe}), all the $\alpha_{\hat i}=\beta_{\hat i}=0$. So no inserted Majorana fermion is needed.

\item Determine all the $\tilde P_i$.
\begin{gather*}
  \tilde P_0 =\sqrt{2} \left(\frac{1-i\gamma_{125B}^{g_0^{-1}g_1}\gamma_{012A}^e}{2}\right)\left(\frac{1-i\gamma_{123A}^{g_0^{-1}g_1}\gamma_{123B}^{g_0^{-1}g_1}}{2}\right) \\
  \tilde P_1 = 1 \\
  \tilde P_2 = 1 \\
  \tilde P_3 = \sqrt{2}\left(\frac{1-i\gamma_{125B}^{g_0^{-1}g_1}\gamma_{123A}^{g_0^{-1}g_1}}{2}\right)\left(\frac{1-i\gamma_{124A}^{g_0^{-1}g_1}\gamma_{124B}^{g_0^{-1}g_1}}{2}\right) \\
  \tilde P_4 = \sqrt{2} \\
  \tilde P_5 = \sqrt{2}\left(\frac{1-i\gamma_{124B}^{g_0^{-1}g_1}\gamma_{012A}^{e}}{2}\right)
\end{gather*}

\item Substitute everything into Eq.(\ref{eqn:O5}) and expand all the terms to calculate one by one. As shown in the Table \ref{tab:derivation0}, there are 8 paths with $1$ contribution. Therefore, we have
\begin{align}
e^{i\theta_\gamma}=2^2\times \left(\frac{1}{2}\right)^5 \times (8\times 1)=1.
\label{eq:example_main}
\end{align}
for such $n_2$-$\omega_2$-$s_1$ configuration. 
\end{enumerate}

Finally, we make a remark. In Ref.\cite{Wang2020}, for unitary symmetry (or more precisely when $\tilde \beta=0$) it is conjectured that  equivalence of the $O_5^\gamma$ can be obtained by only considering a typical (shortest) path. So the most fractionalized phase factor in $O_5^\gamma=e^{i\theta_\gamma}$  is $\theta_\gamma=\frac{\pi}{2}$ there (recall other components of $O_5$ is at most $-1$).  However, it does not work for anti-unitary symmetry, and  we have to quantum average all the possible paths.  In this case, there exists more possibility for the fractionalized phase of $O_5^\gamma$. For example, the whole quantum average may involve the same number of  $1$ and $i$ paths, so that the $O_5^\gamma$  is given by more general fractionalized phase  $\theta_\gamma=\frac{\pi}{4}$, as shown in the above illustrated example.




\begin{table*}
    \centering
    \begin{tabular}{|c|c|c|c|c|c|c|c|}\hline
        \multirow{2}{*}{  {\footnotesize $n$}} &  \multicolumn{2}{c|}{$\tilde P_{3}$}&  \multicolumn{2}{c|}{$\tilde P_{0}$}&  $\tilde P_{5}$& \multirow{2}{*}{  Simplified Term $R_n$ }& \multirow{2}{*}{ {\footnotesize  $\langle \psi_0|R_n|\psi_0\rangle$}}\\
         \cline{2-6}& $(1-i\gamma_6\gamma_2)$& $(1-i\gamma_4\gamma_5)$& $(1-i\gamma_6\gamma_1)$& $(1-i\gamma_2\gamma_3)$& $(1-i\gamma_5\gamma_1)$& &\\
         \hline
         1&  &  &  &  &  &  1&1\\\hline
         2&  &  &  &  &  $(-i\gamma_5\gamma_1)$&  $(-i\gamma_5\gamma_1)$&0\\\hline
         3&  &  &  &  $(-i\gamma_2\gamma_3)$&  &  $(-i\gamma_2\gamma_3)$&$0$ \\\hline
         4&  &  &  $(-i\gamma_6\gamma_1)$&  &  &  $(-i\gamma_6\gamma_1)$&0\\\hline
         5&  &  $(-i\gamma_4\gamma_5)$&  &  &  &  $(-i\gamma_4\gamma_5)$&0\\\hline
         6&  $(-i\gamma_6\gamma_2)$&  &  &  &  &  $(-i\gamma_6\gamma_2)$&0 \\\hline
         7&  &  &  &  $(-i\gamma_2\gamma_3)$&  $(-i\gamma_5\gamma_1)$&  $(-i\gamma_5\gamma_2)$ $(-i\gamma_3\gamma_1)$& $1$\\\hline
         8&  &  &  $(-i\gamma_6\gamma_1)$&  &  $(-i\gamma_5\gamma_1)$&  $(-i\gamma_5\gamma_6)$ &0\\\hline
         9&  &  $(-i\gamma_4\gamma_5)$&  &  &  $(-i\gamma_5\gamma_1)$& $(-i\gamma_4\gamma_1)$  &0\\\hline
 10& $(-i\gamma_6\gamma_2)$& & & & $(-i\gamma_5\gamma_1)$&  $(-i\gamma_6\gamma_2)$    $(-i\gamma_5\gamma_1)$&0\\\hline
 11& & & $(-i\gamma_6\gamma_1)$& $(-i\gamma_2\gamma_3)$& &  $(-i\gamma_6\gamma_1)$ $(-i\gamma_2\gamma_3)$&0\\\hline
 12& & $(-i\gamma_4\gamma_5)$& & $(-i\gamma_2\gamma_3)$& &  $(-i\gamma_4\gamma_5)$ $(-i\gamma_2\gamma_3)$ &0\\\hline
 13& $(-i\gamma_6\gamma_2)$& & & $(-i\gamma_2\gamma_3)$& &  $-\gamma_6\gamma_3$&0\\\hline
 14& & $(-i\gamma_4\gamma_5)$& $(-i\gamma_6\gamma_1)$& & &  $(-i\gamma_4\gamma_5)$ $(-i\gamma_6\gamma_1)$&0\\\hline
 15& $(-i\gamma_6\gamma_2)$& & $(-i\gamma_6\gamma_1)$& & &  $\gamma_2\gamma_1$&0\\\hline
 16& $(-i\gamma_6\gamma_2)$& $(-i\gamma_4\gamma_5)$& & & & $(-i\gamma_6\gamma_4)$ $(-i\gamma_5\gamma_2)$ &$1$\\\hline
 17& & & $(-i\gamma_6\gamma_1)$& $(-i\gamma_2\gamma_3)$& $(-i\gamma_5\gamma_1)$& $(\gamma_6\gamma_5)$ $(-i\gamma_2\gamma_3)$ &0\\\hline
 18& & $(-i\gamma_4\gamma_5)$& & $(-i\gamma_2\gamma_3)$& $(-i\gamma_5\gamma_1)$& $(-i\gamma_4)$$(-i\gamma_2\gamma_3)$$(-i\gamma_1)$ &0\\\hline
 19& $(-i\gamma_6\gamma_2)$& & & $(-i\gamma_2\gamma_3)$& $(-i\gamma_5\gamma_1)$&$(-i\gamma_6)$$(-i\gamma_3)$$(-i\gamma_5\gamma_1)$  &0\\\hline
 20& & $(-i\gamma_4\gamma_5)$& $(-i\gamma_6\gamma_1)$& & $(-i\gamma_5\gamma_1)$& $(-i\gamma_6\gamma_4)$&1\\\hline
 21& $(-i\gamma_6\gamma_2)$& & $(-i\gamma_6\gamma_1)$& & $(-i\gamma_5\gamma_1)$&$(-i\gamma_5\gamma_2)$ &$1$\\\hline
 22& $(-i\gamma_6\gamma_2)$& $(-i\gamma_4\gamma_5)$& & & $(-i\gamma_5\gamma_1)$& $(-i\gamma_6\gamma_2)$$(-i\gamma_4)$$(-i\gamma_1)$ &0\\\hline
\multirow{2}{*}{ 23}& & \multirow{2}{*}{$(-i\gamma_4\gamma_5)$}& \multirow{2}{*}{$(-i\gamma_6\gamma_1)$}& \multirow{2}{*}{$(-i\gamma_2\gamma_3)$}& &  $(-i\gamma_6\gamma_4)$ $(-i\gamma_5\gamma_2)$ &1\\
&&&&&&$(-i\gamma_3\gamma_1)$&\\
\hline
 24& $(-i\gamma_6\gamma_2)$& & $(-i\gamma_6\gamma_1)$& $(-i\gamma_2\gamma_3)$& &  $(-i\gamma_3\gamma_1)$ &1\\\hline
 25& $(-i\gamma_6\gamma_2)$& $(-i\gamma_4\gamma_5)$& & $(-i\gamma_2\gamma_3)$& &  $(-i\gamma_6)$$(-i\gamma_4\gamma_5)$  $(-i\gamma_3)$&0\\\hline
 26& $(-i\gamma_6\gamma_2)$& $(-i\gamma_4\gamma_5)$& $(-i\gamma_6\gamma_1)$& & & $(\gamma_2\gamma_1)$ $(-i\gamma_4\gamma_5)$ &0\\\hline
 27& & $(-i\gamma_4\gamma_5)$& $(-i\gamma_6\gamma_1)$& $(-i\gamma_2\gamma_3)$& $(-i\gamma_5\gamma_1)$&$(-i\gamma_6\gamma_4)$$(-i\gamma_2\gamma_3)$  &0\\\hline 
 28& $(-i\gamma_6\gamma_2)$& & $(-i\gamma_6\gamma_1)$& $(-i\gamma_2\gamma_3)$& $(-i\gamma_5\gamma_1)$& $(\gamma_3\gamma_5)$ &0\\\hline
 29& $(-i\gamma_6\gamma_2)$& $(-i\gamma_4\gamma_5)$& & $(-i\gamma_2\gamma_3)$& $(-i\gamma_5\gamma_1)$&  $(-i\gamma_6\gamma_4)$ $(-i\gamma_3\gamma_1)$ &1\\\hline
 30& $(-i\gamma_6\gamma_2)$& $(-i\gamma_4\gamma_5)$& $(-i\gamma_6\gamma_1)$& & $(-i\gamma_5\gamma_1)$& $(\gamma_2\gamma_4)$ &0\\\hline
 31& $(-i\gamma_6\gamma_2)$& $(-i\gamma_4\gamma_5)$& $(-i\gamma_6\gamma_1)$& $(-i\gamma_2\gamma_3)$& & $(-i\gamma_4\gamma_5)$ $(-i\gamma_3\gamma_1)$ &0\\\hline
 32& $(-i\gamma_6\gamma_2)$& $(-i\gamma_4\gamma_5)$& $(-i\gamma_6\gamma_1)$& $(-i\gamma_2\gamma_3)$& $(-i\gamma_5\gamma_1)$& $(-i\gamma_3\gamma_4)$  &0\\\hline
 \end{tabular}
    \caption{Derivation of Eq.(\ref{eq:example_main}): Expansion  and evaluation of each term in the expectation of state $|GS\rangle_\gamma$. For convenience, we denote $\gamma_{012A}$ by $\gamma_1$,  $\gamma_{123A}$ by $\gamma_2$, $\gamma_{123B}$ by $\gamma_3$, $\gamma_{124A}$ by $\gamma_4$, $\gamma_{124B}$ by $\gamma_5$, and $\gamma_{125B}$ by $\gamma_6$. }
    \label{tab:derivation0}
\end{table*}

\subsection{Mathematical framework: spectral sequence}
\label{sec:math_ss}
Here we review the formal spectral sequence framework and how our results fit into this formal framework. Mathematically, the DWD construction of the FSPT is related to the generalized cohomology in the framework of the Atiyah-Hirzebruch (AH) spectral sequence \cite{Gaiotto2019, wang2021domain}. A briefly introduction of spectral sequence is presented in Appendix \ref{sec:ss}.  For the fermionic symmetry in Eq.(\ref{eq:ses_Gf}),  the classification of 3+1D FSPT using the data $(n_1, n_2, n_3, \nu_4)$ are organized into the $E_2$ page of the spectral sequence:
\begin{align}
\bigoplus_{p+q=4} E_2^{p,q}=\bigoplus_{p+q=4} H^p(G_b, h^q_F) 
\end{align}
where $h^q_F$ is the classification of $q-1$ spatial dimensional invertible topological phases with only $\Z_2^f$ symmetry.  More concretely, for lower dimensional cases, they are given  as follows:
\begin{align}
h_F^0=U(1), \,\,\, h_F^1=\Z_2, \,\,\, h_F^2=\Z_2, \,\,\, h_F^3=\Z,\,\,\, h_F^4=\Z_1.
\end{align}
The decorated states characterized by the data $(n_1, n_2, n_3)$ on the $p$-dimensional ($ p=1,2,3 $) $G_b$-symmetry domain walls correspond to layers of the $p+ip$ superconductor ($h^3_F$), the Kitaev chain ($h^2_F$), and the complex fermion ($h^1_F$), respectively. The $p=4$ layer just corresponds to the bosonic SPT layer and $p=0$ is trivial. In general, $G_b$ can be either unitary or anti-unitary. For the latter, $G_b$ can have a non-trivial action on $h_F^q$. In particular, the anti-unitary element will act on $h_F^3$ by mapping its elements $\nu$ to its inverse $-\nu$. This action is easy to understand because the anti-unitary symmetry will change the sign of the chiral central charge of chiral superconductors. 

However, not every piece of data in $E_2$ will result in a valid FSPT. They should satisfy a set of consistency conditions, given by the obstruction functions discussed above.  Mathematically, this means they should subject to several layers of differentials, 
\begin{align}
d_r: E_r^{p,q}\rightarrow E_r^{p+r, q-r+1}.
\end{align}
The  $d_r$ maps a decorated domain wall construction to the one in one dimensional higher systems. Mathematically, the complete explicit differential maps, especially the higher ones, are very difficult and unknown, which hinders the thorough and scalable application of the AH spectral sequence in classifying the 3D FSPT. Our efforts in this work finally pave all fundamental obstacles in their application. 

For the case of $p=1$, the corresponding decoration is the $p+ip$ layer, which is represented by the term $E_2^{1,3}$. We first examine the differential map:
\begin{align}
    d_2: E_2^{1,3}\rightarrow E_2^{3,2}\in H^3(G_b, \Z_2).
\end{align}
The image of $d_2$ is given exactly by $O_3$. In general, there might be higher differentials: the maps $d_3$ and $d_4$. Our arguments in Sec.\ref{sec:crystalZ2T}   and Sec.\ref{sec:p+ip_infinite} suggest that the final classification can be obtained even without explicit expressions for the $d_3$ and $d_4$ differentials. More remarkably, for finite groups, the differentials $d_3$ and $d_4$ from $E^{1,3}$ are trivial. Consequently, the only non-trivial differential map we need to consider for $E^{1,3}$ is $d_2$.

Let's consider the case $p=2$. There are two relevant differentials:
\begin{align}
&d_2: E^{2,2}_2\rightarrow E_2^{4,1},\\
&d_3: E^{2,2}_3\rightarrow E_3^{5,0}.
\end{align}
The map $d_2$ sends a 3D Kitaev chain decoration to a 4D complex fermion decoration. If its image is non-trivial, the corresponding Kitaev chain decoration is obstructed, and there is no need to proceed to $d_3$. If, however, the image of $d_2$ is trivial, we must then examine whether $d_3$ is trivial. The image of $d_2$ is given by the obstruction $O_4$, while that of $d_3$ is given by $O_5$. Our results now provide the complete formula for $O_5$, which is of fundamental importance both for classifying FSPT phases and within the AH spectral sequence theory.

The case $p=3$ involves only a single differential
\begin{align}
&d_2: E^{3,1}_2\rightarrow E_2^{5,0}.
\end{align}
This differential corresponds to the original supercohomology condition for FSPT phases, which is given by the obstruction $O_5^{c}$ with $n_2=0$. A non-trivial image under $d_2$ indicates an obstruction to the corresponding complex fermion decoration; otherwise, it yields a valid 3D FSPT phase.

On the other hand, a non-trivial differential $d_r$ not only signals an obstruction for the lower-dimensional decoration but also induces a trivialization of the higher-dimensional one. This construction corresponds precisely to anomalous SPT states \cite{Wang2020}. For example, if $d_2 n_2$ (with Kitaev chain decoration $n_2 \in E^{2,2}_2$ in 3D FSPT) is non-trivial, then the complex fermion decoration $n_4 = d_2 n_2$ in a 4D FSPT with the same symmetry is trivialized. This occurs because the obstructed decoration in 3D itself serves as a symmetric, short-range-entangled boundary state for the would-be 4D FSPT phase, thereby indicating that the corresponding 4D bulk must be trivial. For our present purpose, we must consider the following trivializations originating from 2D: 
\begin{align}
&d_2: E_2^{0,3}\rightarrow E_2^{2,2},\,\,\,  E_2^{1,2}\rightarrow E_2^{3,1},\,\,\, 
E_2^{2,1}\rightarrow E_2^{4,0} \\
&d_3: E_3^{0,3}\rightarrow E_3^{3,1},\,\,\,  E_3^{1,2}\rightarrow E_3^{4,0}\\
&d_4: E_4^{0,3}\rightarrow E_4^{4,0}
\end{align}
Specifically, except for the $p+ip$ layer (i.e., $n_1 \in E_2^{1,3}$), each layer of the decoration $(n_2, n_3, \nu_4)$ can be trivialized by a 2D symmetric short-range-entangled state. These trivializing states correspond exactly to $\Gamma_2$, $\Gamma_3$, and $\Gamma_4$ in Sec.~\ref{sec:framework_for_internal}, respectively. Consequently, we must quotient out such states to obtain the final classification of FSPT phases that are free of both obstructions and trivializations.

\section{Classification of Interacting Topological Crystalline Superconductors}

The properties of solid state materials in laboratories are mainly determined by the electrons of their constituent atoms and the crystalline symmetries according to which the atoms are organized. Commonly, their topological aspects need to consider the electron spin-dependent effect, such as spin-orbit coupling,   which enforces that the crystalline operations not only acts on spatial coordinates but also on the internal spin 1/2 space. So we need to consider the double groups of the crystalline symmetries for such spin one-half  systems.

In the double space groups,  the two-fold rotation or mirror acts twice will not do anything but leave a minus sign on a single particle. Such minus sign originates from the fact that the electron carries spin one half degree of freedom.  When acting on many-body states, the minus sign will be replaced by the fermion parity operator $P_f$.
Previous works on 2D crystalline systems and 3D point-group systems showed
the classification of the interacting crystalline topological phases  in electronic systems is in one-to-one correspondence to that of interacting internal FSPT in spinless fermionic systems.


Here we focus on 3D double space groups. More precisely, the double space groups $G_\mathrm{dSG}$ is a group extension of space group $G_\mathrm{SG}$ by $\Z_2^f$ just as Eq.\eqref{eq:ses_Gf}
\begin{align}
     1 \rightarrow \Z_2^f\rightarrow G_\mathrm{dSG} \rightarrow  G_\mathrm{SG} \rightarrow 1
\end{align}
The corresponding $\omega_2\in H^2(G_\mathrm{SG},\Z_2)$ for this extension is 
 characterized by a cohomology class $w_2+w_1\cup w_1$. So according to the fermionic CEP proposed in Eq.(\ref{eq6.2}), the spatial fermionic symmetry ($G_\mathrm{SG},\omega_2,s_1=0$) together with $w_{1,2}$ will be mapped to on-site fermionic symmetry ($G_\mathrm{SG}^\mathrm{onsite}$, $\omega_2^\mathrm{eff}$, $s_1^\mathrm{eff}$) where $\omega_2^\mathrm{eff}=0 \in H^2(G_\mathrm{SG}^\mathrm{onsite},\Z_2^f)$,  and $s_1^\mathrm{eff}$ is identified as $w_1$. 
That is to say, the spin one-half system with double space symmetry groups is mapped to a spinless system with on-site symmetry group $\Z_2^f\times G_{\mathrm{SG}}^{\mathrm{onsite}}$. So, the classification of the interacting topological superconductors   protected by symmetry group $G_\mathrm{dSG}$ in electronic systems is in one-to-one correspondence to that of interacting topological superconductors protected by $\Z_2^f\times G_{\mathrm{SG}}^{\mathrm{onsite}}$ in spinless fermionic systems. 
  In Ref.\cite{Zhang25_point}, we have confirmed such a correspondence for 3D point-group topological superconductors. Here we generalize to 3D space group as an extension of the point group by translation group.

\subsection{Classification}
\subsubsection{ $p+ip$ layer}
\label{sec:p+ip_cry}
According to Sec.\ref{sec:decoupling_pplusip}, whether  the $p+ip$ layer from anti-unitary structure is obstructed or not depends on whether $\omega_{2}^\mathrm{eff}+s_1^\mathrm{eff}\cup s_1^\mathrm{eff}$ (equivalently $s_1^\mathrm{eff}\cup s_1^\mathrm{eff}$) is non-trivial or not. Now we show that for almost all the double space group (except No.81 and 82) containing orientation reversing operation,   $s_1^\mathrm{eff}\cup s_1^\mathrm{eff}$ is non-trivial.  So the $p+ip$ states from the orientation reversing subgroup are obstructed for all double space groups, except for the No.81 and 82 whose  point group is $S_4$. We note that $S_4$ now is mapped to internal group $\Z_4^T$ where $T$ is the generator of the group and then the $s_1^\mathrm{eff}\cup s_1^\mathrm{eff}$ is a trivial cocycle in $H^2(S_4, \Z_2)$.   

On the other hand, there is another origin of $p+ip$ layer that is related to the translation symmetry of the space group. According to Sec.\ref{sec:decoupling_pplusip}, on the one hand, we can also directly calculate the $H^1(G_{SG}, \Z)$ for all space groups using the algorithm in Ref.\cite{QiGithub} and see whether there is torsion free part. One the other hand, we can also check the group structure of the space groups to see whether there is torsion free part. Namely, for a given translation group $\Z$, if all elements of the point group $Q$ that act non-trivially on this translation group satisfy that (1) group conjugation will map the translation element to it inverse  and (2) they are orientation-reversing,  then this translation group $\Z$ would contribute a $\Z$ classification.  We can check all three translations $\Z$ separately to obtain the final classification from the total translation group.  We find that among the seven crystal systems, except the cubic and orthorhombic systems, the other five (triclinic, monoclinic, tetragonal, trigonal, and hexagonal systems) can support the weak SPT from $p+ip$ layer. More explicitly, the two space groups (No.1 and No.2) corresponding to triclinic systems can support $\Z^3$ classification, while the others that only support $\Z$ classification   are listed as follows. 
        \begin{align}
    &\mathrm{Monoclinic: } 
    \,\, C_2\, (\mathrm{No.}3-5),C_s\,(\mathrm{No.}6-9),\,C_{2h}\, (\mathrm{No.}6-9)\nonumber\\
    &\mathrm{Tetragonal :} \,\,C_4\, (\mathrm{No.}75-80),S_4\, (\mathrm{No.}81-82), \,  \nonumber\\
    &\qquad\qquad\,\,\,\, \quad C_{4h} \, (\mathrm{No.83-88}) \nonumber\\
    &\mathrm{Trigonal: } \,\, C_3\, (\mathrm{No.}143-146),S_6\, (\mathrm{No.}147-148) \nonumber\\
    &\mathrm{Hexagonal: }\,\, C_6\, (\mathrm{No.}168-173),C_{3h}\, (\mathrm{No.}174), \,
    \nonumber\\
    &\qquad\,\, \qquad \quad C_{6h} \, (\mathrm{No.175-176}) \nonumber
\end{align}
The simplest example is the  No.2 space group, whose point group is the inversion group $C_i$ generated by $I$. The three translation generators $T_i$ ($i=x,y,z$) all act non-trivially on the inversion $I$: $IT_iI^{-1}=T_i^{-1}$. So, three translation symmetries can lead to the classification $(\Z)^3$ corresponding to the weak SPT by stacking $p+ip$ state in three different translation directions. Another example is the No.78 space group, whose point group is the four fold rotational symmetry generated by $C_4$.  For the translation direction parallel to the rotation axis, the rotation acts trivially on the translation, so stacking $p+ip$ states along this direction will yield a $\Z$ classification of weak SPT. For the other two directions of translation, the rotation $C_4$ acts non-trivially on the two translation operations, i.e., it exchanges the two translations, so they do not contribute any weak SPT. Therefore, the weak SPT from stacking $p+ip$ state for the No.78 space group is $\Z$ classification.  

\subsubsection{Kitaev chain, complex fermion and bosonic SPT layers}
Now we move to calculate the other layers of decoration, from KC, complex fermion layer to the bosonic SPT layers.  With the classification formulas discussed in Sec.\ref{sec:antiuni_SPT}, we can now calculate them in principle. In practice, we apply the algorithm developed in Ref.\cite{Ouyang2021}.  The final results are listed in Table \ref{tab:TCSc} and \ref{tab:TCSc2}. 

The general scheme is as follows: (1) For KC layer, we first calculate the $H^2(G_{\mathrm{SG}}, \Z_2)$ for all possible $n_2$, then  for each $n_2$, check whether it is $O_4$ obstructed. If it is $O_4$ obstructed, we rule out it in the classification. If it is not, then we can solve the solution for $n_3$, which in general is a torsor  over $H^3(G_{\mathrm{SG}}, \Z_2)$. If there is one $n_3$ solution for the given $n_2$ such that  $O_5$ obstruction vanishes, then the given $n_2$  is obstruction-free. Furthermore, we check whether it is trivialized by checking whether this $n_2$ and $\Gamma_2=\omega_2$ are in the same class in $H^2(G_{\mathrm{SG}}, \Z_2)$. If they are not in the same class, then $n_2$ is trivialization-free. 
The obstruction-free and trivialization-free $n_2$ will result in the final classification from KC layer of decoration, denoted as $\mathcal{C}_{2}$. (2) For the complex fermion layer, we first calculate the $H^3(G_{\mathrm{SG}},\Z_2)$ for all possible $n_3$. For each  $n_3$, if it is $O_5$ obstructed, we rule it out in classification. If it is not obstructed, we go ahead to check whether it is trivialized by whether this $n_3$ is in the same class as any of $\Gamma_3[n_1, n_0]$ for all $n_1, n_0$. The obstruction-free and trivialization-free $n_3$ will contribute to the classification  from complex fermion layer, denoted as $\mathcal{C}_3$. (3) For the bosonic layer, we first calculate the $H^4(G_{\mathrm{SG}}, U(1))$ for all possible $\nu_4$, we further check which can be trivialized. The trivialization-free $\nu_4$ will contribute to the classification from the bosonic layer, denoted as $\mathcal{C}_4$. Together with the classification from $p+ip$ layer $\mathcal{C}_1$, we obtain the final classification for the crystalline topological phases, which are collected in the corresponding column, named as $E_\mathbf{2D}$,$E_\mathbf{1D}$,$E_\mathbf{0D}$ and $E_\mathbf{b}$, and the total number of these phases in the column named Total is given by denoted as $|\mathcal{C}_1|\times |\mathcal{C}_2|\times  |\mathcal{C}_3|\times |\mathcal{C}_4|$ in Table.\ref{tab:TCSc} and \ref{tab:TCSc2} .

Let's illustrate how the calculation is performed by the No.188 group. The data for KC layer  should first be a 2-cocycle in $H^2(G_{188}, \Z_2)=\Z_2^3$. However, only one $\Z_2$ of them are $O_4$ obstruction free and further $O_5$ obstruction free. As the $\Gamma_2=\omega_2\cup n_0$ is trivial, there is no trivialization for the KC decoration. So the classification from the KC layer is $\Z_2^2$. On the other hand, the complex fermion decoration is characterized by the 3-cocycles in  $H^3(G_{188}, \Z_2)=\Z_2^4$, but only two of then are obstruction free and trivialization free. So the classification from complex fermion layer is $\Z_2^2$.  Finally, the bosonic layer is characterized by $H^4(G_{188}, U(1))=\Z_2^3$, which corresponds to the classification of bosonic topological crystalline phases protected by No.188 space groups, among which, however, one $\Z_2$ will be trivialized. So the classification from bosonic layer is $\Z_2^2$. We note that  the classification  from
$p+ip$ layer is  trivial.   Therefore, the total classification of No.188 space group is classification by $ \Z_1, \Z_2, \Z_2^2, \Z_2^2$ and the total number is 32. An interesting point is to make a comparison to the corresponding point group $D_{3h}$, which only has trivial classification \cite{Zhang25_PRX_point}. Therefore, all the non-trivial classifications need the translation symmetry to protect. We see that the presence of translation symmetry greatly enrich the phases of matter landscape.

\subsection{Examples} 
With the above classification, a natural interesting question is how to realize these topological quantum phases, either in ideal construction or in  experimental realizations. The generic aspects of this question are left for future study. However, we can illustrate them for some simple examples. Below we first consider how to construct all states for the simplest space group (No.1) which has only three translation symmetries. Furthermore, we discuss some examples of space-group topological crystalline superconductors that are likely related to  experiments

\begin{figure}
    \centering
    \includegraphics[width=1\linewidth]{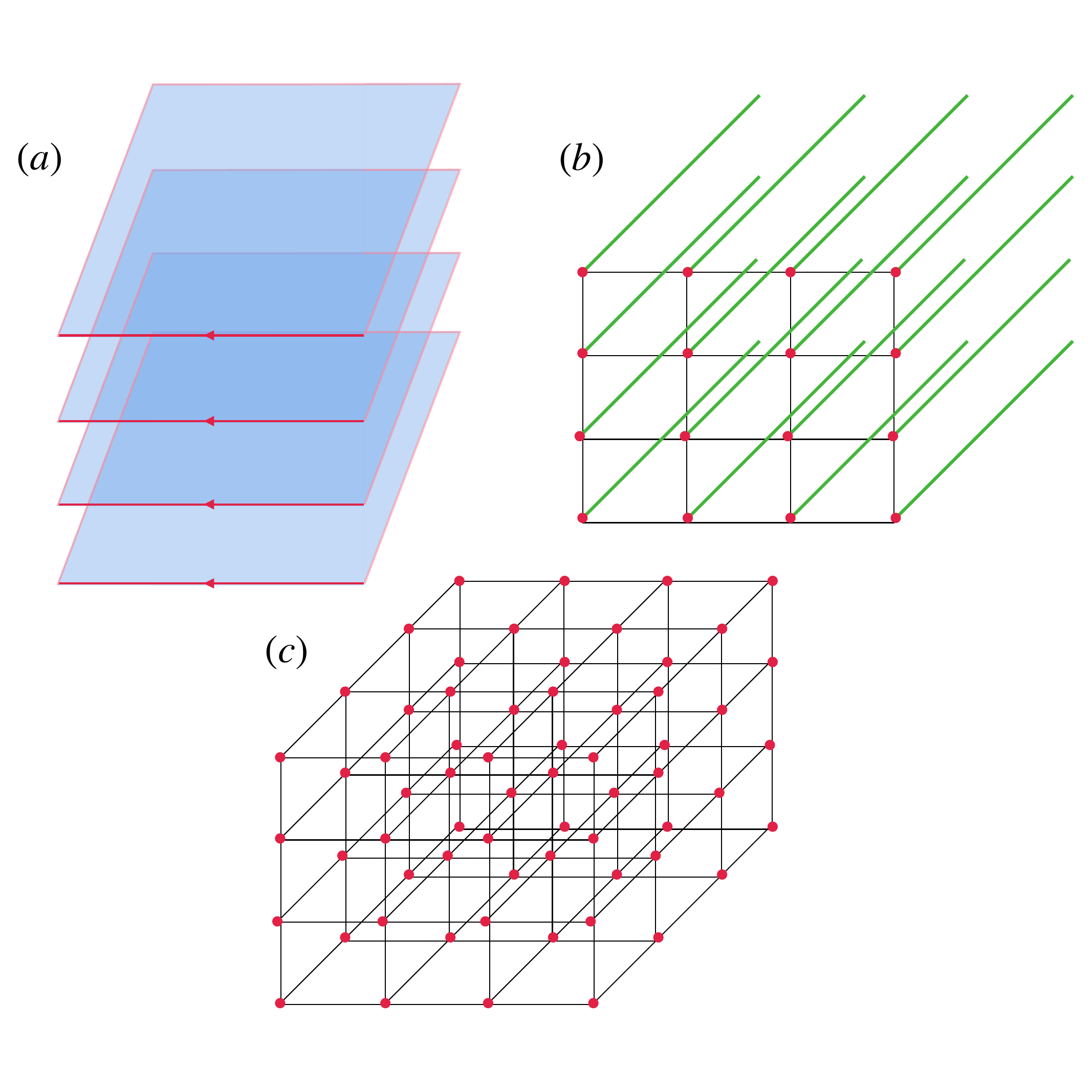}
    \caption{The construction of the topological crystalline superconductors protected by No.1 space group. (a) Stacking  $p+ip$ state along $z$ directions. Along the other two $y,z$ direction give another two root states. The three root states generate a $\Z^3$ classification (b) Stacking the Kitaev chain parallel to the $x$  direction and perpendicular to the $yz$ plane. Parallel to the other two directions gives another two root states. The three root states genereate $\Z_2^3$ classification. (c) Stacking complex fermion for each unit cell, giving rise to a $\Z_2$ classification.  } 
    \label{fig:No1}
\end{figure}

The classification of the No.1 space group is $\Z^3, \Z_2^3, \Z_2, \Z_1$ for the $p+ip$, KC, complex fermion and bosonic SPT layers, respectively. The first $\Z^3$ can be realized by stacking $\nu \in \Z$ copies of $p+ip$ states along three translation direction, as in Fig.\ref{fig:No1}(a). The states exhibit stable chiral boundary modes with ideal prototype as a bunch of 1+1D chiral Majorana fermions. The second $\Z_2^3$ corresponds to stacking Kitaev chain perpendicular to $xy$, $xz$, and $yz$ plane, as in Fig.\ref{fig:No1}(b). These states have single Majorana mode in each unit cell, which ultimately has some non-trivial feature on the boundary, for example at least two-fold degeneracy on the ground state \cite{Tarun16}, in analogue to Kramers and Lieb-Schultz-Mattis theorems. The last $\Z_2$ corresponds to decorating complex fermion to unit cell, as in Fig.\ref{fig:No1}(c). This state is a product state but cannot be smoothly deformed to the vacuum state (with no complex fermion) while preserving the translation symmetries.

Next we turn to discuss some possible insights on the experiments.
Some  3D unconventional superconductors discovered in experiments have a strong correlated nature when the  $d$ or $f$ orbital electrons play an important role in the superconductivity. For example, the $d$ orbital electrons in the cuprates lead to the large Mott gaps while in the heavy fermion superconductors, the $f$ orbitals results in the ``heavy" quasi-particles. Intensive efforts make people to widely believe that their mechanism is not dominated by the electron-phonon interaction but the electron-electron interaction,  however, the nature, especially the topological nature of such unconventional superconductors is generically hard to identify due to the strong correlation. For investigating their topological nature, the odd-parity symmetric pairing is usually  viewed as an important diagnosis, which, however, only apply for the centrosymmetric systems. For non-centrosymmetric systems, the parity symmetry is not even a well defined symmetry so that there may be coexistence of even and odd parity pairings. So a comprehensive understanding of interacting topological superconductivity due to the protection of  the crystalline symmetries are very crucial clues for the investigation of their topological nature. 

In particular, here we mainly focus the heavy fermion superconductors and see what interacting topology can be maintained by their crystalline symmetries according to our classification. (For a very recent brief review on the history on heavy fermion superconductors, refer to the Ref.~\cite{RMP_heavy_SC_Yuan} and for another reviews on various examples in Ref.\cite{RMP_heavy_09}) We summarize a few their interesting features. (1) While their topology may be fermionic or bosonic, most of them are bosonic. Namely, among the various examples, their crystalline symmetry protected topology are mostly from the bosonic SPT layer while a few from the complex fermion layer. (2) In the  well-accepted heavy fermion superconducting systems, there is no crystalline symmetry protected topology from the Kitaev chain layer and the $p+ip$ layer. Only in one system that is under debate, there are such topology from the Kitaev chain layer and the $p+ip$ layer. 
More specifically, below we discuss several situations.


\begin{enumerate}

\item  Both fermionic and bosonic topology. 

In this situation, the fermionic topology is only from the complex fermion layer. We specifically focus on one example---$Fd\bar 3 m$ (No.227) which is relevant for the $\mathrm{CeRu}_2$  that is believed to be fully gapped superconductors. According to our results in Table \ref{tab:TCSc2}, this crystalline symmetry can protected eight different classes of interacting topological superconductors---four of them are fermionic contributed by the complex fermion layer and four of them are bosonic SPT. 

We note that there are many space groups for other heavy fermion superconductors that are belong to this situations, namely they can protect both fermionic and bosonic topology, but they are now not considered as fully gapped superconductors.

\item  Pure bosonic topology. 

There are many space groups that can only protect topological superconductor from purely bosonic SPT layer. Among various heavy fermion superconductors, we focus on the space group 
$I4/mmm$ (No.139), relevant  the first discovery of heavy fermion superconductor---$\mathrm{CeCu}_2\mathrm{Si}_2$ which now is considered as fully gapped superconductors \cite{RMP_heavy_SC_Yuan}. The classification is $\Z_2^{10}$, all of which are from the bosonic SPT layer. This symmetry is also relevant for many other heavy fermion superconductors (they are now not considered as fully gapped superconductors), including $\mathrm{CeX}_2\mathrm{Si}_2$ where $\mathrm{X}=\mathrm{Pd}$ and $\mathrm{Rh}$, $\mathrm{CeY}_2\mathrm{Ge}_2$, where $\mathrm{Y}=\mathrm{Cu}$ and $\mathrm{Ni}$, $\mathrm{ZPd}_5\mathrm{Al}_2$ where $\mathrm{Z}=\mathrm{Ce, Np}$, $\mathrm{URu}_2\mathrm{Si}_2$,and also $\mathrm{Sr}_2\mathrm{RuO}_4$ \cite{RMP_SrRuO}.

We remark that such purely bosonic topology in these systems inspires effort on studying the topological diagnosis for this symmetry protected topology, going beyond the odd-parity pairing which might not capture the bosonic nature of the symmetry protected topology.

\item Purely fermionic topology. 

By survey on almost fifty various examples mainly based on the reviews \cite{RMP_heavy_09,RMP_heavy_SC_Yuan}, we do not find any fully gappd superconductor examples for this situation. We find only two examples that may hold node in their gap structure whose space group can protect only fermionic  interacting crystalline topology---one example has  topology only from the complex fermion layer while the other can have both the three fermionic layers, including $p+ip$, Kitaev chain and complex fermions.  
     
The first one is $Pnma$, No.\,62, relevant for some interesting uranium-based heavy-fermion superconductors $\mathrm{URhGe}$ and $\mathrm{UCoGe}$, which exhibit intriguing coexistence of superconductivity and ferromagnetism, making them  promising for realizing a spin-triplet state.  For this space group,
our classification predicts a classification of $\mathbb{Z}_2^2$ contributed by the complex fermion layer. 

The second one is $P2_1$, No.\,4, which is realized in one particularly heavy fermion system---$\mathrm{UIr}$.  This systems emerges superconductivity only in high pressure and very low temperature, which still lacks complete signature for the superconductivity. Nevertheless, 
our classification predicts a highly non-trivial classification protected by this space group: $\mathbb Z$ from the $p+ip$ layer, $\mathbb{Z}_2^3$ from the KC layer, and $\mathbb Z_2$ from the complex fermion layer. It is important to note that this classification contains all non-trivial FSPT states from all fermionic layers, while no non-trivial bosonic SPT state. 

\end{enumerate}
We make a few remarks on the superconducting gap aspects. First, our classification is for fully gapped topological superconductors protected by their crystalline symmetries. Second, we note that previously, the example $\mathrm{CeCu}_2\mathrm{Si}_2$ is considered to have node in it gap structure, but due to the recent development of experimental technique, now researchers find it to be a fully gapped superconductors \cite{RMP_heavy_SC_Yuan}. Finally, we expect some manipulations (pressure, external magnetic field, strain and so on) on  some nodal heavy-fermion superconductors  may make them evolve in fully gapped superconductor for which  our classification theory comes into application.

\section{Summary and Discussion}

We developed a complete  framework for classifying three-dimensional interacting fermionic SPT protected by generic discrete symmetries and adapt to classify topological crystalline superconductors. Our approach marries two pillars:

\begin{enumerate}
\item The domain-wall decoration framework in $3{+}1$D adapted to generic internal discrete symmetries, involving the generic anti-unitary and infinite-order elements and considering 
 all decorations by $p{+}ip$ layers ($n_1$), Kitaev chains ($n_2$), complex fermions ($n_3$), and bosonic SPT $U(1)$ phase ($\nu_4$).


\item The fermionic crystalline equivalence principle (CEP) that bridges the classification of topological  crystalline superconductors to the FSPT protected by an effective onsite fermionic symmetry with precisely shifted anti-unitary and central-extension data. 
\end{enumerate}

 We found two classification criteria for $p+ip$ decorations: (1) when orientation-reversing symmetries allow $p{+}ip$ decorations,  and (2) when a  normal infinite $\Z$ subgroup contributes to valid $\Z$ classification of weak $p{+}ip$ stacking that is fixed solely by the other elements' action on this $\Z$ group. The new $O_5$ structure identifies $\mathbb{Z}_8$-like fermionic obstructions that are intrinsic  to  generic anti-unitary symmetries. From these ingredients we obtained the full  classification of the interacting electronic topological crystalline superconductors protected by all 230 crystallographic space groups.

\paragraph{Conceptual advances.}
Our results close a persistent gap between abstract cobordism classifications and constructive, circuit-level definitions of 3D FSPTs with generic anti-unitary symmetris, through the FSLU-based fixed-point constructions. The clean decoupling classification of the $p{+}ip$ layer from anti-unitary structure of symmetry finally overcomes the obstacle in the classification due to the lack of the fixed point wavefunction of $p+ip$ state. This conceptual advance paves the way for classifying and construction of FSPT phases beyond the space group we focus on here.

\paragraph{Anti-unitary structure and new obstructions.}
The emergence of new  $O_5^\gamma$ underscores that anti-unitary symmetries qualitatively change the obstruction landscape by properly enforcing multiple inserted Majorana  fermions in the FSLU transformation ($F$ move), which will guide the  way for higher dimensional FSPT. The quantum-averaged Majorana-dimer phases are a physical manifestation of  consistency of the FSLU transformation, which  naturally produce $e^{i\pi/4}$ factors, likely unavoidable in 3D anti-unitary FSPT  (cannot be removed by coboundary transformation). Upon gauging the fermion parity symmetry, the $O_5$ will relate to the 't Hooft anomaly formula for bosonic symmetry and the new $O_5$ structure will lead to new landscape of quantum anomaly in 3+1D.

\paragraph{Relation to other frameworks.}
Our DWD framework for organization of decoration data, obstructions, and trivializations matches the layers and differentials in generalized cohomology/AH spectral sequence. The agreement provides dual validation: (i) FSLU-constructible fixed points and mirror SPT construction realize all unobstructed classes predicted by the spectral sequence for discrete symmetries, and (ii) the obstruction formulas $O_3,O_4,O_5$ correspond to higher obstructions/anomalies in the mathematical classification.

\paragraph{Physical implications.}
The classification guides the real-space construction of the crystalline phases and  search for robust boundary phenomena—hinge gapless modes, Majorana corner modes, and symmetry-enforced surface topological orders—that survive strong interactions and disorder. The weak $p{+}ip$ indices derived from translation actions indicate which crystal families can host stackable chiral-layer responses, which is consistent with orientation-reversing symmetries. 
The fixed point wavefunction  will be utilized to realize and manipulate them in programmable quantum processors and simulators
\paragraph{Outlook.}
The combination of FSLU-fixed points, fermionic CEP, and AHSS provides a scalable pathway to classifications and explicit models across broad symmetry landscapes. We anticipate straightforward extensions to magnetic and spin space groups, systematic libraries of commuting-projector Hamiltonians for  FSPTs, and quantitative predictions for boundary transport and anomaly-matching observables. Beyond classification, these tools offer a bridge from rigorous interacting topology to materials and programmable quantum simulators. Our resolved layer of obstruction calculation can also directly apply to the study of classification of average SPT \cite{Ma_2023,Ma_2025}.

\paragraph{Open questions.}
While exhaustive at the level of SPT classifications for spinful electrons with space-group symmetries, several avenues remain:
\begin{itemize}
\item Incorporating continuous onsite symmetries (e.g., $U(1)$ charge), which can account for the space-group topological crystalline insulators. 
\item Stacking group structure, which tells how two decoration data add into another ones, namely what two FSPT stacking together is equivalent to.
\item Generalizing the quantum-averaging formalism to higher dimensional FSPT (4+1D) which can help understanding the generalized symmetries and quantum anomaly in 3+1D.

\item Field theory understanding. The states with exotic decoration, such as KC decoration, are not clear how to understand in  continuous topological field theory framework.

\item  Simple Hamiltonians (like the Kitaev honeycomb model) realizing these interacting FSPT phases. 
\end{itemize}

\section{Acknowledgment} We thank Adrian Po, Chenjie Wang and Weicheng Ye for helpful discussion. This work is supported by funding from Hong Kong’s Research Grants Council (GRF No. 14307621, GRF No. 14308223, RFS2324-4S02 and CRF C7015-24G). S.Q.N. is supported by a CRF from the Research Grants Council of the Hong Kong (No. C7037-22GF). Q.-R. W is supported by National Natural Science Foundation of China (Grant No. 12274250). Y.Q. is supported by National Key R\&D Program of China (Grant No. 2022YFA1403402), by National Natural Science Foundation of China (Grant No. 12174068) and by the Science and Technology Commission of Shanghai Municipality (Grant No. 23JC1400600).

\appendix

\section{Group Cohomology} \label{ap:group cohomology}

In this appendix, we will briefly introduce the mostly used mathematical concepts and formulas in this paper.
We emphasize that our construction is physical, which does not require any prerequisites for any mathematics.

\subsection{$G$-module}

Given a group $G$ and an Abelian group $M$, the $G$-module is defined by a $G$ action on $M$, denoted as $g\triangleright a\in M$ for $g\in G$ and $a,b\in M$, such that the $G$ action is compatible with the group multiplication of $M$  
\begin{equation}
g\triangleright ( ab) =( g\triangleright a)( g\triangleright b) .\label{eq.ap1.1}
\end{equation}
In this paper, we always consider $M$ to be the $U( 1)$ phase. If $G$ contains only unitary operation, then we set the $G$ action to be trivial $g\triangleright a=a$; while if $G$ contains anti-unitary operation such as time-reversal transformation, we define the $G$ action as $g\triangleright a=a^{s_{1}( g)}$, where $s_{1}( g) =-1$ if group element $g$ is anti-unitary, otherwise $s_{1}( g) =1$.

\subsection{Inhomogeneous cochain}

An $n$-cochain $\omega _{n}( g_{1} ,g_{2} ,\cdots ,g_{n})$ is a function of $n$ group elements with values in the $G$-module $M$. In other words, an $n$-cochain is a map $\omega _{n} :G^{n} \rightarrow M$ from the Cartesian product $G\times G\times \cdots \times G$ to $M$. The set of all $n$-cochains is denoted as $\mathcal{C}^{n}( G,M)$.

The coboundary homomorphism $\mathrm{d}_{n} :\mathcal{C}^{n}( G,M) \rightarrow \mathcal{C}^{n+1}( G,M)$, for $n\in \mathbb{Z}^{+}$, is a map from the set of all $n$-cochains $\mathcal{C}^{n}( G,M)$ to the set of all $( n+1)$-cochains $\mathcal{C}^{n+1}( G,M)$. Given an $n$-cochain $\omega _{n}( g_{1} ,g_{2} ,\cdots ,g_{n})$, we apply the coboundary homomorphism $\mathrm{d}_{n}$ to obtain an $( n+1)$-cochain $(\mathrm{d}_{n} \omega _{n})( g_{1} ,g_{2} ,\cdots ,g_{n} ,g_{n+1})$ 
\begin{equation}
\begin{array}{ l }
(\mathrm{d}_{n} \omega _{n})( g_{1} ,\cdots ,g_{n+1})\\
=[ g_{1}\triangleright \omega _{n}( g_{2} ,\cdots ,g_{n+1})] \omega _{n}^{( -1)^{n+1}}( g_{1} ,\cdots ,g_{n}) \\
\times \prod _{i=1}^{n} \omega _{n}^{( -1)^{i}}( g_{1} ,\cdots ,g_{i-1} ,g_{i} g_{i+1} ,g_{i+2} ,\cdots ,g_{n+1})
\end{array} .\label{eq.ap1.2}
\end{equation}
In some cases, we may omit the subscript $n$ of $\mathrm{d}_{n}$ if there is no ambiguity. Using \eqref{eq.ap1.2}, we can explicitly verify that the composition map of $\mathrm{d}_{n}$ and $\mathrm{d}_{n+1}$ satisfies 
\begin{equation}
\mathrm{d}_{n+1} \circ \mathrm{d}_{n} =0, \label{eq.ap1.3}
\end{equation}
where $0$ is the identity element of $M$. This implies that the image $\mathcal{B}^{n}$ of map $\mathrm{d}_{n-1}$ 
\begin{equation}
\begin{array}{ l }
\mathcal{B}^{n}( G,M) =\\
\left\{\omega _{n} \in \mathcal{C}^{n}( G,M) |\omega _{n} =\mathrm{d}_{n-1} \omega _{n-1} |\omega _{n-1} \in \mathcal{C}^{n-1}( G,M)\right\}
\end{array} \label{eq.ap1.4}
\end{equation}
is contained in the kernel $\mathcal{Z}^{n}$ of map $\mathrm{d}_{n}$
\begin{equation}
\mathcal{Z}^{n}( G,M) =\left\{\omega _{n} \in \mathcal{C}^{n}( G,M) |\mathrm{d}_{n} \omega _{n} =0\right\} . \label{eq.ap1.5}
\end{equation}
This allows us to define the cohomology group $\mathcal{H}^{n}( G,M)$ as the quotient 
\begin{equation}
\mathcal{H}^{n}( G,M) =\mathcal{Z}^{n}( G,M) /\mathcal{B}^{n}( G,M). \label{eq.ap1.6}
\end{equation}
An element in $\mathcal{B}^{n}( G,M)$ is called an $n$-coboundary, while an element in $\mathcal{Z}^{n}( G,M)$ is called an $n$-cocycle.

\subsection{Homogeneous cochain}

What we introduced above is referred to as an inhomogeneous cochain $\omega _{n} :G^{n} \rightarrow M$. This inhomogeneous cochain can be transformed into a homogeneous cochain $\nu _{n} :G^{n+1} \rightarrow M$ by the correspondence 
\begin{equation}
\begin{array}{ l }
\omega _{n}( g_{1} ,g_{2} ,\cdots ,g_{n})\\
=\nu _{n}( 1,g_{1} ,g_{1} g_{2} ,g_{1} g_{2} g_{3} ,\cdots ,g_{1} g_{2} \cdots g_{n})\\
\equiv \nu _{n}( 1,\tilde{g}_{1} ,\tilde{g}_{2} ,\cdots ,\tilde{g}_{n})
\end{array} \label{eq.ap1.7}
\end{equation}
with $\tilde{g}_{i} =g_{1} g_{2} \cdots g_{i}$. The $G$-action on the homogeneous cochain satisfies 
\begin{equation}
g\triangleright \nu _{n}( g_{0} ,g_{1} ,\cdots ,g_{n}) =\nu _{n}( gg_{0} ,gg_{1} ,\cdots ,gg_{n}) . \label{eq.ap1.8}
\end{equation}
Using \eqref{eq.ap1.2}, we can check that the coboundary homomorphism $\mathrm{d}_{n} :\nu _{n} \mapsto \nu _{n+1}$ can be reduced to a simplified form 
\begin{equation}
\begin{array}{ l }
(\mathrm{d}_{n} \nu _{n})( g_{0} ,g_{1} ,\cdots ,g_{n+1})\\
=\prod\nolimits _{i=0}^{n+1} \nu _{n}^{( -1)^{i}}( g_{0} ,\cdots ,\hat{g}_{i} ,\cdots ,g_{n+1})
\end{array} , \label{eq.ap1.9}
\end{equation}
where $\hat{g}_{i}$ denotes omit $g_{i}$. The inhomogeneous $n$-cochain $\omega _{n}( g_{1} ,\cdots ,g_{n})$ and the corresponding homogeneous $n$-cochain $\nu _{n}( g_{0} ,g_{1} ,\cdots ,g_{n})$ can be easily distinguished: the former has $n$ input group element, while the latter has $( n+1)$. Therefore, in the main text, we may not always explicitly indicate whether a cochain is inhomogeneous or homogeneous, but the reader should be able to distinguish them easily.

\subsection{Steenrod's higher cup product}

The Steenrod's higher cup product is defined in Ref.~\cite{Steen1974}. Since the definition of higher cup is more involved, here we only list the cup-0 and cup-1, with the assumption of trivial $G$-action on the coefficient $M$. Given two cochain $f_{m} \in \mathcal{C}^{m}( G,M)$ and $h_{n} \in \mathcal{C}^{n}( G,M)$, the cup product (cup-0) is a map $\cup \ :\mathcal{C}^{m}( G,M) \times \mathcal{C}^{n}( G,M) \rightarrow \mathcal{C}^{m+n}( G,M)$ that is defined as 
\begin{align}
( f_{m} \cup h_{n})( g_{1} ,\cdots ,g_{m+n}) =f_{m}( g_{1} ,\cdots ,g_{m})\nonumber\\
 h_{n}( g_{m+1} ,\cdots ,g_{m+n}) . \label{eq.ap1.10}
\end{align}
The cup-$1$ product is a map $\cup_{1} :\mathcal{C}^{m}( G,M) \times \mathcal{C}^{n}( G,M) \rightarrow \mathcal{C}^{m+n-1}( G,M)$ that is defined as 
\begin{equation}
\begin{array}{ l }
( f_{m} \cup _{1} h_{n})( g_{1} ,\cdots ,g_{m+n-1})\\
=\sum _{j=0}^{m-1}( -1)^{( m-j)( n+1)} f_{m}( g_{1} ,\cdots ,g_{j} ,g_{j+n} ,\cdots ,g_{m+n-1})\\
\times h_{n}( g_{j} ,\cdots ,g_{j+n-1})
\end{array} . \label{eq.ap1.11}
\end{equation}
The higher cup products satisfy the following Leibniz's rule 
\begin{equation}
\begin{array}{ l }
\mathrm{d}( f_{m} \cup _{i} h_{n}) =\mathrm{d} f_{m} \cup _{i} h_{n} +( -1)^{m} f_{m} \cup _{i}\mathrm{d} h_{n}\\
+( -1)^{m+n-i} f_{m} \cup _{i-1} h_{n} +( -1)^{mn+m+n} h_{n} \cup _{i-1} f_{m}
\end{array} . \label{eq.ap1.12}
\end{equation}

\section{Spectral sequence: a practical introduction}
\label{sec:ss}
Here we give a quick and operational definition of the (cohomology) spectral sequence based on the Lyndon-Hochschild-Serre (LHS) spectral sequence, which is a special case of the Atiyah-Hirzebruch (AH) spectral sequence.  We will also discuss the generalization to the AH  spectral sequence for FSPT.

From a practical perspective, a spectral sequence consists of a assembly of Abelian groups $E_r^{p,q}$ with non-negative $r,p$ and $q$. For a fixed $r$,  the collection of all $E_r^{p,q}$ are called the $E_r$  {page}. Within each page, there exist  group homomorphisms labeled $p,q$,  called differentials  $d_r^{p,q}$, which map  $E_r^{p,q}$  to $E_r^{p+r, q-r+1}$: 
\begin{align}
d_r^{p,q}: E_r^{p, q}\rightarrow E_r^{p+r,q-r+1}
\end{align}
which  is required to  satisfy the condition that two consecutive differentials should equal to trivial maps (mapping all elements in trivial element), namely 
\begin{align}
d_r^{p+r,q-r+1}\cdot d_r^{p,q}=0.
\label{eq:d2=0}
\end{align} 
Sometimes, for simplicity, we omit the superscript of $d_r^{p,q}$ as $d_r$ without causing confusion, so that the Eq.~(\ref{eq:d2=0}) can be written as $$d_r^2=0.$$ 
Therefore, the image of $d_r^{p-r, q+r-1}$ must be in the kernel of the map $d_r^{p,q}$.  The kernel of $d_r^{p,q}$ quotient out the  image of $d_r^{p-r, q+r-1}$ will arrive at the elements of the next page, namely,
\begin{align}
	E_{r+1}^{p,q}\simeq \frac{\mathrm{ker}(d_r^{p,q})}{\mathrm{img} (d_r^{p-r, q+r-1})}.
\label{eqn:iso_ss}
\end{align}
The family of $\{ E_r^{p,q}\}$ can be constructed iteratively, namely, knowing the $E_r$ page (i.e., $E_r^{p,q}$ for all $p,q$ with  given $r$), one can construct the next page, i.e., $E_{r+1}$-page via the isomorphism (\ref{eqn:iso_ss}).  

Let us take the LHS spectral sequence as an example.  Given a short exact sequence of groups
\begin{equation}
	1\rightarrow A\rightarrow H\rightarrow G\rightarrow 1,
	\label{}
\end{equation}
the LHS spectral sequence computes the group cohomology of $H$ in terms of the group cohomology of the normal subgroup $A$ and the quotient group $G=H/A$. To introduce some notation, we denote the set of $i$-cochains, $i$-cocycles, and $i$-coboundaries  of an group $K$  with  coefficients in $M$ by ${\mathcal C}^i[ K, M]$, ${\mathcal Z}^i[ K, M]$, and ${\mathcal B}^i[ K, M]$ repsectively. 

\begin{figure*}[t]
   \centering
   \includegraphics[width=\textwidth]{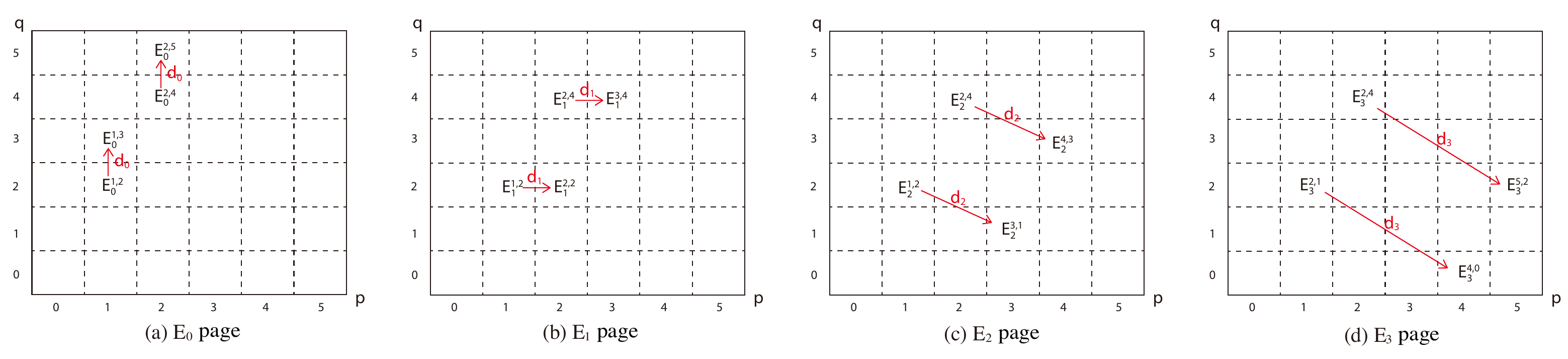} 
   \caption{Examples for illustrating  how $d_n$ differentials map within $E_n$ page.}
   \label{fig:Er_differential}
\end{figure*}

We begin with defining $E_0$-page of the LHS spectral sequence, which is just the group of  cochains  ${\mathcal C}^p[G, {\mathcal C}^q[A, M]]$. The $d_0$ differential map sends a cochain in $E_0^{p,q}={\mathcal C}^p[G, {\mathcal C}^q[A, M]]$ to a cochain in $E_0^{p,q+1}={\mathcal C}^p[G, {\mathcal C}^{q+1}[A, M]]$. See Fig.\ref{fig:Er_differential}(a) for illustration. The kernel of $d_0^{p,q}$ is just ${\mathcal C}^p[G, {\mathcal Z}^q[A, M]]$, while the  image of $d_0^{p,q-1}$ is just ${\mathcal C}^p[G, \mathcal B^q[A, M]]$, hence the $E_1$ page is given by
\begin{align}
	E_1^{p,q}=\frac{\mathrm{ker}(d_0^{p,q})}{\mathrm{img}(d_0^{p,q-1})}={\mathcal C}^p[G, H^q[A, M]].
    \label{eqn:sequence_0}
\end{align}
If we denote the elements of $E_0^{p,q}$ by $w_0^{p,q}$,  the more precise meaning of (\ref{eqn:sequence_0}) is that 
the elements of $E_1^{p,q}$ are equivalence classes of elements in $E_0^{p,q}$ that satisfy the condition $d_0^{p,q}w_0^{p,q}=0$, or more compactly, 
\begin{align}
d_0 w_0=0,
\end{align}
with the equivalence relation given by $w_0\sim w_0d_0c_{-1}$, where $c_{-1}\in {\mathcal C}^p[G, {\mathcal C}^{q-1}[A,M]]$.

 Now consider the $E_1$ page, given by $E_1^{p,q}={\mathcal C}^p[G, H^q[A,M]]$. The differential $d_1$ maps a cohomology class in $E_1^{p,q}$ to a class in $E_1^{p+1,q}$. The kernel of $d_1^{p,q}$ consists of cocycles ${\mathcal Z}^{p}[G, H^q[A, M]]$, and the image of $d_1^{p-1, q}$ consists of coboundaries $\mathcal B^p[G, H^q[A,M]]$. We illustrate $d_1$ in Fig.\ref{fig:Er_differential}(b). The $E_2$ page is defined as the homology of the $E_1$ page:
\begin{align}
	E_2^{p,q}=\frac{\mathrm{ker}(d_1^{p,q})}{\mathrm{img}(d_1^{p-1,q})}=H^p[G, H^q[A, M]].
	\label{eq:E2_page_LHS}
\end{align}

To understand this operationally in terms of explicit cochains, we must distinguish between the spectral sequence differentials $d_r$ and the group coboundary operators. Let $\delta_1$ denote the usual coboundary operator on $G$ (horizontal) and $\delta_0$ denote the usual coboundary operator on $A$ (vertical). In fact $d_0$ and $\delta_0$ are identical.

A representative element of the $E_2$ page is a cochain $w_0^{p,q} \in \mathcal{C}^p[G, \mathcal{C}^q[A,M]]$ that satisfies a set of ``descent equations.''
First, to represent an element in $E_1$, it must be closed under the vertical differential:
\begin{align}
\delta_0 w_0^{p,q} = 0.
\end{align}
This implies $w_0^{p,q}$ defines a class $[w_0^{p,q}] \in E_1$.
Second, to survive to $E_2$, the image of this class under $d_1$ must vanish. On the cochain level, this means that applying the horizontal differential $\delta$ to $w_0^{p,q}$ results in a term that is trivial in the cohomology of $A$. In other words, $\delta_1 w_0^{p,q}$ must be a vertical coboundary:
\begin{align}
\delta_1 w_0^{p,q} = - \delta_0 w_1^{p+1, q-1},
\end{align}
for some cochain $w_1^{p+1, q-1}$.
Thus, the condition for an element to survive to the $E_2$ page is the existence of a pair $(w_0, w_1)$ satisfying:
\begin{align}
&d_0w_0:=\delta_0 w_0 = 0, \\
&d_1 w_0:=\delta_1 w_0 + \delta_0 w_1 = 0. \label{eq:descent_E2}
\end{align}
Here, Eq.~\eqref{eq:descent_E2} represents the vanishing of the $d_1$ differential on the class $[w_0]$.

On the $E_2$ page, we have the $d_2$ differentials:
\begin{align}
d_2^{p,q}: H^p[G,H^q[A,M]]\rightarrow H^{p+2}[G,H^{q-1}[A,M]].
\end{align}
Operationally, the obstruction to ``filtrate" this element $w_0^{p,q}$ to $E_3$ page  corresponds to looking for a new cochain $w_2^{p+2,q-2}$ such that  the next descent equation can be satisfied:
\begin{align}
d_2w_0:=\delta_2 w_0+\delta_1 w_1 + \delta_0 w_2 = 0.
\end{align}
Following the pattern, 
we can continue this operational definition for arbitary $E_r$-page, which formally given by Eq.\ref{eqn:iso_ss} (see the Fig.\ref{fig:Er_differential}(d) for the $d_3$ illustration).
 Starting from $E_0$ page, the representative elements $w_0^{p,q}$ survives to $E_{r+1}$-page must satisfy the following  conditions: there exist a additional multiple cochain $(w_1^{p+1,q-1},w_2^{p+2,q-2},...,w_r^{p+r,q-r})$, such that the following equations satisfy
\begin{align}
&d_0 w_0:=\delta_0w_0=0\\
&d_1 w_0:=\delta_1 w_0 \delta _0 w_1=0\\
&\qquad \qquad ...\\
&
d_r w_0:= \delta_r w_{0}+\delta_{r-1}w_1+...\delta_0 w_0=0.
\label{eqn:sequence_cond_general}
\end{align}
If for a large enough $r$ the condition $d_{r}w_{0}=0$ is satisfied over the entire $E_r$, then the elements in $E_r$ automatically become the elements in $E_{r+1}$ which means that we have $E_{r+1}=E_r$, and all the higher pages are the same. We say that the sequence stablizes at $E_r$, or we have reached the $E_\infty$ page. For the LHS spectral sequence, the $E_\infty$ is isomorphic to the group cohomology $H^n[K, M]$ as a set. In order to fully recover the group $H^n[K, M]$ one also needs to understand the group multiplication structure on $E_\infty$, which is called the extension problem.

For a fixed $n$, a general cocycle in $H^n[K, M]$ can be constructed from cochains in $E_0^{p,n-p}$, which, in order  to satisfy the cocycle condition these cochains, must satisfy conditions which schematically take the form of Eq.~\eqref{eqn:sequence_cond_general}. However, the actual expressions involve ``cochain-level'' differentials instead of differentials mapping between cohomology classes.  Remarkably, all the differentials on the cochain level for the the lower dimensional (up to $n=4$) are explicitly obtained in Ref.\cite{wang2021domain}.

For the AH spectral sequence for FSPT, whose fermionic symmetry are subject to the following short exact sequence
\begin{align}
1\rightarrow \Z_2^f\rightarrow G_f \rightarrow G\rightarrow 1,
\end{align}
 now we start from the $E_2$ page whose elements now are replacing the $H^q(A,M)$ by $h^q(\Z_2^f)$ in Eq.\eqref{eq:E2_page_LHS}, namely 
 \begin{align}
     E_2^{p,q}=H^p[G, h^q(\Z_2^f)],
	\label{eq:E2_page_LHS}
 \end{align}
 where $h^q(\Z_2^f)$ is the classification of $(q-1)$D fermionic invertible topological orders.
 Physically, it replaces the decoration of $A$ bosonic SPT by fermionic  invertible topological phases. For $n=3$, all the differentials $d_r$ on the cochain level with $r\ge 2$ are know. Now in this paper, we obtain all the necessary cochain-level differentials for the $n=4$, especially the explicit form of $d_3^{2,2}$ that is not known before. (Note that the explicit form of $d_3^{1,3}$ and $d_4^{1,3}$ are still  not known but they does not affect us to obtain the classification of 3D FSPT.) Similar to  the $E_\infty$ page of LHS spectral sequence will give the  group cohomology $H^{n}(K, M)$, the infinite page $E_\infty$ page of AH spectral sequence with $h^q(\Z_2^f)$ will give the $(n-1)$D FSPT.

\section{Review of 2+1D Fermionic SPT and trivialization}

In this section, we give a brief review to the 2+1D interacting FSPT (also including the invertible topological order) protected by internal symmetries, which will be used as the trivialization of our 3+1D FSPT.

\subsection{2+1D FSPT}
\label{sec:2DSPT}
\begin{figure}[h]
\centering
   \includegraphics[width=0.8\linewidth]{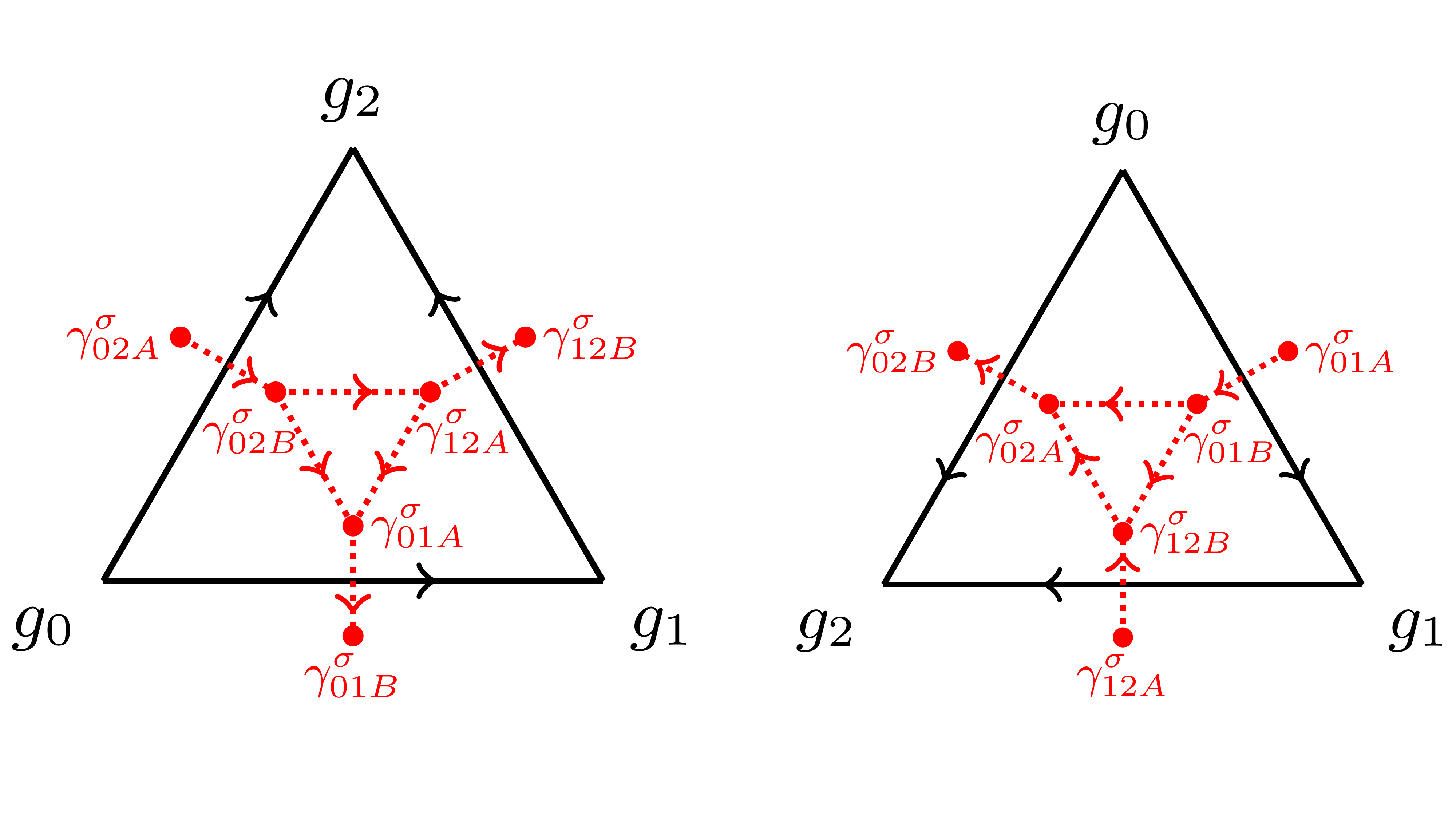} 
    \caption{Kasteleyn orientation corresponding to the positive and negative triangle of $\mathcal{T}$.}
   \label{fig:2DKasteleyn}
\end{figure}

Here we briefly review the decoration domain wall framework to classify the 2+1D FSPT \cite{Wang2020}. Consider a triangulation $\mathcal{T}$ of 2D space manifold, we can obtain a polyhedral decomposition $\mathcal{P}$ of $\mathcal{T}$ by adding a point in the center of each triangle of $\mathcal{T}$ and connecting the three points  around it. Then we can obtain the dual lattice  $\tilde{\mathcal{P}}$ of $\mathcal{P}$. We can adding direction to each link of $\tilde{\mathcal{P}}$ so that  $\tilde{\mathcal{P}}$ is Kasteleyn oriented. Differing from 3D case, the 2D $\tilde{\mathcal{P}}$ can have global Kasteleyn orientation, namely for each loops, no matter how large, it is always Kasteleyn oriented.\cite{Wang2018} One can see to the orientation locally in Fig.\ref{fig:2DKasteleyn}. With these geometry information at hand, now we can discuss the domain wall decoration for 2+1D FSPT with internal symmetry discrete $G_f = (G_b, \omega_2, s_1)$.

\begin{figure}[h]
\centering
   \includegraphics{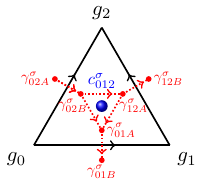} 
   \caption{Degree of freedom for 2+1D FSPT}
   \label{fig:2Ddof}
\end{figure}

\textit{Degree of freedom} As see in Fig.\ref{fig:2Ddof}, for each vertex of $\mathcal{T}$, we assign $|G_b|$ states that is label by $|g\rangle$ for $g \in G_b$. On each center of the triangle of $\mathcal{T}$, we assign $|G_b|$ species of complex fermions, denoted as $c_{ijk}^\sigma$ for $\sigma \in G_b$. For each link of $\mathcal{T}$ we also assign $|G_b|$ species of complex fermions $a_{ij}^\sigma$, each of which can be decomposed into two Majorana femions $a_{ij}^{\sigma }=( \gamma _{ij,A}^{\sigma } +i\gamma _{ij,B}^{\sigma })/2$. We always assign the $A$(B)-type Majorana at the bottom (top) to the red arrow that is dual to the link $\langle ij \rangle$. 

\textit{Symmetry transformation} These degrees of freedom should transform suitably under the symmetry action $U( g)$ defined as follows for any $g, \sigma \in G_{b}$: 
\begin{align}
U( g)\ket{g_{i}} &=\ket{gg_{i}} \label{eq:2D_symmetry1}\\
U( g) c_{ijk}^{\sigma } U^{\dagger }( g) &=( -1)^{\omega _{2}( g,\sigma )} c_{ijk}^{g\sigma }\label{eq:2D_symmetry2}\\
U( g) a_{ij}^{\sigma } U^{\dagger }( g) &=( -1)^{\omega _{2}( g,\sigma )} a_{ij}^{g\sigma }.\label{eq:2D_symmetry3}
\end{align}
The last one leads to the transformation for the Marjoana fermion as follows.
\begin{align}
U( g) \gamma _{ij,A}^{\sigma } U^{\dagger }( g) &=( -1)^{\omega _{2}( g,\sigma )} \gamma _{ij,A}^{g\sigma } ,\label{eq:sym trans Majorana A 2D}\\
U( g) \gamma _{ij,B}^{\sigma } U^{\dagger }( g) &=( -1)^{\omega _{2}( g,\sigma ) +s_{1}( g)} \gamma _{ij,B}^{g\sigma } .\label{eq:sym trans Majorana B 2D}
\end{align}

\textit{Decoration rules} With these degrees of freedoms, we specify the decoration rules. There are two layers of fermionic decoration, that is Kitaev chain and complex fermion decoration. Given a patterns of $G_b$ vertices, we use the $n_1(g_i,g_j)$ for specify whether there is a Kitaev chain passing through the link $\langle ij\rangle$. If $n_1(g_i,g_j)=0$, all the Majorana fermions at the link $\langle ij\rangle $ are in vacuum pairing, while if $n_1(g_i,g_j)=1$, all the species of Majorana fermions at the link $\langle ij\rangle $ are in vacuum pairing except the species $\gamma_{ijA}^{g_i}$ and $\gamma_{ijB}^{g_i}$ that will form a Kitaev chain with the Majorana fermions from other links. 

 According to above decoration rules, the number of decorated Kitaev chains going through the links of a given triangle $\langle ijk \rangle$ is given by
\begin{align}
dn_1(g_i,g_j,g_k)=n_1(g_j,g_k)-n_1(g_i,g_k)+n_1(g_i,g_j)
\end{align}
As we want construct a gapped FSPT state without intrinsic topological order, there should not be dangling Majorana zero mode in each triangle. This will lead to the consistent condition
\begin{align}
dn_1=0
\end{align}

To specify the position of two Majorana fermions in a non-trivial Majorana pairing, we first consider the standard  triangle whose first vertex has identity element in $G_b$. In this case, we can pair the Majorana fermions inside the triangle according to the links that connect them, namely put the Majorana fermions at the bottom of the link in front in the pairing. For the nonstandard triangles, we abtain them from the standard one by symmetry transformation as follows
 \begin{align}
    \raisebox{-0.5\height}{\includegraphics[width=0.20\textwidth]{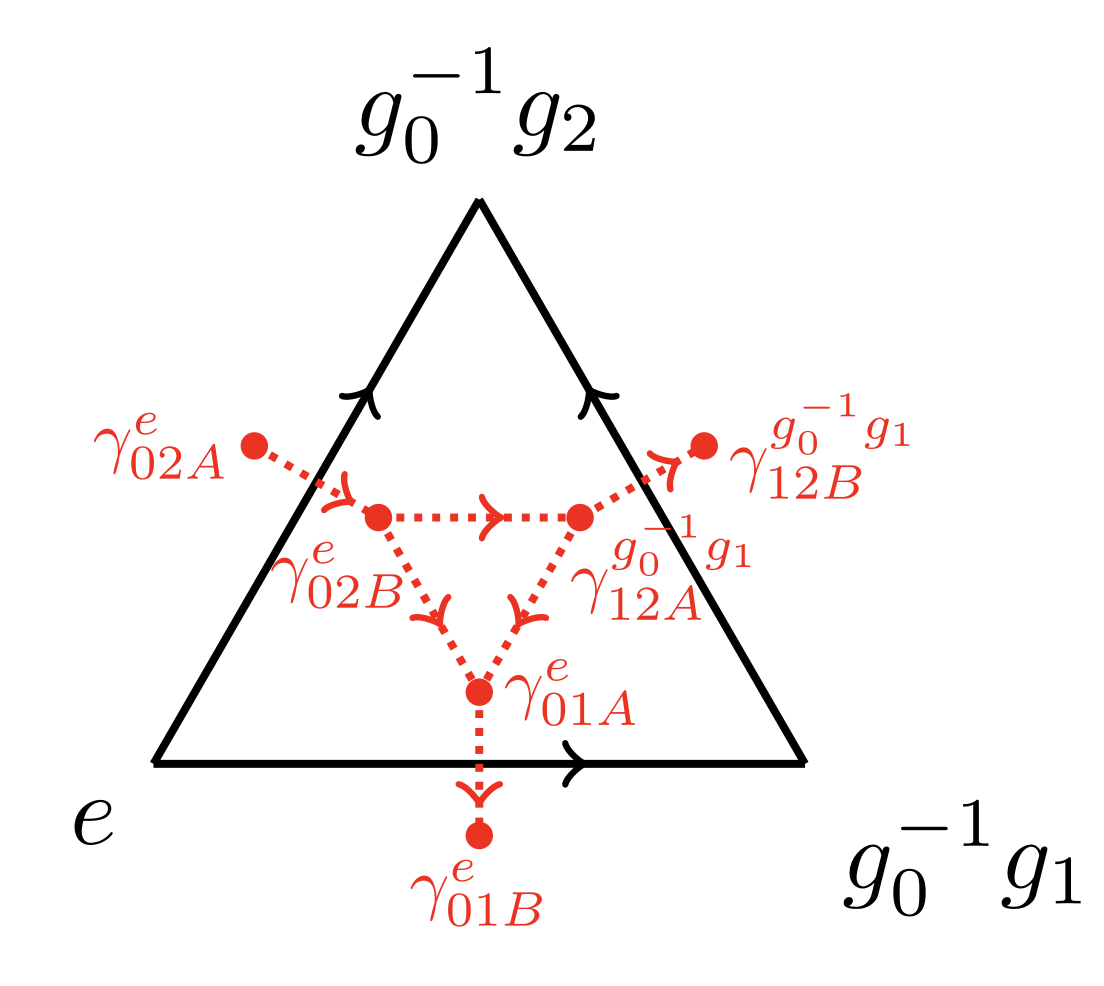}} \underrightarrow{\ \ U(g_0)\ \ } \raisebox{-0.5\height}{\includegraphics[width=0.20\textwidth]{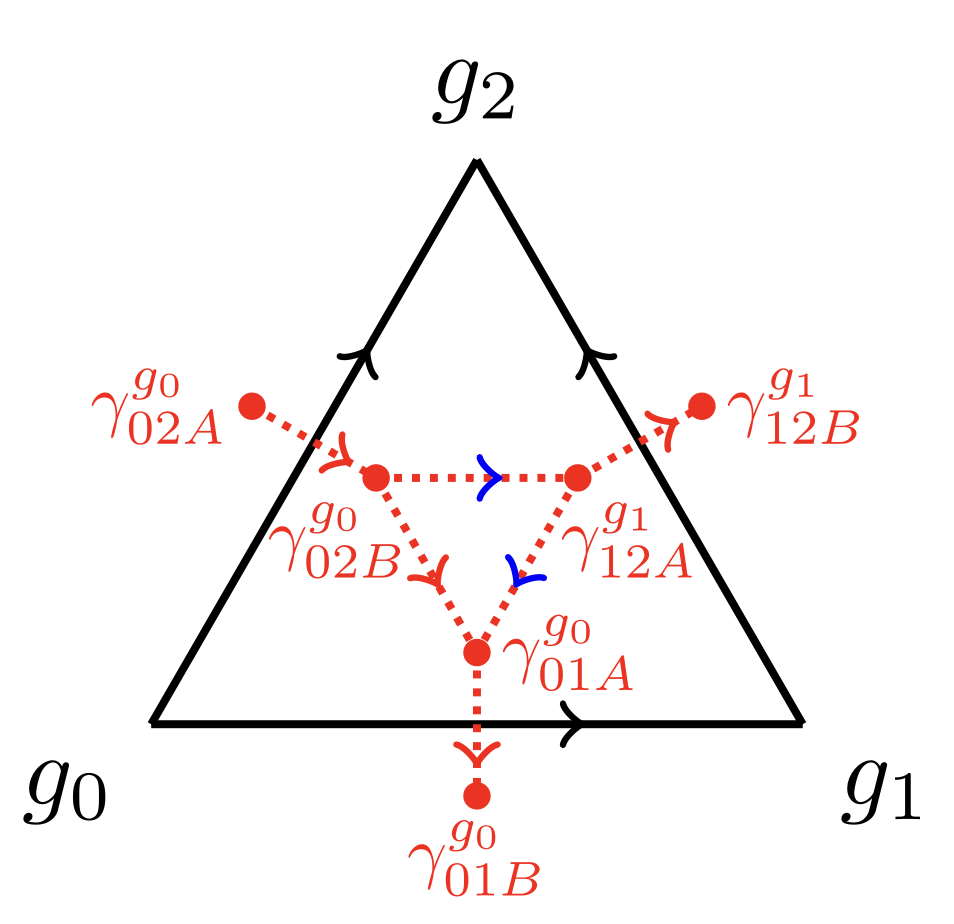}}
    \label{eq:2D_symm_tranf}
\end{align}
According to the symmetry transformation Eq.(\ref{eq:sym trans Majorana A 2D}) and Eq.(\ref{eq:sym trans Majorana B 2D}), there would be some minus sign in the pairing operators as indicated by the blue arrow  in (\ref{eq:2D_symm_tranf}). In general, there are three possible non-trivial pairings in the triangle $\langle 012\rangle$, characterized by Majorana pairing projection operators
\begin{widetext}
\begin{align}\label{2D:02B01A}
P_{02B,01A}^{g_0,g_0} &= U(g_0) P_{02B,01A}^{e,e} U(g_0)^\dagger
= \frac{1}{2}\,\big(1-i\gamma_{02B}^{g_0}\gamma_{01A}^{g_0}\big),\\
P_{02B,12A}^{g_0,g_1} &= U(g_0) P_{02B,12A}^{e,g_0^{-1}g_1} U(g_0)^\dagger
= \frac{1}{2} \left[1-(-1)^{\omega_2(g_0,g_0^{-1}g_1)}i\gamma_{02B}^{g_0}\gamma_{12A}^{g_1}\right],\\\label{2D:12_01}
P_{12A,01A}^{g_1,g_0} &= U(g_0) P_{12A,01A}^{g_0^{-1}g_1,e} U(g_0)^\dagger
= \frac{1}{2} \left[1-(-1)^{\omega_2(g_0,g_0^{-1}g_1)+s_1(g_0)}i\gamma_{12A}^{g_1}\gamma_{01A}^{g_0}\right].
\end{align}
\end{widetext}
Among these non-trivial pairings, the last two may change their directions in the non-standard triangle, which  are indicated by blue arrows in the right-hand-side of (\ref{eq:2D_symm_tranf}).

For each triangle $\langle ijk\rangle \in \mathcal{T}$, we can specify the decoration of complex fermion on it by the function $n_2(g_i,g_j,g_k)$. If $n_2(g_i,g_j,g_k)=1$, there is one complex fermion $c_{ijk}^{g_i}$ is occupied on the center of the triangle and others are in vacuum. If $n_2(g_i,g_j,g_k)=0$, all complex fermions are in vacuum.

\textit{Fermionic F move}  As we want to construct the fixed point wavefunction for 2+1D FSPT,  the retriangulation process (i.e., Pachner move) should induce a FSLU transformation between the two states before and after the move, which lead to the so-called $F$ move. We can also first consider the standard $F$ move as given by
\begin{widetext}
\begin{align}
    \psi\left(\raisebox{-0.5\height}{\includegraphics[width=0.20\textwidth]{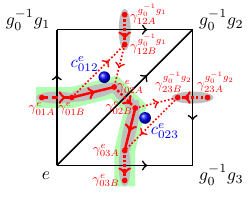}}\right)=F(e,\bar 0 1, \bar 0 2, \bar 0 3) \psi\left(\raisebox{-0.5\height}{\includegraphics[width=0.20\textwidth]{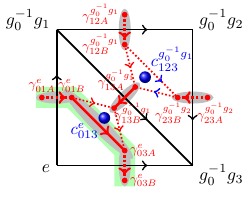}}\right)
    \label{eq:2DFmove}
\end{align}
\begin{align}
F(e,\bar 0 1, \bar 0 2, \bar 0 3)
= \nu_3(\bar 01,\bar 12,\bar 23) \big(c_{012}^{e\dagger}\big)^{ n_2(012)} \big(c_{023}^{e\dagger}\big)^{ n_2(023)} \big(c_{013}^{e}\big)^{n_2(013)} \left(c_{123}^{g_0^{-1}g_1}\right)^{n_2(123)} X_{0123}[n_1]
\end{align}
where  $\nu_3(\bar 01,\bar 12,\bar 23):= \nu_3(e,\bar 0 1, \bar 0 2, \bar 0 3)$  and we used $\bar ij$ to denote $g_i^{-1}g_j$
\end{widetext}
and \begin{align}
X_{0123}[n_1]
& = P_{0123}[n_1] \cdot \left(\gamma_{23B}^{g_0^{-1}g_2}\right)^{\alpha(0123)}
\end{align}
The normalized projection operator $ P_{0123}[n_1]$ is similar to that in Eq.(\ref{eq:X}) and now there is only one inserted Majorana fermions for the flipping the fermion parity sectors of  the fermions of the transition loop. 

The $F$ move should preserve the total fermion parity. The fermion parity change from the complex fermion decoration is given by $dn_2(0123)$ while that from the Kitaev chain decoration by counting through the transition loop is given by $\alpha(0123)=[\omega_2\cup n_1 +s_1\cup n_1 \cup n_1](0123)$. Therefore, we have the following consistent condition
\begin{align}
dn_2(0123)=[\omega_2\cup n_1 +s_1\cup n_1 \cup n_1](0123).
\end{align}

The nonstandard $F$ move where the first vertex $g_0\neq e$ can be obtained by applying the symmetry transformation according to Eq.(\ref{eq:2D_symmetry1})-(\ref{eq:2D_symmetry3}).  
\begin{align}\label{2D:Fg}
F(g_0,g_1,g_2,g_3) = { }^{g_0}F(e,g_0^{-1}g_1,g_0^{-1}g_2,g_0^{-1}g_3) \nonumber \\
:= U(g_0) F(e,g_0^{-1}g_1,g_0^{-1}g_2,g_0^{-1}g_3) U(g_0)^\dagger.
\end{align}
The $F$ moves constructed in this way are automatically symmetric, because we can derive the transformation rule
\begin{align}
F(gg_0,gg_1,gg_2,gg_3) = U(g) F(g_0,g_1,g_2,g_3) U(g)^\dagger,
\end{align}
using the fact $U(g)$ form the linear representation of $G_f$ and $F$ is fermion parity even.

After a $U(g_0)$-action on the standard $F$ operator Eq.(\ref{eq:2DFmove}), we can obtain the generic $F$ symbol expression:
\begin{widetext}
\begin{align}\label{2D:F}
F(g_0,g_1,g_2,g_3)
= \nu_3(g_0,g_1,g_2,g_3) \big(c_{012}^{g_0\dagger}\big)^{ n_2(012)} \big(c_{023}^{g_0\dagger}\big)^{ n_2(023)} \big(c_{013}^{g_0}\big)^{n_2(013)} \big(c_{123}^{g_1}\big)^{n_2(123)} X_{0123}[n_1].
\end{align}
\end{widetext}

From on-site nature of the symmetry and the decoration rules of Majorana fermions and complex fermions, $n_1$ and $n_2$ should be invariant under symmetry actions. The generic homogeneous cochain $\nu_3$ in $F(g_0,g_1,g_2,g_3)$ is a combination of the inhomogeneous $\nu_3$ in the standard $F$ move and the $\pm 1$ signs appearing in the symmetry action. So we have the following symmetry conditions for $n_1$, $n_2$ and $\nu_3$:
\begin{widetext}
\begin{align}
n_1(g_0,g_1) &= n_1(e,g_0^{-1}g_1) = n_1(g_0^{-1}g_1),\\
n_2(g_0,g_1,g_2) &= n_2(e,g_0^{-1}g_1,g_0^{-1}g_2) = n_2(g_0^{-1}g_1,g_1^{-1}g_2),\\\label{2D:nu3_symm}
\nu_3(g_0,g_1,g_2,g_3) &= { }^{g_0}\nu_3(g_0^{-1}g_1,g_1^{-1}g_2,g_2^{-1}g_3)
= \nu_3(g_0^{-1}g_1,g_1^{-1}g_2,g_2^{-1}g_3)^{1-2s_1(g_0)} \cdot \mathcal O_4^\mathrm{symm}(g_0,g_1,g_2,g_3).
\end{align}
The symmetry sign $\mathcal O_4^\mathrm{symm}$ in the last equation is given by
\begin{align}
\mathcal O_4^\mathrm{symm}(g_0,g_1,g_2,g_3)
&= (-1)^{(\omega_2\cup n_2 + s_1\cup d n_2 )(g_0,g_0^{-1}g_1,g_1^{-1}g_2,g_2^{-1}g_3) +\omega_2(g_0,g_0^{-1}g_2) d n_2(g_0^{-1}g_1,g_1^{-1}g_2,g_2^{-1}g_3)},
\label{2D:Osymm}
\end{align}
\end{widetext}

\begin{figure*}
   \centering
   \includegraphics{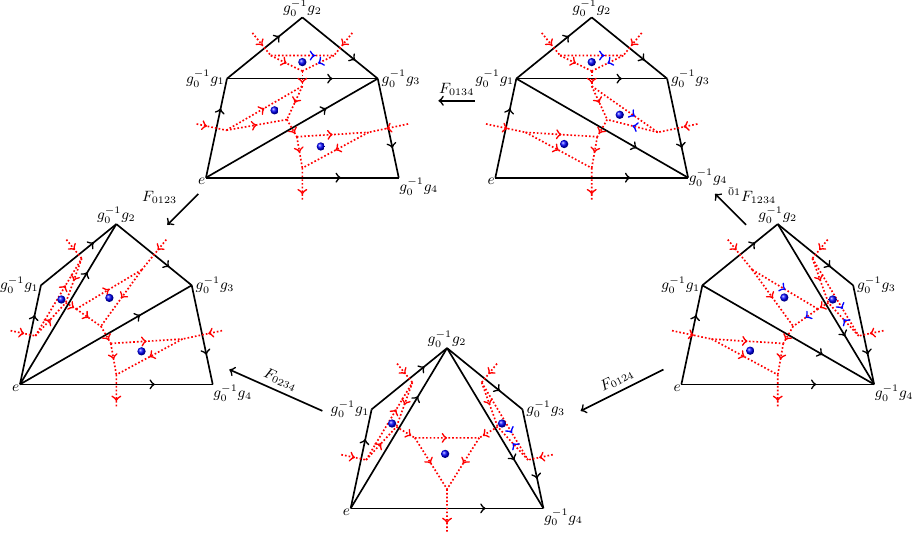} 
   \caption{Superpentagon equation for 2+1D fermionic $F$ move}
   \label{fig:2Dpentagon}
\end{figure*}

\textit{Superpentagon condition} Finally, we need to consider the super-pentagon condition for the fermionic $F$ move to ensure the consistency. Again, we can only need to consider the standard one with the first vertex equal to identity since other nonstandard $F$ move can be obtain via symmetry transformation. The standard super-pentagon equation is algebraically given by (as see Fig.\ref{fig:2Dpentagon})
\begin{align}
&F(e,\bar 02,\bar 03,\bar 04) \cdot F(e,\bar 01,\bar 02,\bar 04)\nonumber \\
&= F(e,\bar 01,\bar 02,\bar 03) \cdot F(e,\bar 01,\bar 03,\bar 04) \cdot F(\bar 01,\bar 02,\bar 03,\bar 04)\nonumber \\
&= F(e,\bar 01,\bar 02,\bar 03) \cdot F(e,\bar 01,\bar 03,\bar 04) \cdot { }^{\bar 01}F(e,\bar 12,\bar 13,\bar 14),
\end{align}
 Note that only the last $F$ symbol is non-standard in the above equation. The will lead the condition on $\nu_3$ as 
\begin{align}
d \nu_3 = \mathcal O_4[n_2]=\mathcal O_4^\mathrm{symm}[n_2] \cdot \mathcal O_4^{c}[n_2] \cdot \mathcal O_4^{c\gamma}[\dd n_2] \cdot \mathcal O_4^{\gamma}[d n_2].
\end{align}
The first term comes from the nonstandard $F$ move of 
 ${ }^{\bar 01}F(e,\bar 12,\bar 13,\bar 14)$ which have factor determined by Eq.(\ref{2D:Osymm}).  The second and third terms come from the anticommutation relation between the  fermion operators. And the last term comes from the Majorana fermions configuration and similarly to the $O_5^\gamma$ we derive here. In fact, using our general strategy discussed in Sec.\ref{sec:derivationO5} to derive it. However, the result can be showed to be equivalent to that in Ref.\cite{Wang2020} by counting only the typical (shortest) path by a coboundary transformation. The total expression of $\mathcal O_4[n_2]$ is given by
\begin{align}
&\mathcal O_4[n_2]\nonumber \\
&=( -1)^{[ \omega _{2} \cup n_{2} +n_{2} \cup n_{2} +n_{2} \cup _{1}\mathrm{d} n_{2} +\mathrm{d}( s_{1} \cup n_{2} +n_{2} \cup _{2}\mathrm{d} n_{2})]( 01234)}\nonumber\\
&\quad\times ( -1)^{\omega _{2}( 013)\mathrm{d} n_{2}( 1234) +\mathrm{d} n_{2}( 0124)\mathrm{d} n_{2}( 0234)}\nonumber \\
&\quad\times ( -i)^{\mathrm{d} n_{2}( 0123)[ 1-\mathrm{d} n_{2}( 0124)] (\mathrm{mod}\ 2)}.
\end{align}

\subsubsection{Classification of 2+1D FSPT}
To summarize, the classification of $2+1$D interacting FSPT protected by internal discrete symmetry $G_f = (G_b, \omega_2, s_1)$ is characterized by a triplet $(n_1,n_2,\nu_3)$:  
\begin{enumerate}
\item  $n_1 \in C^1(G_b, \mathbb Z_2)$ represents the 1+1D Kitaev chain decoration; 
\item  $n_2 \in C^2(G_b, \mathbb Z_2)$ represents the 0+1D complex fermion decoration;
\item  $\nu_3 \in C^3(G_b, U(1))$ represents the 2+1D bosonic SPT phase.
\end{enumerate}
The triplet $( n_{1} ,n_{2} ,\nu _{3})$ must satisfy the following conditions:
\begin{align}
\mathrm{d} n_{1} &=0 \mod 2 ,\\
\mathrm{d} n_{2} &=\omega _{2} \cup n_{1} +s_{1} \cup n_{1} \cup n_{1} \mod 2 ,\\
\mathrm{d} \nu _{3} &=\Gamma_{4}^{\mathrm{SPT}}[ n_{2}]( 01234)\nonumber\\\nonumber
&=( -1)^{[ \omega _{2} \cup n_{2} +n_{2} \cup n_{2} +n_{2} \cup _{1}\mathrm{d} n_{2} +\mathrm{d}( s_{1} \cup n_{2} +n_{2} \cup _{2}\mathrm{d} n_{2})]( 01234)}\\\nonumber
&\quad\times ( -1)^{\omega _{2}( 013)\mathrm{d} n_{2}( 1234) +\mathrm{d} n_{2}( 0124)\mathrm{d} n_{2}( 0234)}\\
&\quad\times ( -i)^{\mathrm{d} n_{2}( 0123)[ 1-\mathrm{d} n_{2}( 0124)] (\mathrm{mod}\ 2)}.
\end{align}

\subsection{2+1D fermionic invertible topological order}

The classification of $2+1$D interacting FSPT protected by internal discrete symmetry $G_f = (G_b, \omega_2)$ is characterized by a quadruplet $(n_0, n_1,n_2,\nu_3)$, where $n_0 \in C^0(G_b, \mathbb Z)$ represents the 2+1D $p+ip$ superconductor decoration, while the physical meaning of the remaining cochains are the same as above.

For odd $n_0$, the first layer of obstruction is 
\begin{align}
dn_1=\omega_2
\end{align}
This obstruction is non-trivial as long as $\omega_2$ is non-trivial in $H^2(G_b, \Z_2)$.  On the other hand, if $\omega_2$ is trivial, so this obstruction vanish and higher obstruction are possible, given by
\begin{align}
&dn_2=0\\
&d\nu_3=(-1)^{n_2\cup n_2}
\end{align} 

For $n_0=2k$, the layer of obstruction is given by
\begin{align}
&dn_1=0\\
&dn_2=\omega_2\cup n_1+\frac{n_0}{2}\omega_2\cup_1 \omega_2\\
&d\nu_3=\Gamma_4^{\mathrm{iTO}}[n_0,n_1,n_2]\nonumber \\
&\,\,\,\quad= (-1)^{\mathcal{P}(n_2)+n_2\cup \omega_2+n_1\cup[(n_1\cup \omega_2)\cup_2 \omega_2+n_1\cup \omega_2]}\nonumber \\
&\qquad\quad i^{n_1\cup n_1\cup \omega_2+n_0(n_1\cup \omega_2)\cup_2 (\omega_2\cup_1 \omega_2} e^{\frac{i\pi n_0}{8} \mathcal{P}(\omega_2)}.
\end{align}

\subsection{Trivialization}

The reason why we review the 2+1D obstruction functions is that they serve as the trivialization for the 3+1D FSPT. We refer the reader to Ref.\cite{Wang2020} for more detailed discussions. Here we present one example to illustrate the idea. Let's consider the case that the boundary anomalous FSPT with obstructed Kitaev chain decoration that trivialize the complex fermion decoration of 3+1D FSPT, given by 
\begin{align}
n_3=\omega_2\cup n_1+s_1\cup n_1\cup n_1.
\label{eq:n3_trivialization}
\end{align}
The 2+1D anomalous FSPT with Kitaev chain decoration can be constructed similarly in Sec.\ref{sec:2DSPT}. Through each links, we put almost one Kitaev chain specified by the decoration data $n_1\in H^1(G_b, \Z_2)$. We can count that the fermion parity change of the Majorana fermions under the boundary 2D $F$ move as that in Eq.(\ref{eq:2DFmove}) is again given by
\begin{align}
\alpha(g_0,g_1,g_2,g_3)=[\omega _{2} \cup n_{1} +s_{1} \cup n_{1} \cup n_{1}](g_0,g_1,g_2,g_3).
\end{align}
If $\alpha$ is a non-trivial 3-cocycle in $H^3(G_b, \Z_2)$, then this Kitaev chain decoration is obstructed in 2+1D FSPT since the total fermion parity is always broken in 2+1D no matter how we assign the complex fermion decoration. However, we can resolve it by putting it on the boundary of a 3+1D FSPT with complex fermion decoration $n_3\in H^3(G_b,\Z_2)$ specified by Eq.(\ref{eq:n3_trivialization}).  To see this, consider the $g_0,g_1,g_2,g_3$ are all connected to the both vertex $g_*$. So we can count the bulk complex fermion change in the 2+1D boundary $F$ move as
\begin{align}
n_3(g_*,g_0,g_1,g_2)+n_3(g_*,g_0,g_2,g_3)\nonumber \\
+n_3(g_*,g_0,g_1,g_3)+n_3(g_*,g_1,g_2,g_3)
\end{align} 
which exactly equals to $n_3(g_0,g_1,g_2,g_3)$ mod 2 using the condition $dn_3=0$ mod 2. So in total the boundary and bulk fermion parity are invariant under the boundary $F$ move. Since we construct a gapped symmetric ASPT state without topological order on the boundary of the 3D FSPT state, we conclude that the bulk FSPT with $n_3$ given by Eq.(\ref{eq:n3_trivialization}) is trivialized.




\bibliography{reference}

\end{document}